\documentclass[useAMS,usenatbib,a4paper]{mn2e}

\usepackage{amssymb,epsfig,color}
\usepackage{amsmath}
\usepackage{graphicx}

\usepackage[hyperindex,breaklinks=true, colorlinks, citecolor=blue]{hyperref}  
 \usepackage{flushend} 

\def\reff@jnl#1{{\rm#1\/}}

\def\aj{\reff@jnl{AJ}}                  
\def\araa{\reff@jnl{ARA\&A}}            
\def\apj{\reff@jnl{ApJ}}                
\def\apjl{\reff@jnl{ApJ}}               
\def\apjs{\reff@jnl{ApJS}}              
\def\ao{\reff@jnl{Appl.Optics}}         
\def\apss{\reff@jnl{Ap\&SS}}            
\def\aap{\reff@jnl{A\&A}}               
\def\aapr{\reff@jnl{A\&A~Rev.}}         
\def\aaps{\reff@jnl{A\&AS}}             
\def\azh{\reff@jnl{AZh}}                        
\def\baas{\reff@jnl{BAAS}}              
\def\jrasc{\reff@jnl{JRASC}}            
\def\memras{\reff@jnl{MmRAS}}           
\def\mnras{\reff@jnl{MNRAS}}            
\def\nar{\reff@jnl{New Astronomy Reviews}}            
\def\pra{\reff@jnl{Phys.Rev.A}}         
\def\prb{\reff@jnl{Phys.Rev.B}}         
\def\prc{\reff@jnl{Phys.Rev.C}}         
\def\prd{\reff@jnl{Phys.Rev.D}}         
\def\prl{\reff@jnl{Phys.Rev.Lett}}      
\def\pasa{\reff@jnl{PASA}}              
\def\pasp{\reff@jnl{PASP}}              
\def\pasj{\reff@jnl{PASJ}}              
\def\qjras{\reff@jnl{QJRAS}}            
\def\skytel{\reff@jnl{S\&T}}            
\def\solphys{\reff@jnl{Solar~Phys.}}    
\def\sovast{\reff@jnl{Soviet~Ast.}}     
 \def\ssr{\reff@jnl{Space~Sci.Rev.}}     
\def\zap{\reff@jnl{ZAp}}                        
\def\nat{\reff@jnl{Nature}}             

\title[C-BASS: Constraining diffuse radio emission in the NCP region]{The C-Band All-Sky Survey (C-BASS): Constraining diffuse Galactic radio emission in the North Celestial Pole region}
\author[C. Dickinson et al.]{C.\,Dickinson,$\!^{1,2}$\thanks{\url{E-mail: Clive.Dickinson@manchester.ac.uk}} A.\,Barr,$\!^{1}$ H.\,C.\,Chiang,$\!^{3,4}$ C.\,Copley,$\!^{5,6,7}$ R.\,D.\,P.\,Grumitt,$\!^{7}$ 
\newauthor  S.\,E.\,Harper,$\!^{1}$ H.\,M.\,Heilgendorff,$\!^{4}$ L.\,R.\,P.\,Jew,$\!^{7}$  J.\,L.\,Jonas,$\!^{5,6}$  Michael\,E.\,Jones,$\!^{7}$ 
\newauthor J.\,P.\,Leahy,$\!^{1}$ J.\,Leech,$\!^{7}$  E.\,M.\,Leitch,$\!^{2}$ S.\,J.\,C.\,Muchovej,$\!^{2}$ T.\,J.\,Pearson,$\!^{2}$ M.\,W.\,Peel,$\!^{8,1}$ 
\newauthor A.\,C.\,S.\,Readhead,$\!^{2}$ J.\,Sievers,$\!^{3,9}$ M.\,A.\,Stevenson,$\!^{2}$ Angela\,C.\,Taylor,$\!^{7}$  \\
$^{1}$Jodrell Bank Centre for Astrophysics, Alan Turing building, School of Physics and Astronomy, The University of Manchester, \\
Oxford Road, Manchester, M13 9PL, Manchester, U.K. \\
$^{2}$Cahill Centre for Astronomy and Astrophysics, California Institute of Technology, Pasadena, CA 91125, USA \\
$^{3}$Department of Physics, McGill University, 3600 Rue University, Montr\'{e}al, QC H3A 2T8, Canada \\
$^{4}$Astrophysics \& Cosmology Research Unit, School of Mathematics, Statistics \& Computer Science, University of KwaZulu-Natal, \\
Westville Campus, Private Bag X54001, Durban 4000, South Africa \\
$^{5}$Department of Physics and Electronics, Rhodes University, Grahamstown, 6139, South Africa \\
$^{6}$South African Radio Astronomy Observatory, 2 Fir Road, Observatory, Cape Town, 7925 , South Africa \\
$^{7}$Sub-department of Astrophysics, University of Oxford, Denys Wilkinson Building, Keble Road, Oxford OX1 3RH, U.K. \\
$^{8}$Departamento de F\'{i}sica Matematica, Instituto de F\'{i}sica, Universidade de S\~{a}o Paulo, Rua do Mat\~{a}o 1371, S\~{a}o Paulo, Brazil \\
$^{9}$School of Chemistry and Physics, University of KwaZulu-Natal, Westville Campus, Private Bag X54001, Durban 4000, South Africa
}

\begin{document}

\date{Accepted XXX. Received YYY; in original form ZZZ}

\pagerange{\pageref{firstpage}--\pageref{lastpage}} \pubyear{2014}

\maketitle

\label{firstpage}


\begin{abstract}
The C-Band All-Sky Survey (\mbox{C-BASS}) is a high-sensitivity all-sky radio survey at an angular resolution of 45\,arcmin and a frequency of 4.7\,GHz. We present a total intensity map of the North Celestial Pole (NCP) region of sky, above declination $>+80^{\circ}$, which is limited by source confusion at a level of $\approx 0.6$\,mK rms. We apply the template-fitting (cross-correlation) technique to {\it WMAP} and {\it Planck} data, using the C-BASS map as the synchrotron template, to investigate the contribution of diffuse foreground emission at frequencies $\sim 20$--40\,GHz. We quantify the anomalous microwave emission (AME) that is correlated with far-infrared dust emission. The AME amplitude does not change significantly ($<10\,\%$) when using the higher frequency \mbox{C-BASS} 4.7\,GHz template instead of the traditional Haslam 408\,MHz map as a tracer of synchrotron radiation. We measure template coefficients of $9.93\pm0.35$ and $9.52\pm0.34\,$K per unit $\tau_{353}$ when using the Haslam and \mbox{C-BASS} synchrotron templates, respectively. The AME contributes $55\pm2\,\mu$K rms at 22.8\,GHz and accounts for $\approx 60\%$ of the total foreground emission. Our results show that a harder (flatter spectrum) component of synchrotron emission is not dominant at frequencies $\gtrsim 5$\,GHz; the best-fitting synchrotron temperature spectral index is $\beta=-2.91\pm0.04$ from 4.7 to 22.8\,GHz and $\beta=-2.85 \pm 0.14$ from $22.8$ to 44.1\,GHz. Free-free emission is weak, contributing $\approx 7\,\mu$K rms ($\approx 7\%$) at 22.8\,GHz. The best explanation for the AME is still electric dipole emission from small spinning dust grains.
\end{abstract}

\begin{keywords}
surveys -- radiation mechanism: non-thermal -- radiation mechanism: thermal -- diffuse radiation --  radio continuum: ISM.  
\end{keywords}


\section{Introduction}

Diffuse radio foreground emission is a useful tool for studying the various components of the interstellar medium (ISM), including cosmic ray electrons via synchrotron radiation \citep{Lawson1987,Strong2011,Orlando2013} and warm ionized medium (WIM) via free-free emission \citep{Davies2006,Jaffe2011,PEP_XXI,Alves2012}. Understanding their detailed spatial and spectral characteristics is also important for removing them from cosmic microwave background (CMB) data \citep{Leach2008,Dunkley2009a,Delabrouille2009,Armitage-Caplan2012,Errard2012,PIP_XII,Planck2015_IX,Remazeilles2016}. 

An additional component, referred to as Anomalous Microwave Emission (AME), has  been detected at frequencies $\sim$10--60\,GHz \citep{Kogut1996,Leitch1997,Banday2003,Lagache2003,deOliveira-Costa2004,Davies2006,Kogut2011,Gold2011,Ghosh2012}; see \cite{Dickinson2018} for a recent review. This emission does not appear to correlate with low radio frequency data, such as the 408\,MHz map by \cite{Haslam1982}, which rules out steep-spectrum synchrotron emission as a cause. Similarly, analyses have shown that AME does not strongly correlate with H$\alpha$ data (e.g., \citealt{Dickinson2003,Finkbeiner2003}), which rules out free-free emission from warm ($T_e \approx 10^4$\,K) ionized gas. However, AME is remarkably well-correlated with far-infrared (FIR) and sub-mm maps \citep{Leitch1997, Miville-Deschenes2005} which trace interstellar dust grains in the ISM. \cite{Draine1998a} revisited the theory of electric dipole radiation from small spinning dust grains (``spinning dust''), originally postulated by \cite{Erickson1957}, and showed that spinning dust can naturally account for AME and explain the close correlation with FIR emission. Since then, there has been considerable evidence for spinning dust emission from molecular clouds and H{\sc ii} regions \citep{Finkbeiner2002,Casassus2006,Casassus2008,Dickinson2009a,Dickinson2010, Scaife2009}. The best examples are diffuse clouds within the Perseus \citep{Watson2005} and $\rho$~Ophiuchi \citep{Casassus2008} regions, which have high precision spectra showing the characteristic ``bump'' (in flux density) at a frequency of $\approx 30$\,GHz and can be fitted by plausible physical models for the spinning dust grains \citep{PEP_XX,Tibbs2011}. A survey of bright Galactic clouds in the {\it Planck} data \citep{PIP_XV} has detected a number of potential candidates but follow-up observations at higher resolution are required to confirm them.

The origin of the diffuse AME found at high Galactic latitudes is still not clear \citep[e.g.,][]{Hensley2016}. Although spinning dust can readily account for the bulk of the AME \citep[e.g., ][]{Planck2015_X}, other emission mechanisms could be contributing \citep{Planck2015_XXV}. Magnetic dipole radiation from fluctuations in dust grain magnetization could be significant \citep{Draine1999,Draine2013,Hensley2016,Hoang2016}, although upper limits on AME polarization \citep{Lopez-Caraballo2011,Dickinson2011,Macellari2011,Rubino-Martin2012a,Genova-Santos2017} appear to indicate that this cannot account for the majority of the signal. Similarly, a harder (flatter spectrum) component of synchrotron radiation may also be responsible for AME, which was proposed by \cite{Bennett2003b} at the time of the first {\it WMAP} data release. The harder spectrum naturally explains the correlation with dust, since both are related to the process of star-formation. Furthermore, we already know that there are regions that have synchrotron spectral indices\footnote{We use brightness temperature spectral indices, given by the definition $T_b \propto \nu^{\beta}$, which are related to flux density spectral indices by $\alpha=\beta + 2$.} that are at $\beta \approx -2.5$ or flatter, both  supernova remnants \citep{Onic2013} and more diffuse regions such as the {\it WMAP}/{\it Planck} haze \citep{Finkbeiner2004,PIP_IX}.

A harder synchrotron component may have been missed when applying component separation methods to microwave data. The majority of AME detections from fluctuations at high Galactic latitudes have been made using the ``template fitting'' technique, i.e., fitting multiple templates for each foreground component to CMB data, accounting for CMB fluctuations and noise \citep{Kogut1996,Banday2003}. The synchrotron template is traditionally the 408\,MHz all-sky map \citep{Haslam1982}, or another low frequency template. However, data at these frequencies will naturally be sensitive to the softer (steeper spectrum) synchrotron emission, which has a temperature spectral index ($T \propto \nu^{\beta}$) $\beta \approx -3.0$ at frequencies $\gtrsim 5$\,GHz \citep{Davies1996,Davies2006,Kogut2007,Dunkley2009b,Gold2011}. This leads to a significant AME signal at $\sim$10--60\,GHz that is correlated with FIR templates, which cannot be accounted for by the Rayleigh-Jeans tail of dust emission. 

A hard synchrotron component of AME can be constrained (or ruled out) by using a higher radio frequency template of synchrotron emission. \cite{Peel2012} used the 2.3\,GHz southern-sky survey of \cite{Jonas1998} as a synchrotron template for the {\it WMAP} data and found that the dust-correlated AME component changed by only $\approx 7$\,\%, compared to using the 408\,MHz template. This suggests that the the bulk of the diffuse high latitude synchrotron emission is indeed steep ($\beta \approx -3.0$) above 2.3\,GHz, resulting in little change to the AME at 20--40\,GHz.

The C-Band All-Sky Survey (\mbox{C-BASS}) is a survey of the entire sky at 5\,GHz, in intensity and polarization, at a resolution of $45$\,arcmin \citep{King2010,Jones2018}. The frequency  chosen is ideal for CMB component separation studies, being much closer to observing frequencies used by CMB experiments, typically at 30\,GHz and higher. The northern survey \citep{King2014} observations are complete and will be described in forthcoming papers. First results from the bright Galactic plane emission have been presented by \cite{Irfan2015}.

In this paper we present a preliminary \mbox{C-BASS} intensity map at 5\,GHz of the North Celestial Pole (NCP) region, primarily based on the joint-fitting of spatial templates at a given frequency \citep[e.g.,][]{Davies2006}. The NCP area is known to have significant AME and FIR emission from the Polaris flare region, sometimes known as ``the duck" \citep{Davies2006}, having been studied in the first identification of AME \citep{Leitch1997}. The AME is relatively bright, while the synchrotron and free-free components appear to be weak, and their morphology is distinct from the AME/FIR emission. Furthermore, the older surveys of \cite{Haslam1982} at 408\,MHz and \cite{Reich1986} at 1.4\,GHz are clearly affected by varying zero-levels, which appear as stripes in these maps. Also, due to the C-BASS scan strategy, the NCP region is observed very deeply by the \mbox{C-BASS} northern telescope, and has an almost negligible level of instrumental noise.

Section~\ref{sec:obs} describes the \mbox{C-BASS} observations and data analysis. Maps are presented in Section~\ref{sec:maps}. The template-fitting results and foreground spectral energy distributions (SEDs) are presented in Section~\ref{sec:template_fitting}. A discussion of the results for each component is given in Section~\ref{sec:discussion}. Conclusions are summarized in Section~\ref{sec:conclusions}. 


\section{Observations and data analysis}
\label{sec:obs}

The \mbox{C-BASS} project is surveying the entire sky, in intensity and polarization, at a nominal frequency of 5\,GHz and an angular resolution of $\approx 45$\,arcmin \citep{Jones2018}. It uses two telescopes to obtain full-sky coverage; the northern telescope is a 6.1\,m Gregorian antenna situated at the Owens Valley Radio Observatory, California,  and the southern telescope is a 7.6\,m Cassegrain antenna situated at Klerefontein, the MeerKAT/SKA South Africa support base. The northern instrument was commissioned in 2009--2012 (Muchovej et al., in prep.) and observed routinely until 2015 April, after which the receiver was decommissioned. The southern instrument is currently carrying out the southern part of the survey.

\subsection{Observations}

The \mbox{C-BASS} scan strategy consists of 360\degr\ scans in azimuth at a constant elevation, at a speed of $ \approx 4\degr$ azimuth per
second. The majority of the northern survey data were taken at elevations of $37.\!^{\circ}2$ (therefore passing through the NCP) and $47.\!^{\circ}2$, with additional scans at $67.\!^{\circ}2$ and $77.\!^{\circ}2$ to improve coverage at intermediate declinations. For the NCP region, only data at the two lower elevations are relevant, since higher elevations do not pass through the NCP region. However, the higher elevation data are useful in reducing the effects of drifts in the data due to $1/f$ noise (Taylor et al., in prep.). 

The \mbox{C-BASS} receiver
is a continuous-comparison radiometer {\citep{King2014}}, which measures
the difference between sky brightness temperature and a stabilised 
resistive load. This architecture reduces receiver 1/\emph{f} noise, which would otherwise be indistinguishable
from variations on the sky. The receiver also implements a correlation polarimeter. However, no polarization data are used in this present analysis. In this paper we use data taken from the \mbox{C-BASS} North survey, covering the period 2012~November to 2015~March. 

\subsection{Data analysis}

The data have been processed using the latest version of the standard \mbox{C-BASS} data reduction pipeline. A detailed
description of the pipeline will be given in forthcoming papers. In brief, the reduction pipeline identifies and flags out events of radio
frequency interference, performs amplitude and polarization
calibration, and applies atmospheric opacity corrections. It also removes contamination due to microphonics caused by the cryocooler cycle, which introduce oscillations in the output signal at a frequency of 1.2\,Hz and harmonics thereof. Residual 1.2\,Hz contamination is estimated to be at a level comparable to the thermal noise. The pipeline also removes large-scale ground spillover. Ground templates are made for each day of observations by subtracting the current best sky model from the time-ordered data, masking the strongest regions of Galactic emission, and binning in azimuth. This azimuth template is then subtracted from the time-ordered data before mapping. As well as removing the majority of the ground emission, this process inevitably removes a small fraction of the right-ascension-averaged sky signal. This is a significant effect on large angular scales but is not expected to be a major issue in the small region at the NCP analysed here. This effect will be mitigated in future analyses by including data from the southern survey, which have very different ground contamination at a given sky position. The current analysis also uses data taken only when the Sun is below the horizon, and all data within $5^{\circ}$ of the Moon are also flagged. 

The northern receiver includes a noise diode which, when fired, injects a signal of
constant temperature. The pipeline calibrates the
intensity signal onto the noise diode scale, and the noise diode is
subsequently calibrated to the astronomical sources Cas\,A and Tau\,A. The calibrator flux densities are calculated from the spectral 
forms given in {\citet{Weiland2011}}. The noise diode temperature is stable to within \mbox{1 per cent} over
time periods of several months and atmospheric opacity corrections are typically $<1\,\%$, resulting in relative calibration uncertainties of $\approx 1\,\%$ \citep{Irfan2014}.

The northern receiver has a nominal bandpass of 4.5--5.5\;GHz
but in-band filters to remove terrestrial RFI reduce the effective
bandwidth to 0.499\;GHz with a central frequency of 4.783\;GHz. The effective 
observing frequency depends on the calibrator source and the source spectrum being observed
but we do not attempt to colour correct the data considered
in this paper as these corrections are of order 1 per cent. Since our main calibrator is Tau\,A ($\beta=-2.3$), and the bulk of the emission seen by \mbox{C-BASS} is steeper-spectrum synchrotron ($\beta \approx -3$), the effective frequency will be slightly lower than this by $\approx 0.05$--0.1\,GHz. We therefore assume an effective frequency of 4.7\,GHz throughout this paper. 

The initial absolute calibration of the \mbox{C-BASS} maps is computed in terms of antenna temperature, which corresponds to the sky convolved with the full \mbox{C-BASS} beam. Since there is a significant ($\approx 25\,\%$; \citealt{Holler2013}) amount of power outside of the main beam, the conversion to brightness temperature depends on the angular scale of interest. In previous work, a common way around this issue has been to correct the scale to the ``main beam'' (point source) or ``full beam'' (extended source) scale to account for power lost to the far sidelobes \citep[e.g., ][]{Reich1982,Jonas1998}. For a previous \mbox{C-BASS} paper \citep{Irfan2015}, we used a single factor of 1.124 to convert to an effective full-beam temperature scale for angular scales of a few degrees. 

For the analysis in this paper, we use the theoretical beam  calculated using the GRASP physical optics package \citep{Holler2013} to deconvolve the sidelobe response while smoothing to an effective Gaussian resolution of $1^{\circ}$ full width at half maximum (FWHM). This  produces maps with the correct resolution and smoothing function and  it also results in a brightness temperature scale that is not dependent on angular scale.  We find that the flux densities of bright sources agree with previous measurements to within $\approx 5\,\%$. Since in this analysis we do not take into account the effects of a finite bandpass (colour corrections), we will adopt $5\,\%$ as our absolute calibration uncertainty. With future \mbox{C-BASS} data releases, and taking into account colour corrections, we expect to be able to reach $\approx 1\,\%$ accuracy.

The \mbox{C-BASS} maps are made by the DEStriping CARTographer,
\textsc{Descart} {\citep{Sutton2010}}. \textsc{Descart} performs a maximum
likelihood fit to the contribution of 1/\emph{f} noise in our 
timestreams with a series of offsets to the signal baseline, in 
this case 5\,s long. Different scans are given the same optimal
$\sigma^{-2}$ weighting as is used in full maximum likelihood mapping,
and are treated as independent. Variances were estimated from the 
power spectrum of the data. White-noise correlation between the channels
is neglected for this work. The \mbox{C-BASS} data are
made into {\tt{HEALPix}} \citep{Gorski2005}\footnote{\url{http://healpix.sourceforge.net}} maps at \mbox{$N_{\rm side} = 256$}, corresponding to pixels $\approx 13.4$\,arcmin on-a-side.


\section{Maps}
\label{sec:maps}

\subsection{\mbox{C-BASS} NCP map}

\begin{figure}
\begin{center}
\includegraphics[width=0.4\textwidth,angle=0]{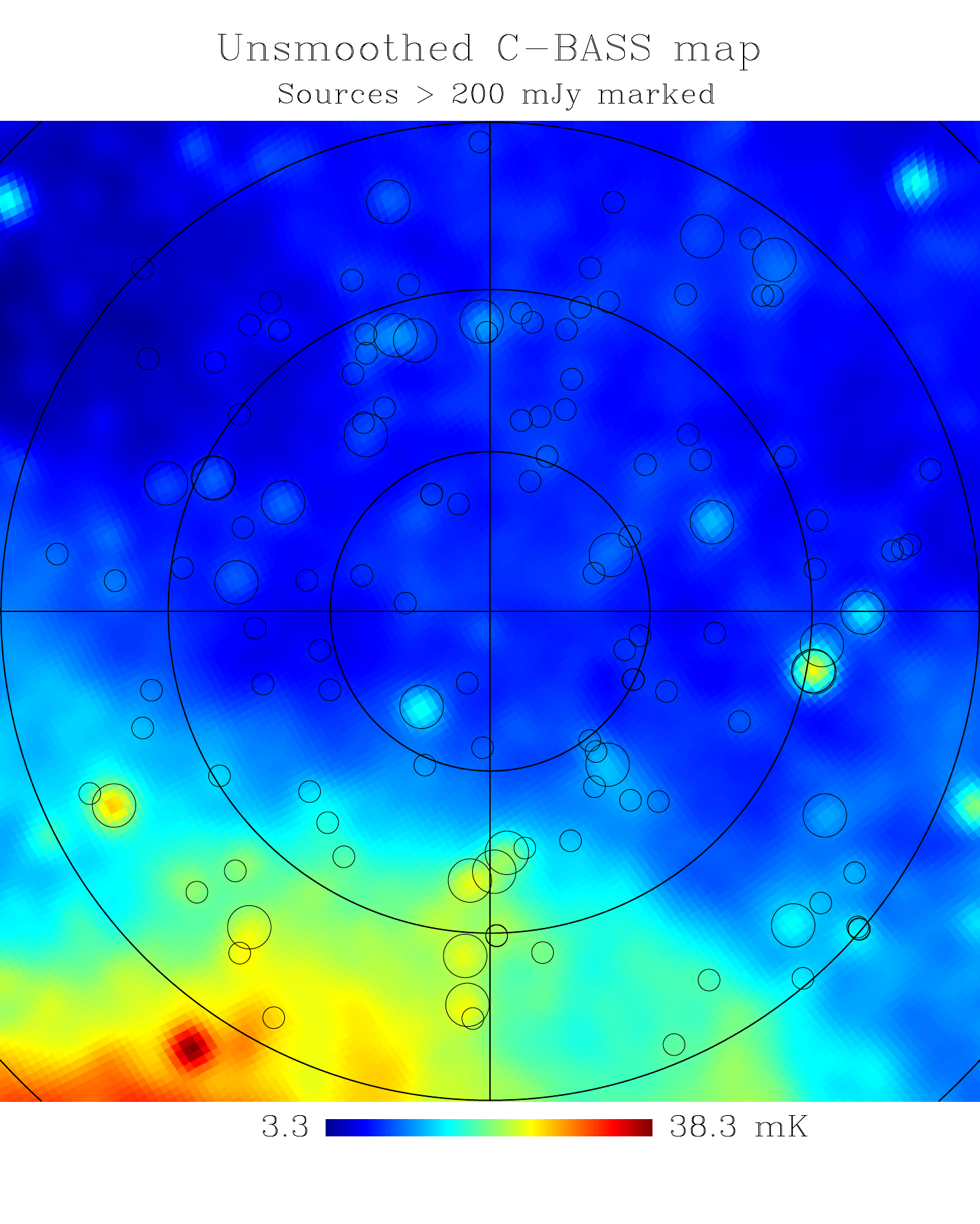}
\includegraphics[width=0.4\textwidth,angle=0]{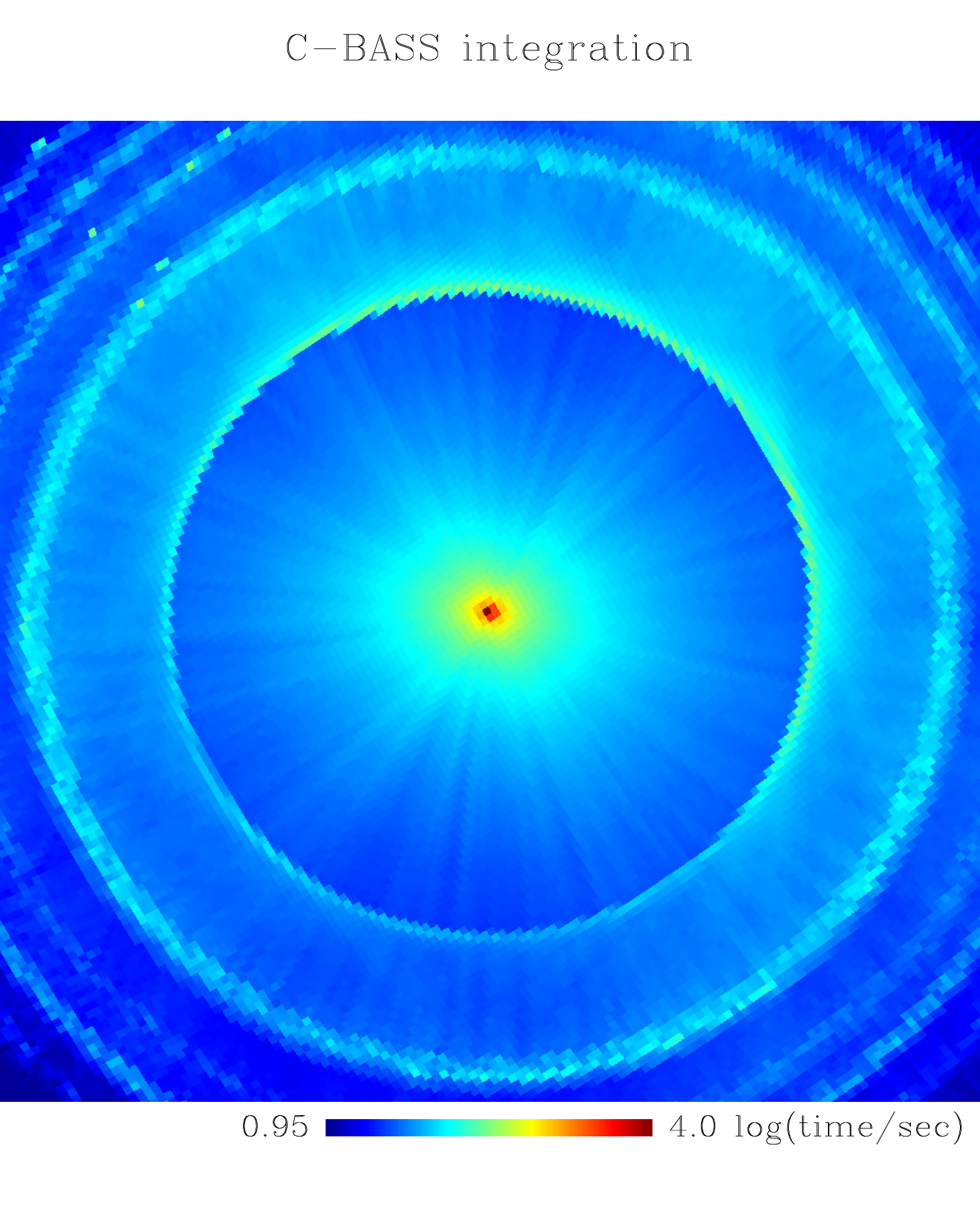}
\caption{{\it Top}: Unsmoothed \mbox{C-BASS} 5\,GHz \mbox{$N_{\rm side}=256$} map of the NCP region at an angular resolution of 45\,arcmin FWHM. The map covers a $30^{\circ} \times 30^{\circ}$ region centred at the NCP ($\delta=+90^{\circ}$) with R.A.=$0^{\circ}$ at the bottom of the map (R.A.=$180^{\circ}$ at the top) and increasing clockwise. The circular graticule lines mark $5^{\circ}$ intervals in declination. The colour scale is linear between the minimum/maximum values shown. Radio sources with flux densities $>200$\,mJy and $\delta > +75^{\circ}$ from the \protect\cite{Mingaliev2007} catalogue are indicated by circles, with larger circles for flux densities $>600$\,mJy. {\it Bottom}: Map of integration time (sec) per \mbox{$N_{\rm side}=256$} pixel on a logarithmic scale. The NCP region has the deepest integration in the \mbox{C-BASS} data. The circular rings correspond to the higher declination limits where the telescope is turning around, resulting in higher hit counts.}
\label{fig:cbassmap}
\end{center}
\end{figure}

The unsmoothed \mbox{C-BASS} 5\,GHz map of the NCP region is shown in the top panel of Fig.~\ref{fig:cbassmap}. The image is a gnomonic projection of the {\tt HEALPix} \mbox{$N_{\rm side}=256$} map, in units of mK (Rayleigh-Jeans), and with an angular resolution of 45\,arcmin FWHM. An approximate zero level has been removed by subtracting the global minimum in the map to ensure positivity; the results are insensitive to this global offset. The map contains hundreds of radio sources superposed onto diffuse Galactic emission; sources brighter than 200\,mJy at 4.85\,GHz and at $\delta > +75^{\circ}$ are marked with small circles. The majority of the radio sources are extragalactic and will be discussed further in Section~\ref{sec:sources}. The diffuse emission is expected to be primarily synchrotron radiation, with a contribution of free-free emission from our Galaxy. The Galactic plane is located towards the bottom of the map where the brightest emission is located.

The integration time for each \mbox{$N_{\rm side}=256$} ($\approx 13.4$\,arcmin on-a-side) pixel is shown in the lower panel of Fig.~\ref{fig:cbassmap}. The integration time in the NCP region is very high everywhere due to the constant elevation scans that repeatedly cross the NCP region. The median integration at $\delta >+80^{\circ}$ is $50$\,s per \mbox{$N_{\rm side}=256$} pixel, and much higher in the centre of the map. For the analyses at \mbox{$N_{\rm side}=64$} ($\approx 55.0$\,arcmin pixels), the integration time is effectively 16 times this value, i.e., $\approx 800$\,s. For the typical \mbox{C-BASS} rms noise level of $\approx 2.0\,$mK\,s$^{1/2}$ in intensity, this corresponds to a map rms noise level of $\approx 0.07$\,mK  or better. Compared to the Galactic signal of several mK, the instrumental noise is therefore effectively negligible in this region. The typical rms confusion noise from fluctuations in the background of radio sources at this angular resolution is $\sim0.6$\,mK \citep{Jones2018}. The effects of a large number of weak sources can be seen in the colder regions of the \mbox{C-BASS} map (Fig.~\ref{fig:cbassmap}). Bright ($ \gtrsim 200$\,mJy) extragalactic sources will be removed from the \mbox{C-BASS} map before analysis (see Sect.~\ref{sec:sources}).

The analysis was repeated on several versions of the \mbox{C-BASS} map, using a variety of data cuts and analysis procedures. Visual inspection of the maps showed the same structures at the same intensity level. Difference maps revealed low-level artifacts that were typically a few per cent or less of the signal of interest. The main contaminant was the Sun (via the far-out sidelobes of the \mbox{C-BASS} beam), which produced large-scale emission and scatter across the map, comparable to the Galactic emission in the NCP region. We therefore chose to use night-time only data, where this effect was eliminated. Nevertheless, we found that the results were consistent within the quoted uncertainties. 


\subsection{Extragalactic radio sources in the NCP region}
\label{sec:sources}

It is clear that there is a significant contribution from extragalactic sources in the \mbox{C-BASS} data. The large beam of \mbox{C-BASS} means that the point-source flux density sensitivity is modest and we are limited by source confusion. The brightest sources are easily discernible in the \mbox{C-BASS} map (Fig.~\ref{fig:cbassmap}). Fortunately, a number of high angular resolution radio surveys have been made of the region, and they can be used to either mask or remove the brightest sources.

The \cite{Kuhr1981b} S5 survey mapped the NCP region ($\delta>+70^{\circ}$) with the Effelsberg 100-m telescope, with 476 detected sources, and is complete down to 250\,mJy. Unfortunately, an electronic version of this catalogue is not currently available. The \cite{Kuhr1981a} 5\,GHz catalogue of bright ($>1$\,Jy) sources contains only 7 sources at $\delta >+80^{\circ}$. The Green Bank 6-cm radio source catalog \citep{Gregory1996} is a blind survey complete to $\approx 18$\,mJy but is limited to $\delta<+75^{\circ}$ and therefore is not useful here. The NRAO VLA All-Sky Survey \citep[NVSS;][]{Condon1998} at 1.4\,GHz is the highest frequency unbiased radio survey that covers $\delta>+88^{\circ}$; it contains hundreds of radio sources but the majority are $\sim 3$--100\,mJy and cannot be accurately extrapolated to 5\,GHz. \cite{Healey2009} surveyed the region at 4.85\,GHz but pre-selected sources that were flat spectrum. They measured 3 sources in this relatively unobserved region ($\delta>+88^{\circ}$) at 4.85\,GHz with flux densities of 67, 58 and 142\,mJy, respectively. \cite{Ricci2013} made 5--30\,GHz measurements of bright sources detected in the K-band Medicina pilot survey \citep{Righini2012}. 

The only other radio survey of the NCP region ($\delta>+75^{\circ}$) at frequencies above 1.4\,GHz is the RATAN-600 multi-frequency survey of \cite{Mingaliev2007}, which includes measurements at 4.8\,GHz. They observed 504 sources from the 1.4\,GHz NVSS that were located $+75^{\circ} < \delta < +88^{\circ}$ and had a flux density $S_{\rm 1.4\,GHz}>0.2$\,Jy.  We therefore use the 4.8\,GHz catalogue of \cite{Mingaliev2007} for identifying and masking sources in the NCP region.

The locations of radio sources above 200\,mJy from \cite{Mingaliev2007} are over-plotted in Fig.~\ref{fig:cbassmap}. The brightest sources ($>600$\,mJy) at $\delta > +80^{\circ}$ are listed in Table~\ref{tab:brightsources}. Virtually all of these sources at $\delta > +75^{\circ}$ lie on peaks in the \mbox{C-BASS} map. This is reassuring and indicates that we understand well the bright source population at 5\,GHz. However, it should be noted that a number of these sources exhibit variability in their flux density \citep[e.g.,][]{Liu2014}. The brightest source in our field ($\delta>+80^{\circ}$) away is the double-lobed radio galaxy 3C\,61.1, which has a flux density of $\approx 2$\,Jy at 4.7\,GHz \citep{Hargrave1975}.\!\footnote{The \cite{Mingaliev2007} catalogue reports a flux density for 3C\,61.1 of $970\pm80$\,mJy at 4.8\,GHz, which is inconsistent with various measurements of $\approx 1.9$\,Jy at 5\,GHz (the source is not thought to be significantly variable). Since they report 3 individual sources (they do not give the flux densities for these due to confusion of multiple sources hence the ``X"s in the catalogue entry) it may be that this is a typographical error in which they have reported one of the contributing sources. Other sources appear to be unaffected.}

\begin{table*}
\caption{List of bright ($>600$\,mJy) sources in the NCP region ($\delta>+80^{\circ}$), ordered in decreasing flux density. Flux densities at 4.8\,GHz are taken from \protect\cite{Mingaliev2007}, except for 3C\,61.1$^{*}$ (see text and footnote$^3$), which is taken from \protect\cite{Hargrave1975}. The uncertainties are dominated by $3\%$ calibration errors.}
\begin{tabular}{lcccccl}
\hline
Source    &Flux density   &(R.A.,Dec.) [J2000]       &$(l,b)$    &Alternate name   \\
         &[mJy]          &[deg.]                    &[deg.]     &        \\ \hline
115312+805829 & 2130 &(178.30,+80.97) &(125.72,+35.84) &S5\,0014+81 \\
0222XX+86XXXX$^{*}$ & 1900 &(35.70,+86.31) &(124.49,+23.72) &3C\,61.1 \\
234403+822640 & 1433 &(356.02,+82.44) &(120.61,+19.88) &S5\,1150+81 \\
104423+805439 & 1404 &(161.10,+80.91) &(128.74,+34.74) &S5\,1039+81 \\
001708+813508 & 1190 &(4.29,+81.59) &(121.61,+18.80) &S5\,0014+81 \\
075058+824158 & 1061 &(117.74,+82.70) &(130.99,+28.77) &S5\,0740+82 \\
074305+802544 & 804 &(115.77,+80.43) &(133.60,+28.86) &3C\,184.1 \\
062602+820225 & 751 &(96.51,+82.04) &(131.74,+25.97) &S5\,0615+82 \\
163226+823220 & 749 &(248.11,+82.54) &(115.77,+31.20) &NGC\,6251 \\
074246+802741 & 698 &(115.70,+80.46) &(133.56,+28.85) &3C\,184.1 \\
093923+831526 & 672 &(144.85,+83.26) &(128.81,+31.52) &3C\,220.3 \\
161940+854921 & 670 &(244.92,+85.82) &(119.14,+29.65) &S5\,1631+85 \\
213008+835730 & 656 &(322.54,+83.96) &(117.88,+23.18) &3C\,435.1 \\
235622+815252 & 626 &(359.10,+81.88) &(120.89,+19.23) &S5\,2353+81 \\
105811+811432 & 601 &(164.55,+81.24) &(127.97,+34.75) &S5\,1053+81 \\
\hline
\end{tabular}
\label{tab:brightsources}
\end{table*}

\begin{figure*}
\begin{center}
\includegraphics[width=0.24\textwidth,angle=0]{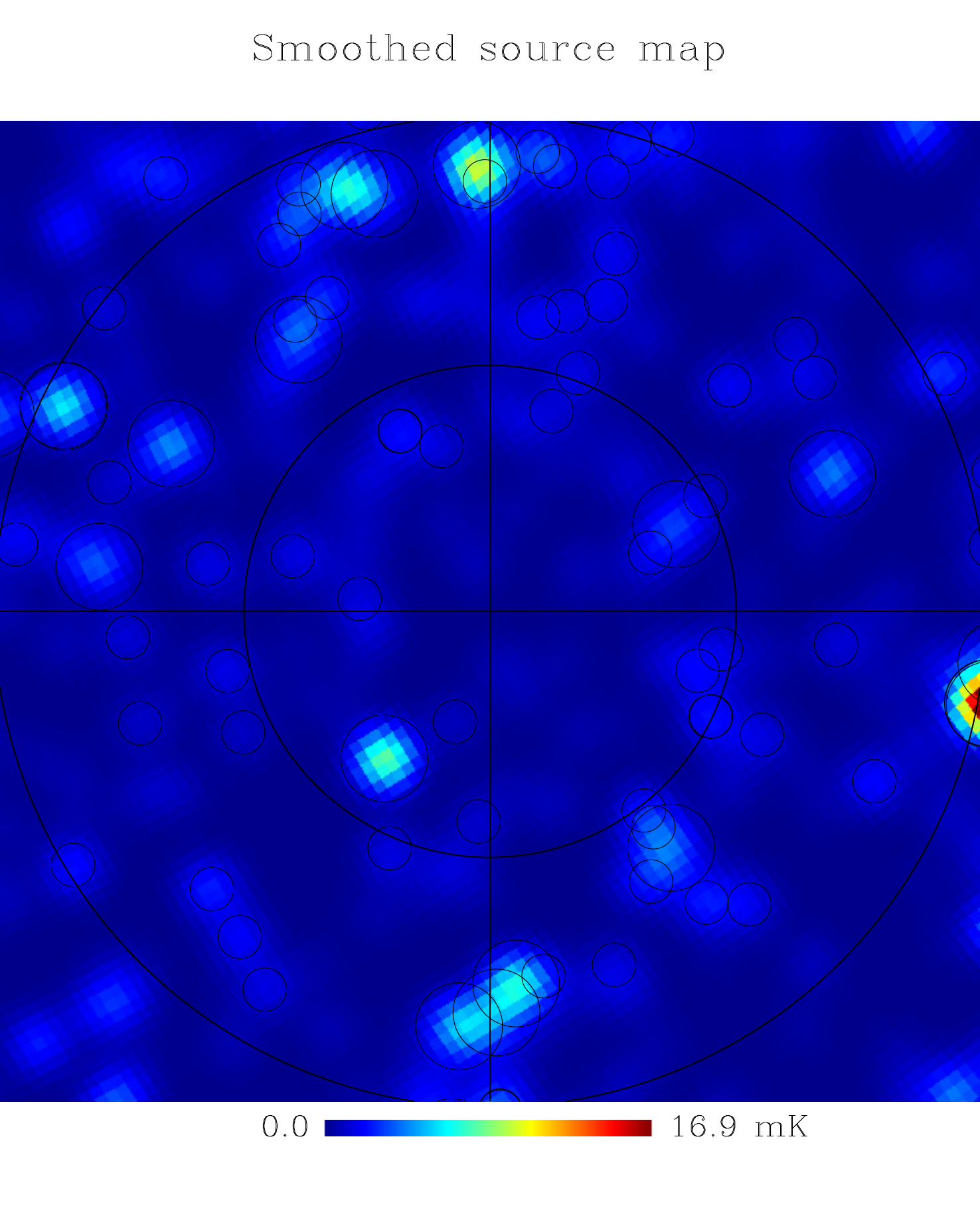}
\includegraphics[width=0.24\textwidth,angle=0]{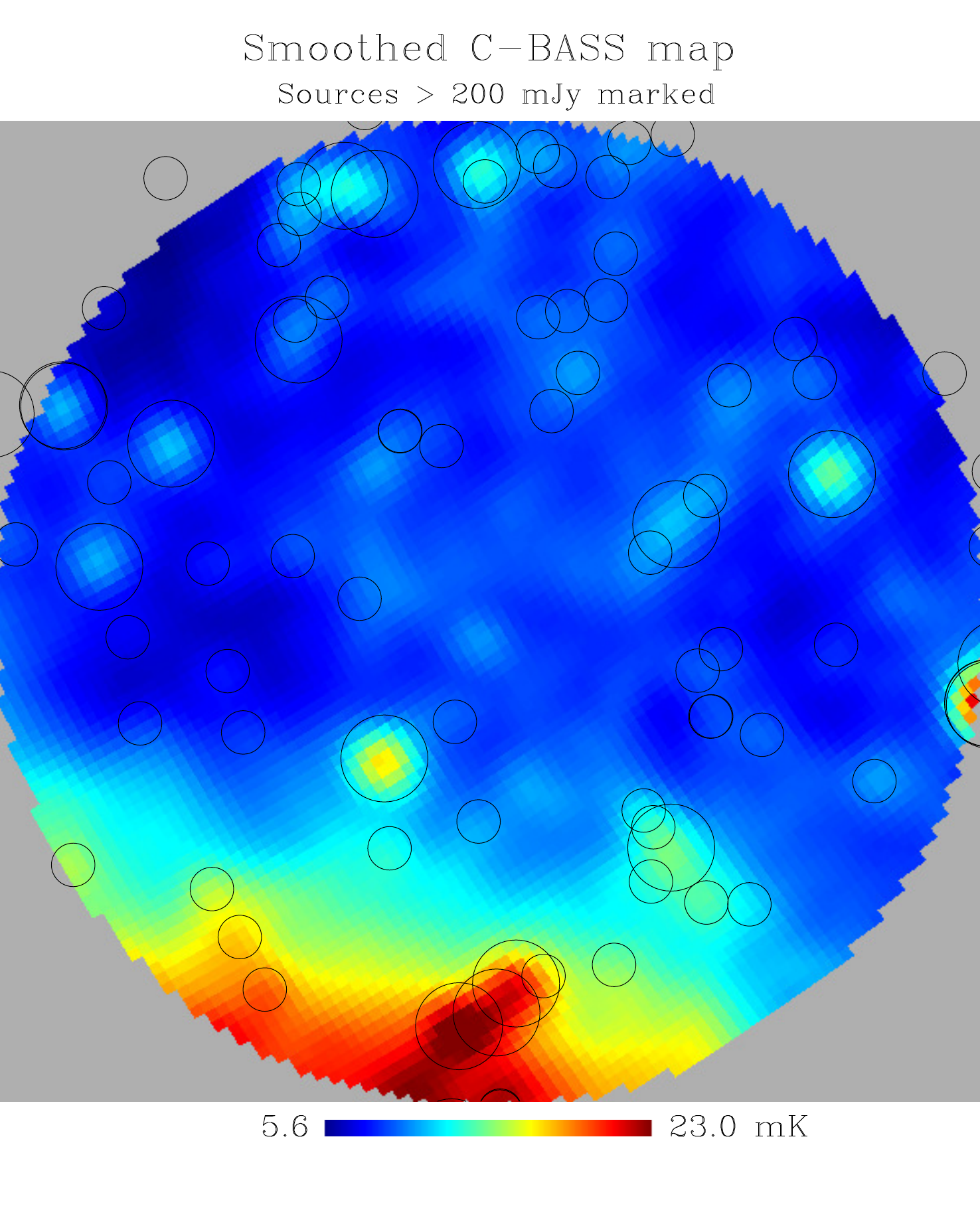}
\includegraphics[width=0.24\textwidth,angle=0]{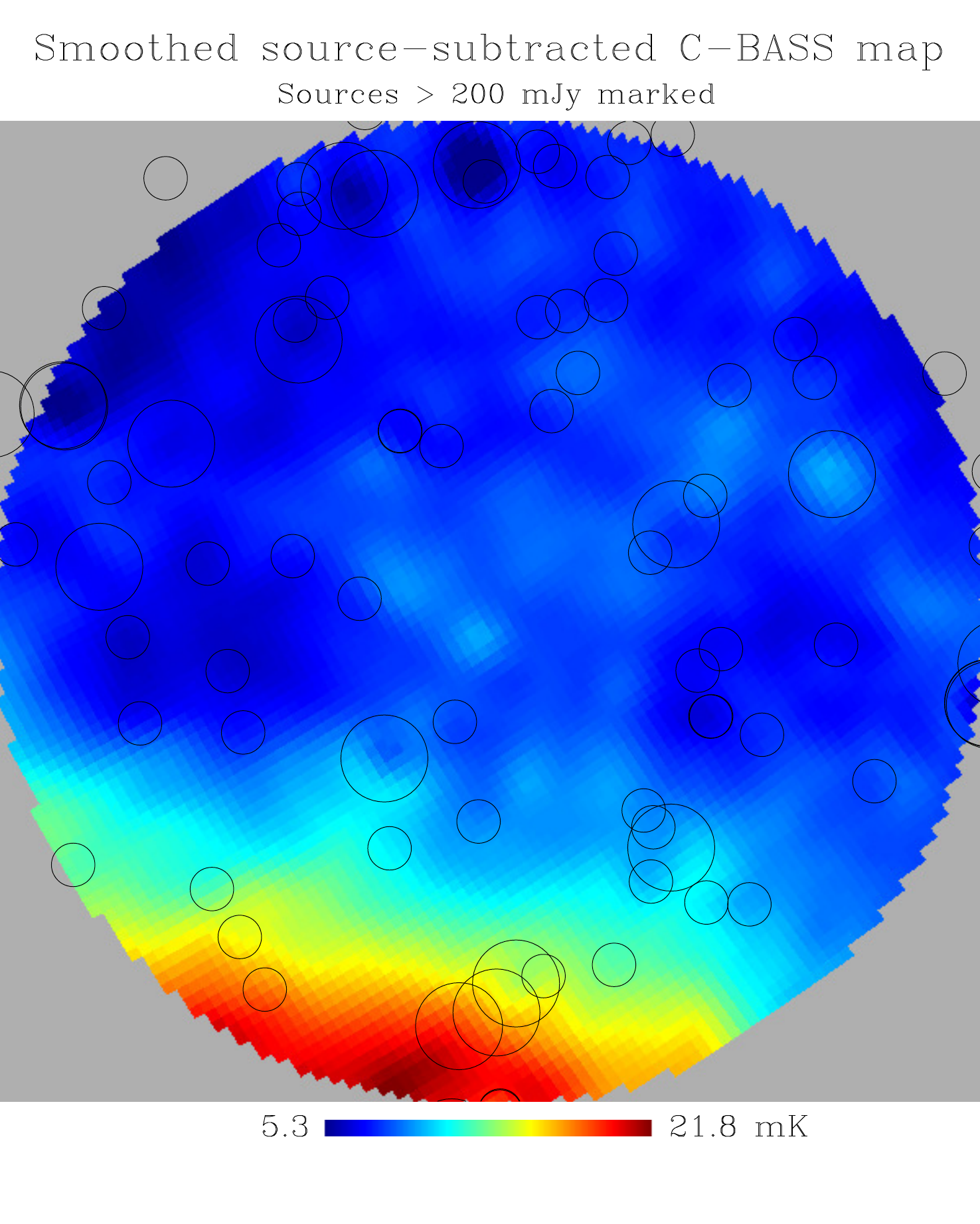}
\includegraphics[width=0.24\textwidth,angle=0]{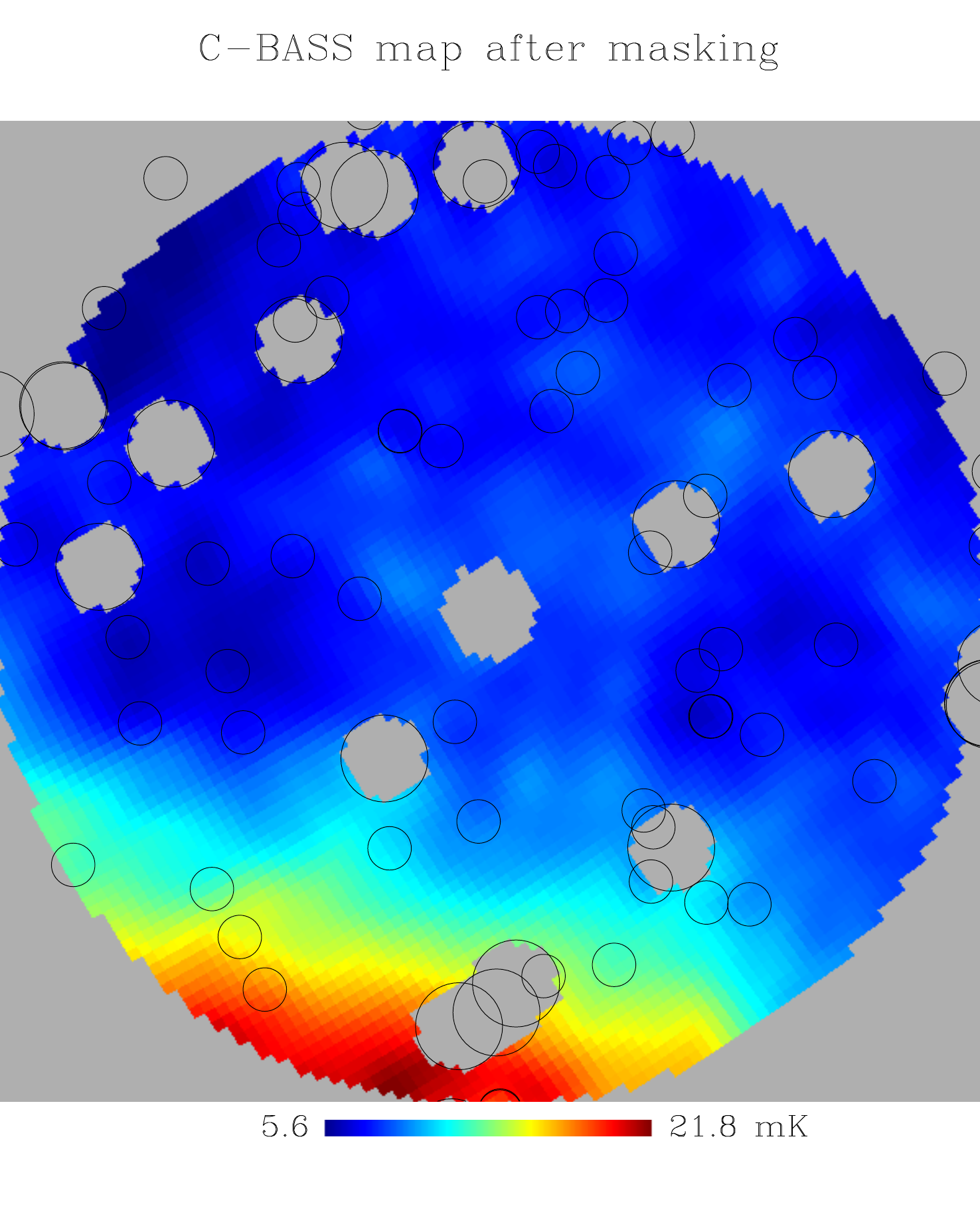}
\caption{From {\it left} to {\it right}: i) Map of bright ($>200$\,mJy) sources from the \protect\cite{Mingaliev2007} catalogue after convolving with a $1\degr$ FWHM Gaussian beam. ii) \mbox{C-BASS} map, smoothed to $1\degr$ FWHM resolution. iii) \mbox{C-BASS} map after subtracting bright extragalactic sources. iv) \mbox{C-BASS} map with masked regions shown as grey ($\delta<+80^{\circ}$, the brightest sources, and for $\delta>+89^{\circ}$; see text). The coordinate system is the same as in Fig.\,\ref{fig:cbassmap} but covering $\delta>+80^{\circ}$. Bright sources above 200\,mJy and 600\,mJy are marked as small and large circles, respectively. The colour scale is linear between the minimum/maximum values shown.}
\label{fig:cbassmap_sourcesub}
\end{center}
\end{figure*}

Fig.~\ref{fig:cbassmap_sourcesub} shows the map of point sources, convolved with a $1^{\circ}$ FWHM Gaussian beam. The fluctuations in brightness temperature are at the level of $\approx 1$--2\,mK away from bright sources. The source 3C\,61.1 has a peak brightness temperature of $\approx 9$\,mK above the background. In the middle panel of Fig.~\ref{fig:cbassmap_sourcesub} we show the \mbox{C-BASS} map after smoothing (and deconvolving) to the common $1^{\circ}$ FWHM resolution. The right panel shows the \mbox{C-BASS} map after subtracting the map of sources. Visual inspection shows that the subtraction has been successful, with most of the obvious sources no longer being visible. For the brightest few sources (3C\,61.1, NGC\,6251) the subtraction is not perfect. For example, NGC\,6251 has been under-subtracted (possibly due to its large angular extent of over $1^{\circ}$). These could be removed by fitting for their flux densities separately. However, we choose to mask the four brightest sources with an aperture of $0.\!^{\circ}7$ radius. We also mask out the NCP itself ($\delta=+90^{\circ}$) because there appears to be a relatively bright source in the \mbox{C-BASS} map, which is not in the majority of source catalogues. The source mask is shown in Fig.~\ref{fig:cbassmap_sourcesub}. Fainter sources can be either masked or subtracted, as a test of point-source contamination. Our results are not strongly sensitive to the effects of residual source contamination (see Sect.~\ref{sec:sources2}). 

At the higher frequencies ($\gtrsim 20$\,GHz) observed by {\it WMAP} and {\it Planck}, some of the weaker radio sources are much fainter than at lower frequencies because many galaxies have spectral indices $\alpha <0$ (flux density $S \propto \nu^{\alpha}$). The main population of radio sources at frequencies $\gtrsim 20$\,GHz, particularly at high flux densities, are galaxies that harbour active galactic nuclei (AGN), which can produce flat-spectrum radiation up to tens of gigahertz or even higher. In the NCP region ($\delta >+75^{\circ}$), the {\it Planck} compact source catalogue, PCCS2 \citep{Planck2015_XXVI}, contains 28 sources above 0.5\,Jy while only 3 sources are above 1\,Jy. Of these, only 1 source (NGC\,6251) is in the $\delta >+80^{\circ}$ region, with a 28.4\,GHz flux density of $1.3 \pm 0.1$\,Jy. Pixels affected by this source will be masked out.


\setlength{\tabcolsep}{3pt}

\begin{table*}
\caption{Summary of multi-frequency data. The {\it top} part of the table represent the various foreground templates that will be fitted to the microwave/sub-mm ({\it WMAP}/{\it Planck}) data, which are listed in the {\it bottom} part of the table. The listed absolute calibration errors are those assumed for deriving spectral indices (see text).}
\small
\begin{tabular}{lcccll}
\hline
Telescope/Survey   		&Freq.   	&Ang. res. &Abs. Cal.   &Reference   	                    &Notes   	\\
&[GHz]		&[arcmin]	    &Error [\%]&		             &		\\ \hline
Haslam				&0.408		&51		&10          &\protect\cite{Haslam1982,Remazeilles2015}	&Synchrotron template \\
Reich				&1.42		&36		&10	     &\protect\cite{Reich1986}	&Synchrotron template \\
C-BASS				&4.7		&45		&5	     &This work			&Synchrotron template \\
F03 H$\alpha$			&...		&6/60		&10	     &\protect\cite{Finkbeiner2003} 	&Free-free template \\
D03 H$\alpha$			&...		&60		&10    	     &\protect\cite{Dickinson2003}	&Free-free template \\
{\it Planck} PR3 HFI 353\,GHz	&353		&4.7		&5	     &\protect\cite{Planck2018_I}    &Dust template (vR3.00)	\\
{\it Planck} PR3 HFI 545\,GHz	&545		&4.73		&5	     &\protect\cite{Planck2018_I}    &Dust template (vR3.00)	\\
{\it Planck} PR3 HFI 857\,GHz	&857		&4.51		&5	     &\protect\cite{Planck2018_I}    &Dust template (vR3.00)	\\
{\it Planck} 353\,GHz optical depth	&353		&5.0		&5	     &\protect\cite{PIP_XLVIII}    &Dust template (vR2.00)	\\
{\it Planck} dust radiance $\Re$		&--		&5.0		&--	&\protect\cite{Planck2013_XI,PIP_XLVIII}  & Dust template \\
{\it IRAS} (IRIS) 100\,$\mu$m 	&2997		&4.3		&13.5        &\protect\cite{Miville-Deschenes2005}  	&Dust template \\
FDS94 model 8				&94		&6.1		&--		&\protect\cite{Finkbeiner1999}			&Dust template		\\
\hline
    {\it WMAP} 9-year K-band	&22.8		&51.3		&3	     &\protect\cite{Bennett2013}    	&$1^{\circ}$-smoothed product	\\
{\it Planck} PR3 LFI 30\,GHz	&28.4		&33.16		&3	     &\protect\cite{Planck2018_I}    &vR3.00	\\
{\it WMAP} 9-year Ka-band	&33.0		&39.1		&3           &\protect\cite{Bennett2013}    	&$1^{\circ}$-smoothed product	\\
{\it WMAP} 9-year Q-band	&40.7		&30.8		&3	     &\protect\cite{Bennett2013}    	&$1^{\circ}$-smoothed product	\\
{\it Planck} PR3 LFI 44\,GHz	&44.1		&28.09		&3	     &\protect\cite{Planck2018_I}    &vR3.00	\\
{\it WMAP} 9-year V-band	&60.7		&30.8		&3	     &\protect\cite{Bennett2013}    	&$1^{\circ}$-smoothed product	\\
{\it Planck} PR3 LFI 70\,GHz	&70.4		&13.08		&3	&\protect\cite{Planck2018_I}    &vR3.00	\\
{\it WMAP} 9-year W-band	&93.5		&30.8		&3	&\protect\cite{Bennett2013}    	&$1^{\circ}$-smoothed product	\\
{\it Planck} PR3 HFI 100\,GHz	&100		&9.59		&5	&\protect\cite{Planck2018_I}    &vR3.00	\\
{\it Planck} PR3 HFI 143\,GHz	&143		&7.18		&5	&\protect\cite{Planck2018_I}    &vR3.00	\\
{\it Planck} PR3 HFI 217\,GHz	&217		&4.87		&5	&\protect\cite{Planck2018_I}    &vR3.00	\\
{\it Planck} PR3 HFI 353\,GHz	&353		&4.7		&5	&\protect\cite{Planck2018_I}    &vR3.00	\\
{\it Planck} PR3 HFI 545\,GHz	&545		&4.73		&5	&\protect\cite{Planck2018_I}    &vR3.00	\\
{\it Planck} PR3 HFI 857\,GHz	&857		&4.51		&5	&\protect\cite{Planck2018_I}    &vR3.00	\\
\hline
\end{tabular}
\label{tab:data}
\end{table*}
\normalsize

\setlength{\tabcolsep}{6pt}

\begin{figure*}
\begin{center}
\includegraphics[width=0.24\textwidth,angle=0]{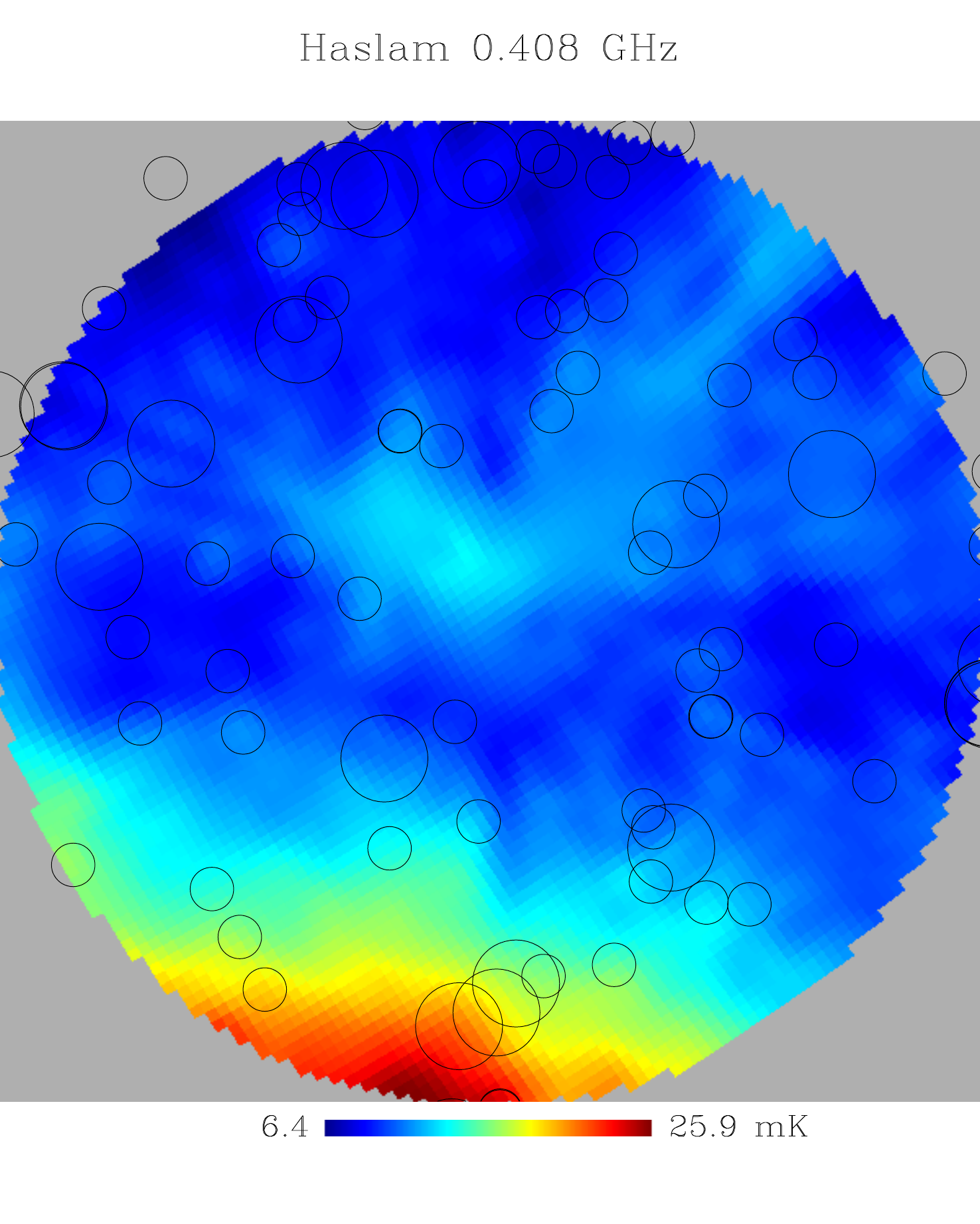}
\includegraphics[width=0.24\textwidth,angle=0]{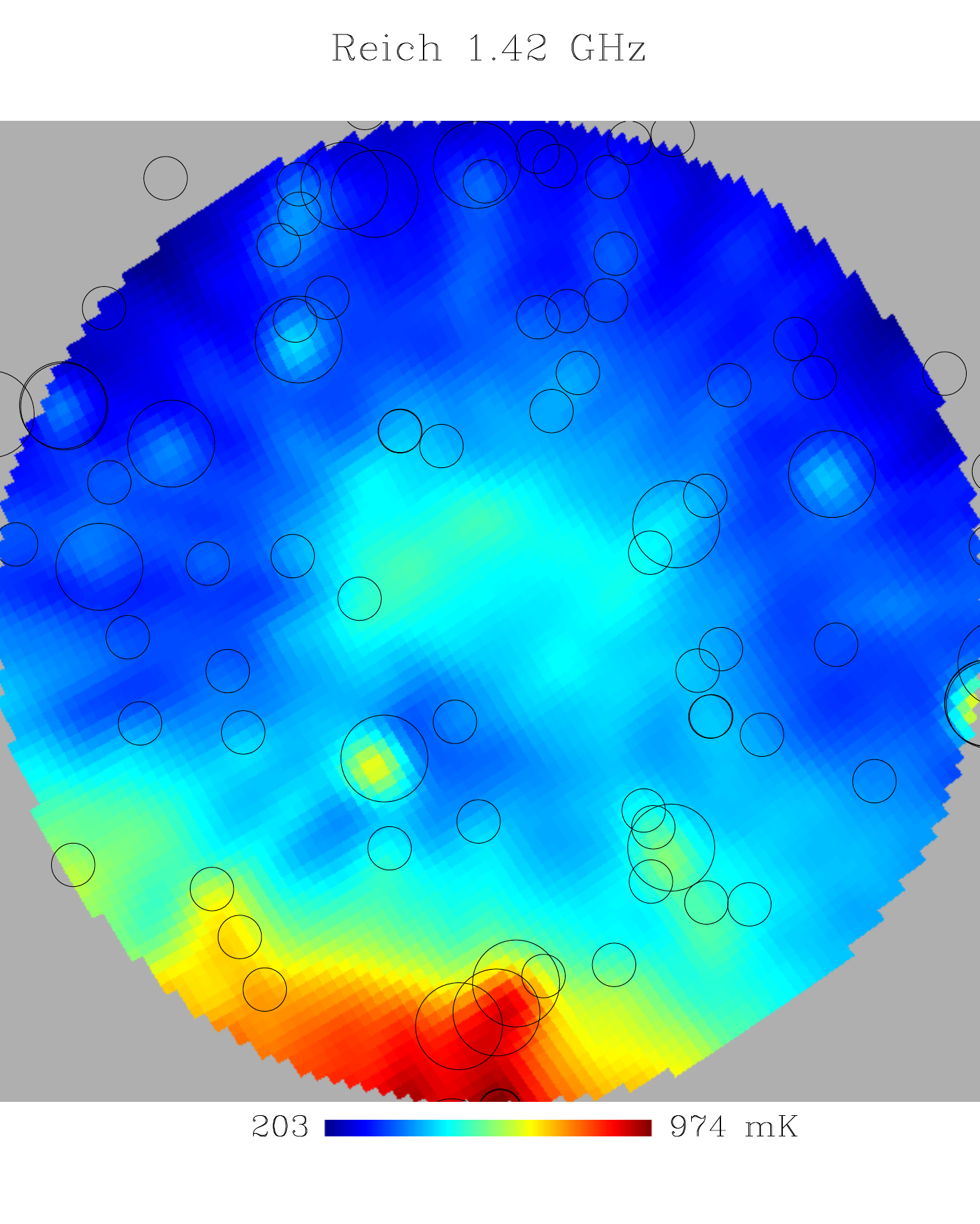}
\includegraphics[width=0.24\textwidth,angle=0]{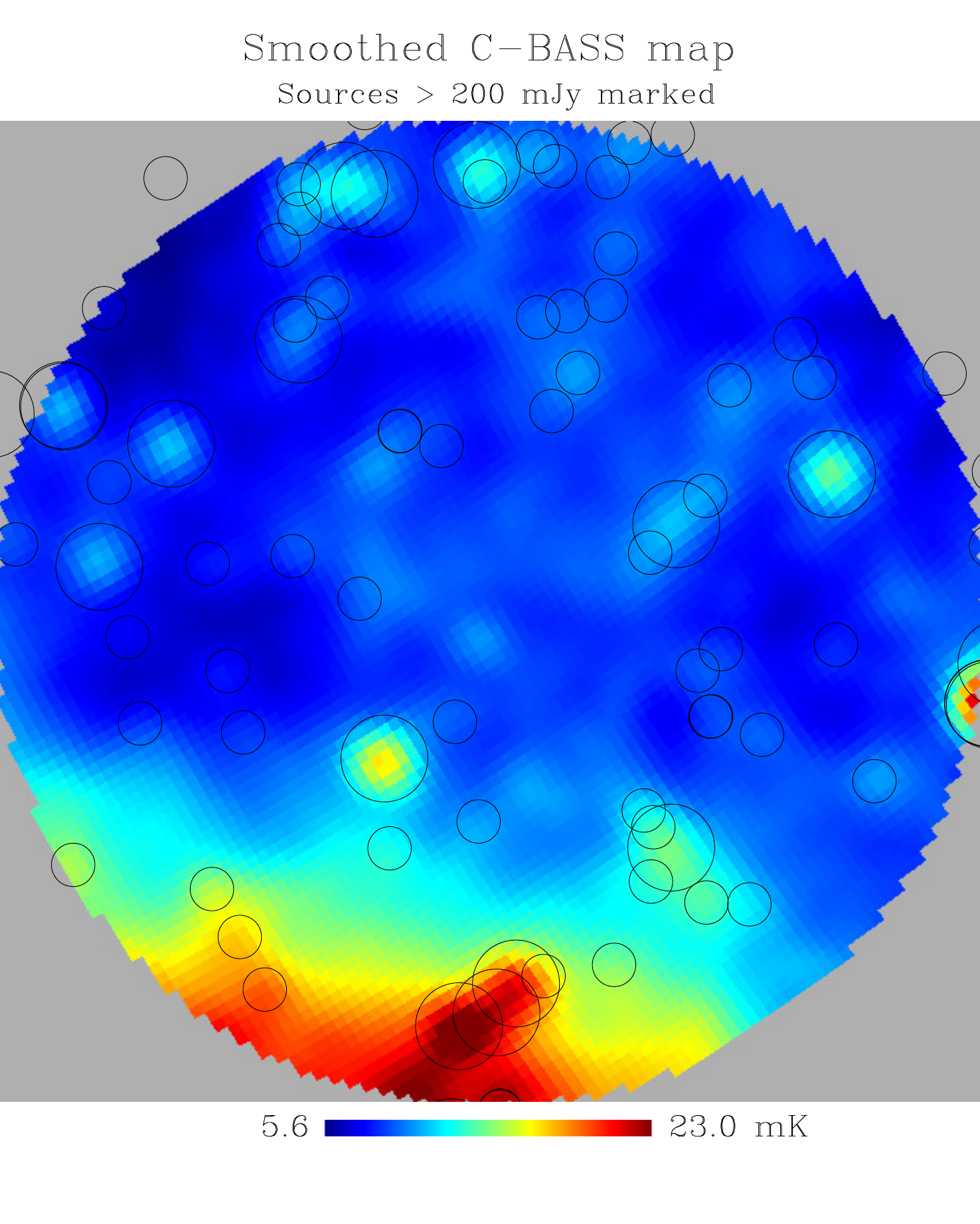}
\includegraphics[width=0.24\textwidth,angle=0]{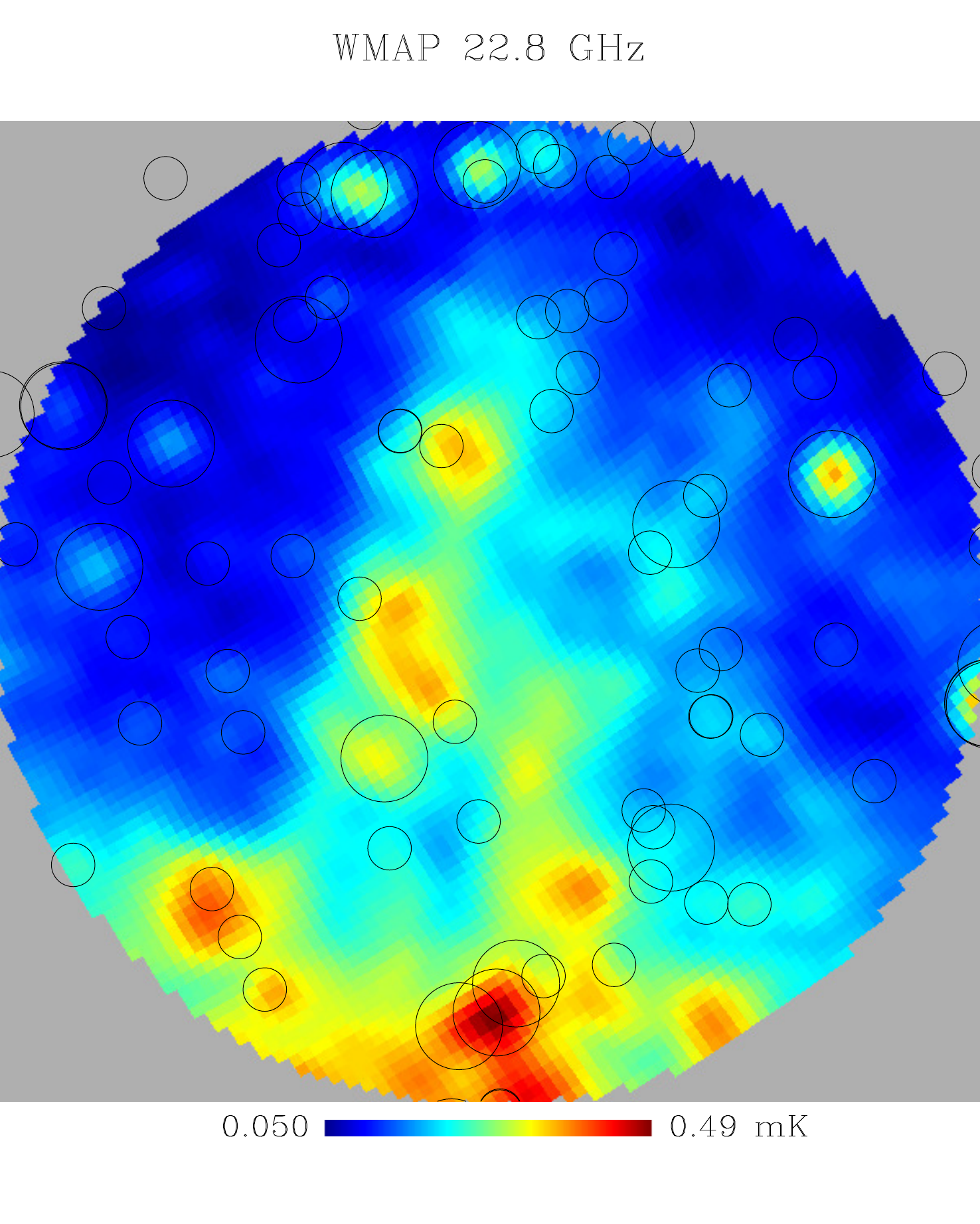} 
\par
\vspace{2mm}
\includegraphics[width=0.24\textwidth,angle=0]{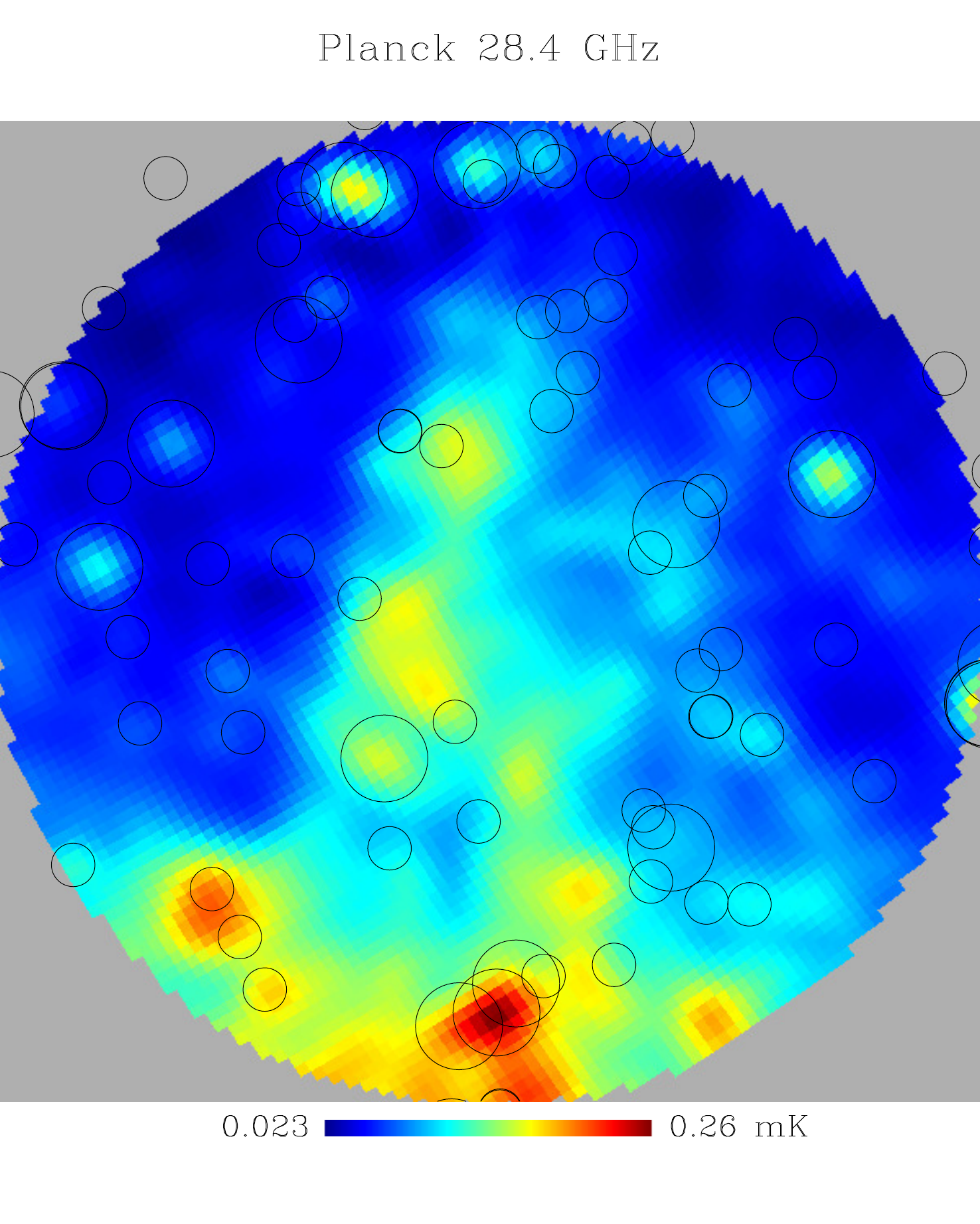}
\includegraphics[width=0.24\textwidth,angle=0]{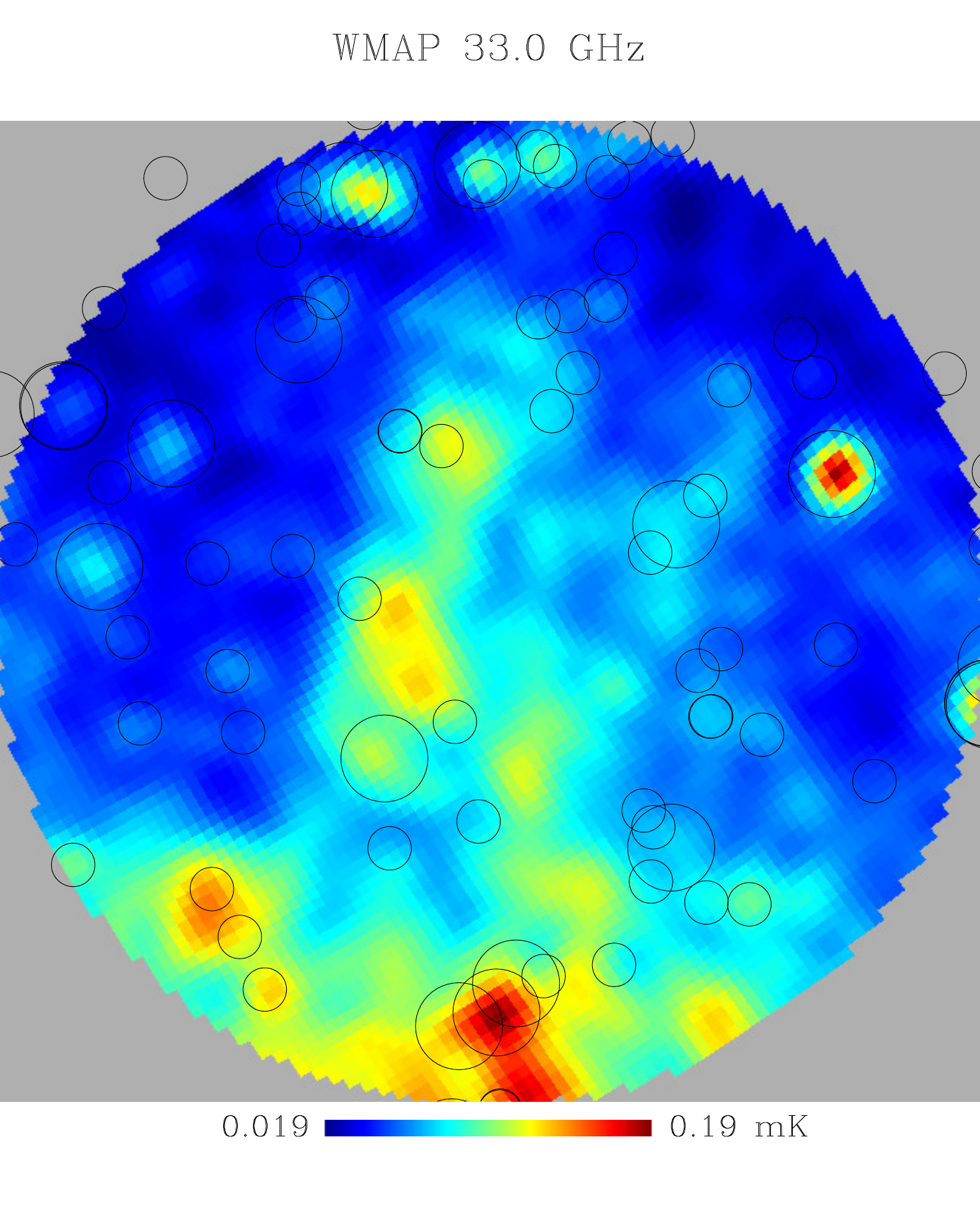}
\includegraphics[width=0.24\textwidth,angle=0]{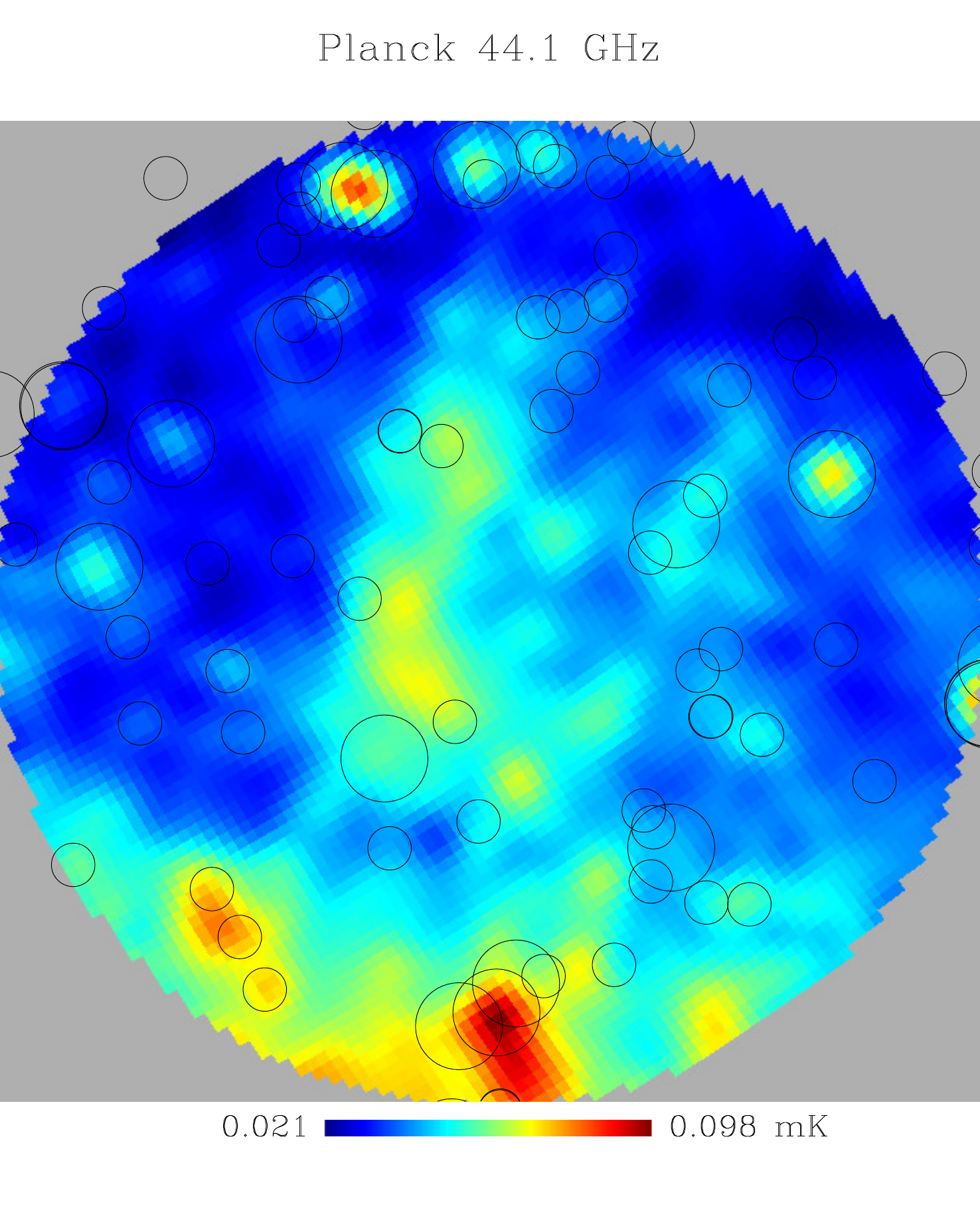}
\includegraphics[width=0.24\textwidth,angle=0]{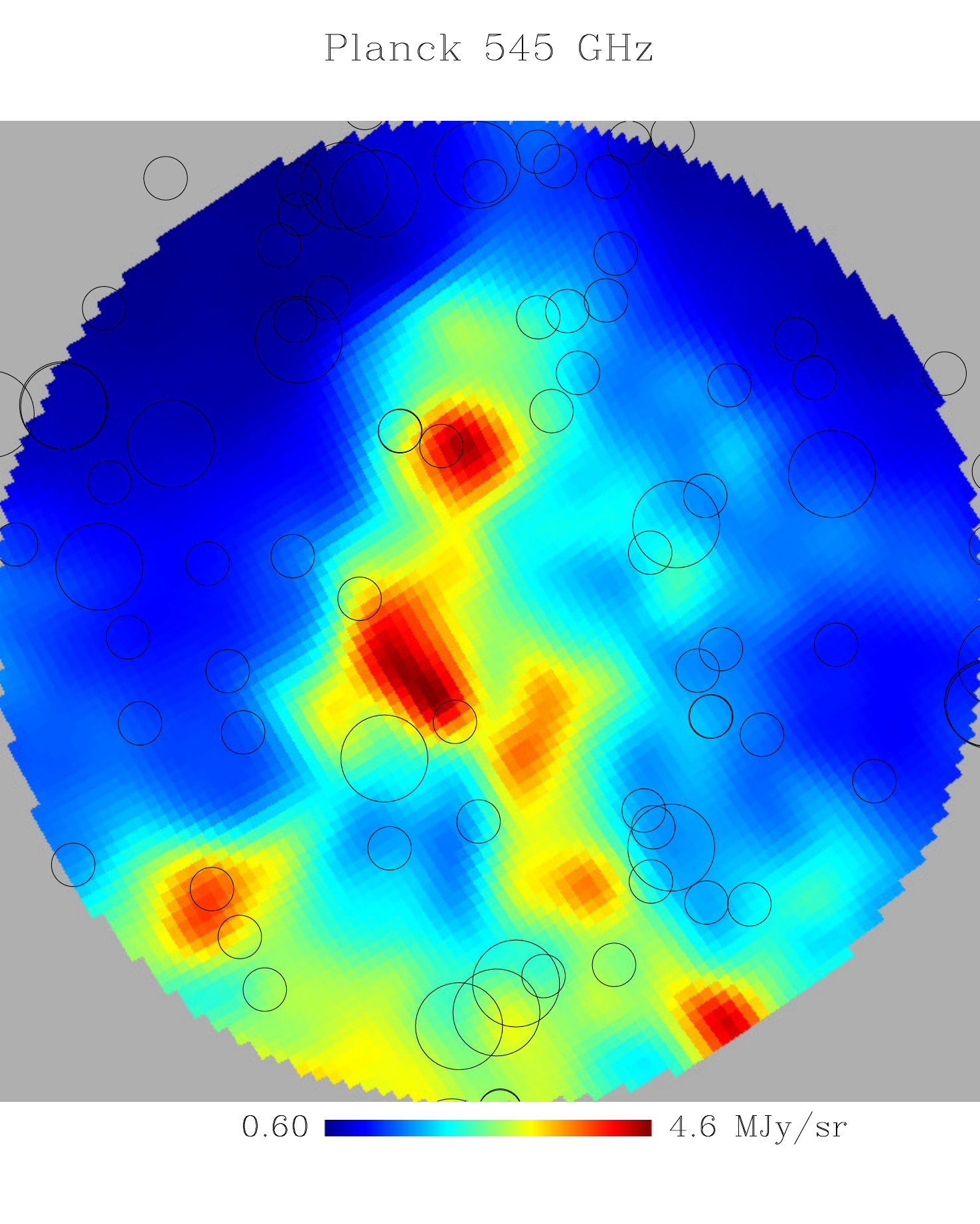}
\par
\vspace{2mm}
\includegraphics[width=0.24\textwidth,angle=0]{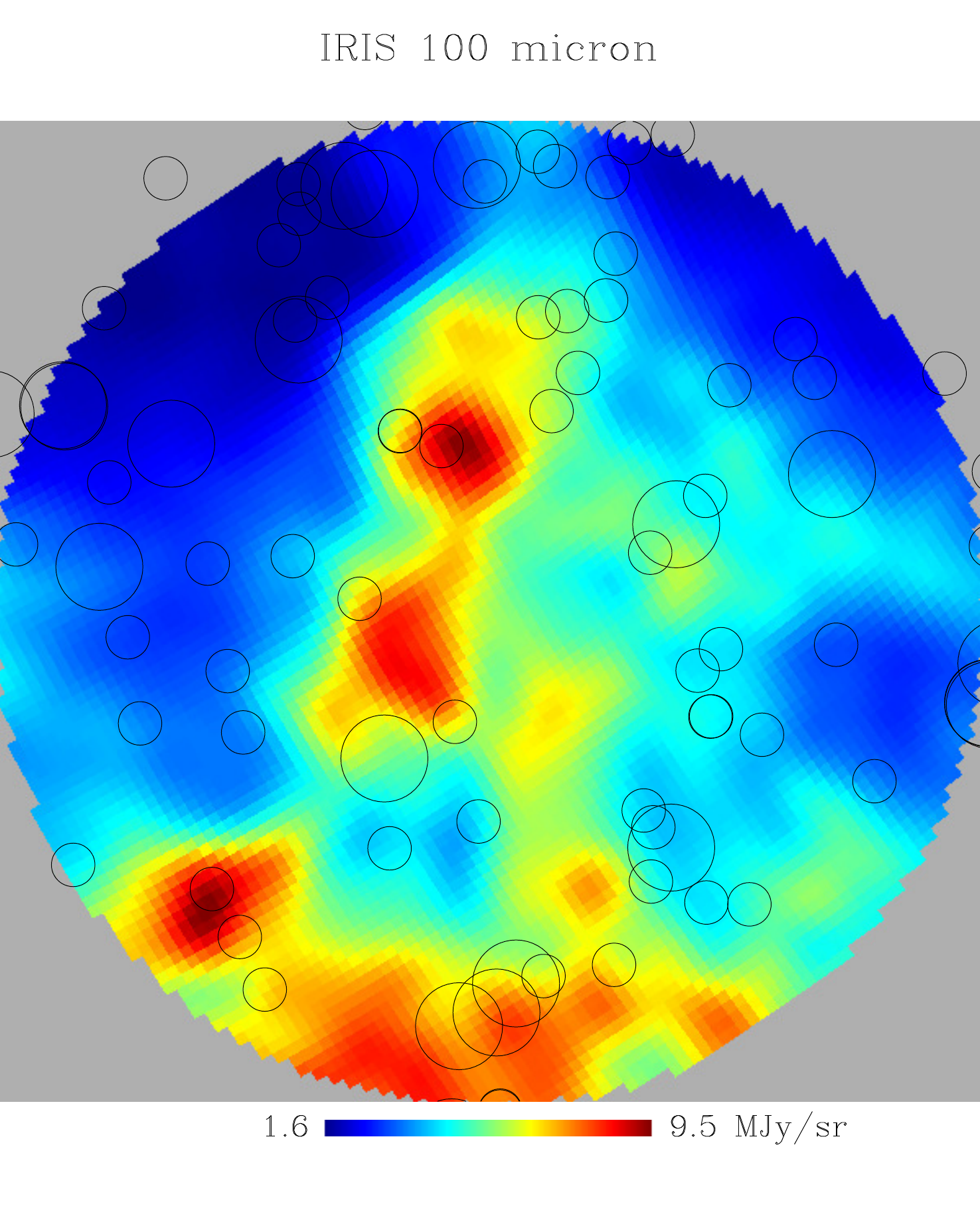}
\includegraphics[width=0.24\textwidth,angle=0]{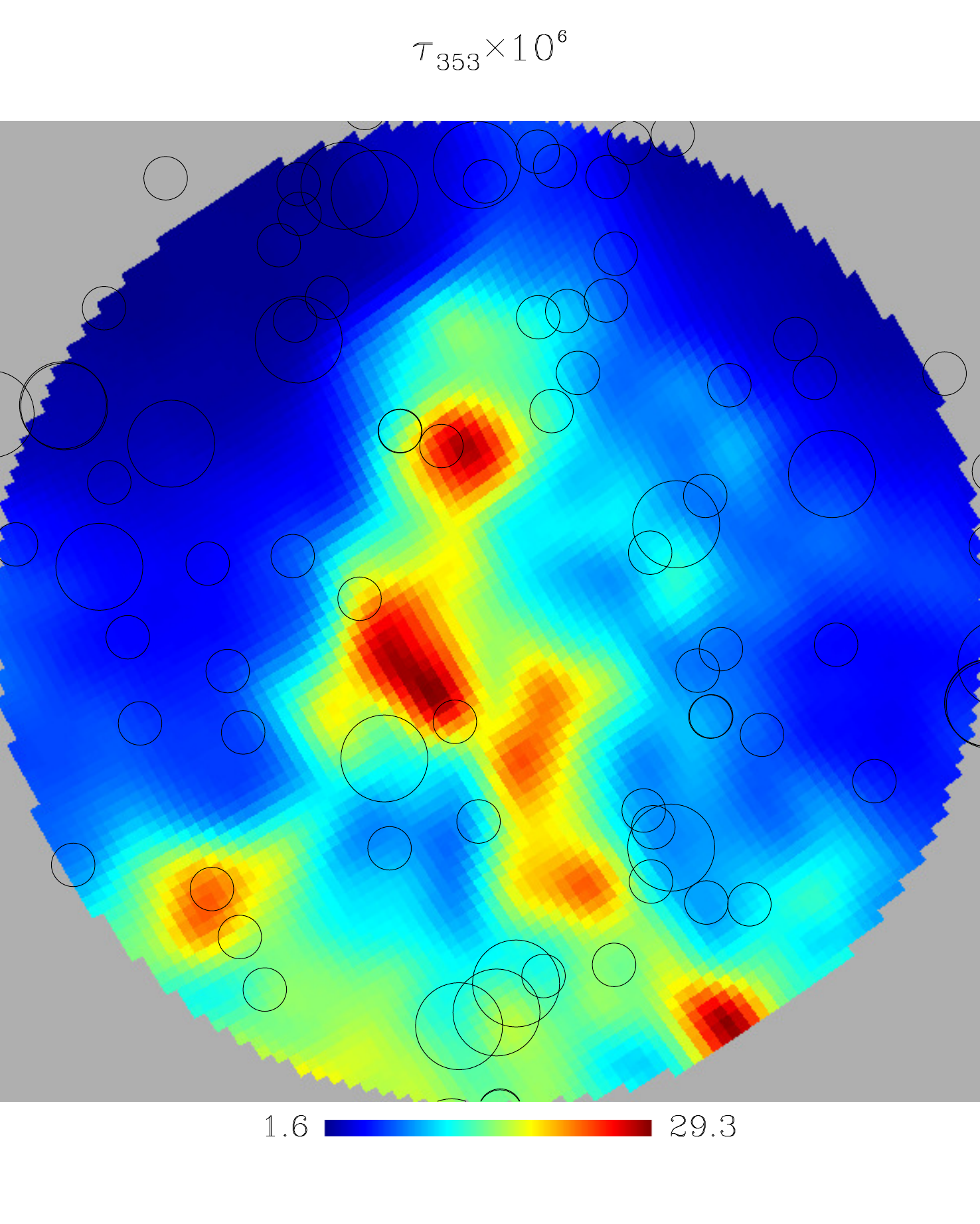}
\includegraphics[width=0.24\textwidth,angle=0]{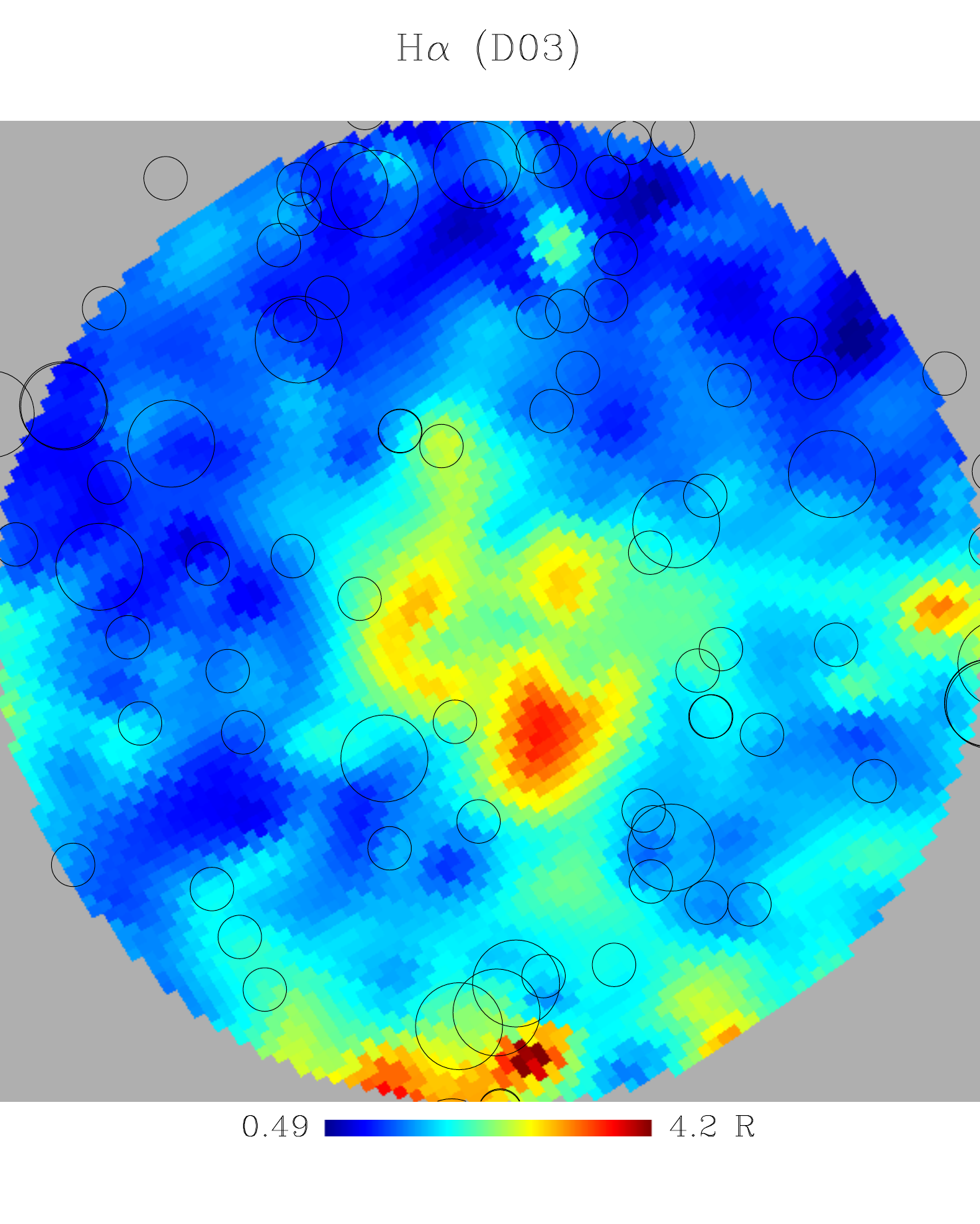}
\includegraphics[width=0.24\textwidth,angle=0]{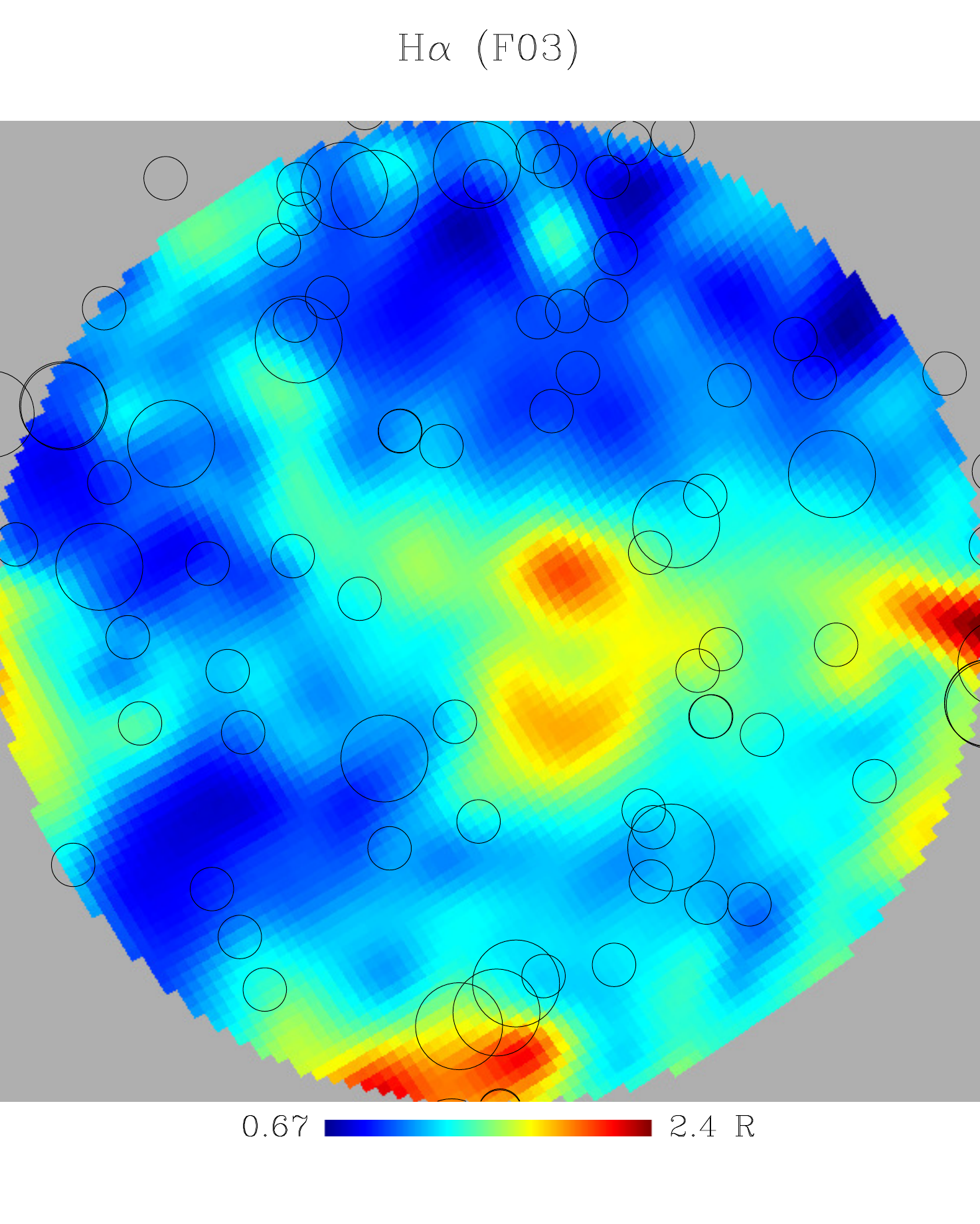}

\caption{Multi-frequency maps of the NCP region at a common resolution of $1^{\circ}$ (see Table\,\protect\ref{tab:data} for details). The panels are arranged in increasing frequency order: 0.4, 1.42, 4.7, 22.8, 28.4, 33.0, 41.0, 545, 3000 GHz ($100\,\mu$m). The last three panels are $\tau_{353}$, followed by two versions of the H$\alpha$ map (D03 and F03). The colour scales are all on a linear stretch between the minimum/maximum values shown. The coordinate system is the same as in Fig.\,\ref{fig:cbassmap} but covering $\delta>+80^{\circ}$. Radio sources are indicated by circles as in previous figures. The dust-correlated AME structure (e.g., at 545\,GHz, 100$\,\mu$m, $\tau_{353}$) is clearly visible at 22.8 and 28.4\,GHz but not at 4.7\,GHz. Striations and other artifacts are also visible in the 0.408/1.42\,GHz maps that are not seen in the \mbox{C-BASS} data.}
\label{fig:maps}
\end{center}
\end{figure*}


\begin{figure*}
\begin{center}
\includegraphics[width=0.24\textwidth,angle=0]{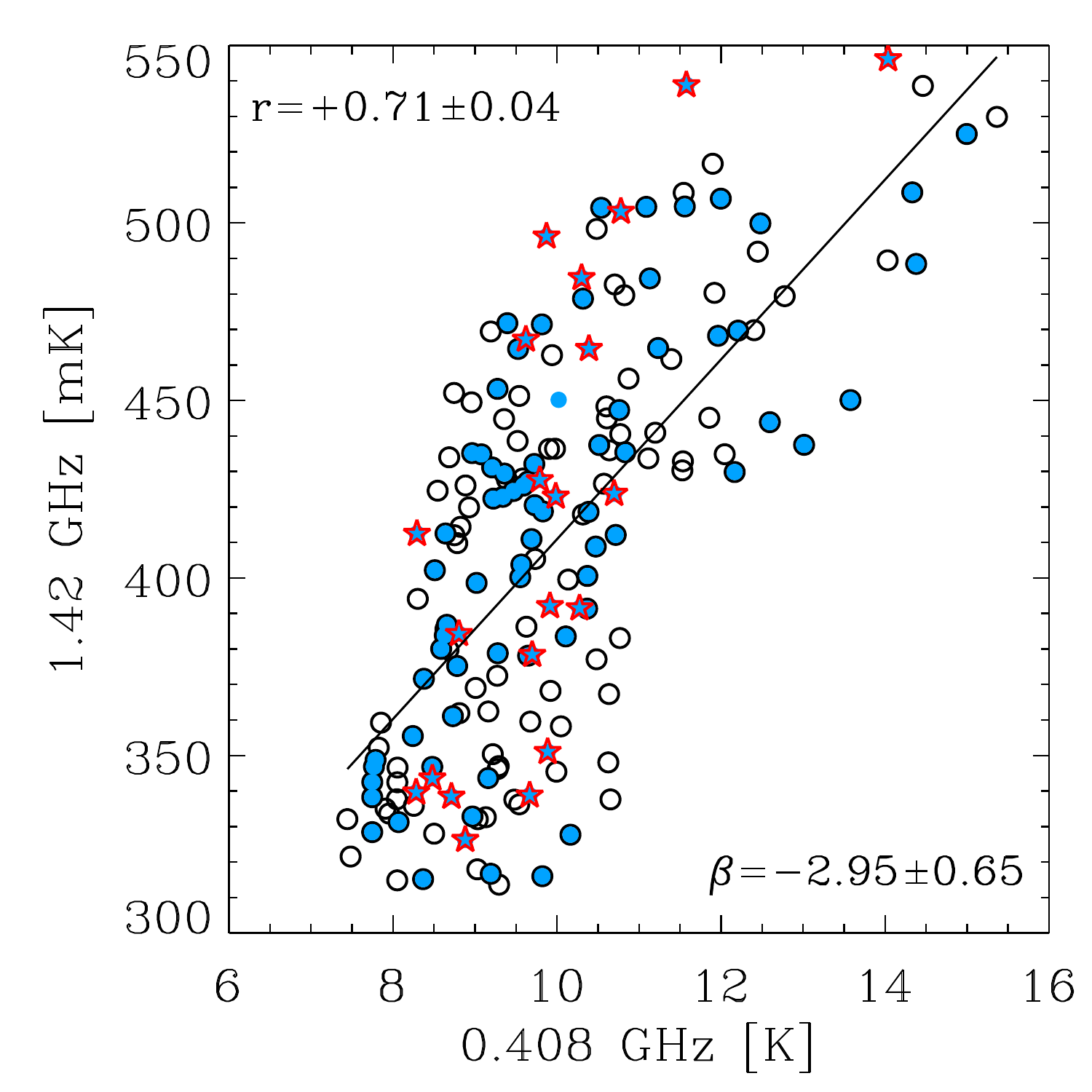}
\includegraphics[width=0.24\textwidth,angle=0]{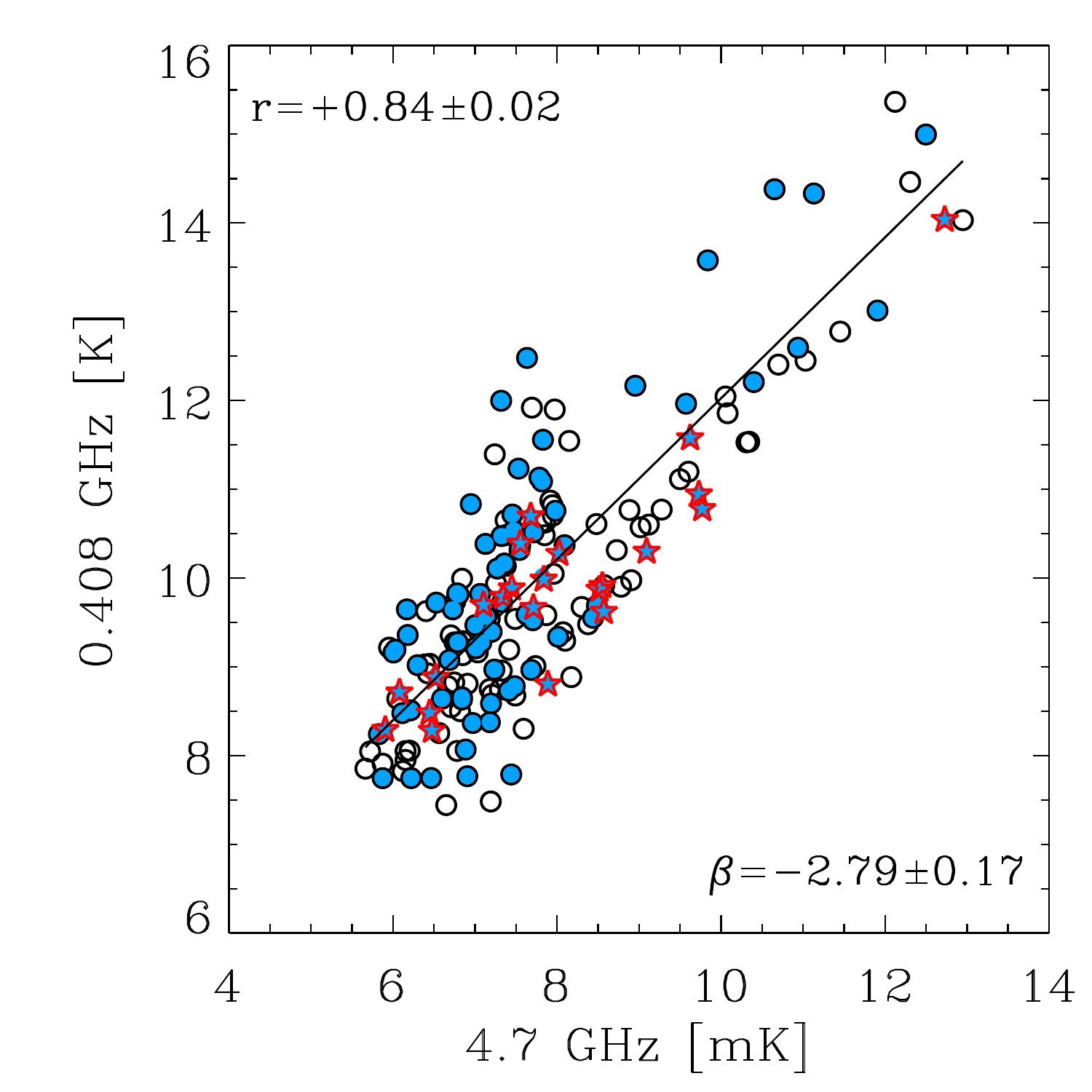}
\includegraphics[width=0.24\textwidth,angle=0]{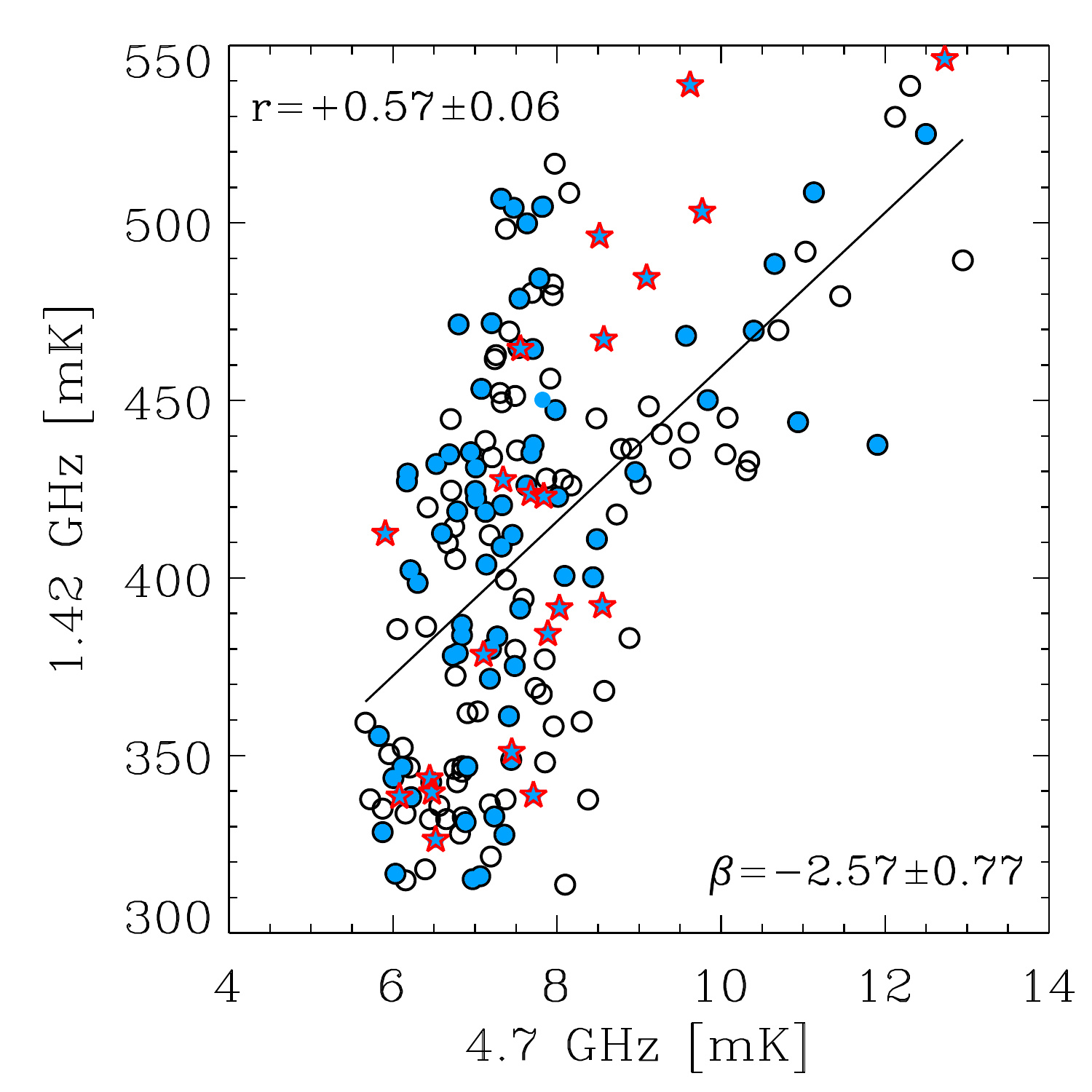}
\includegraphics[width=0.24\textwidth,angle=0]{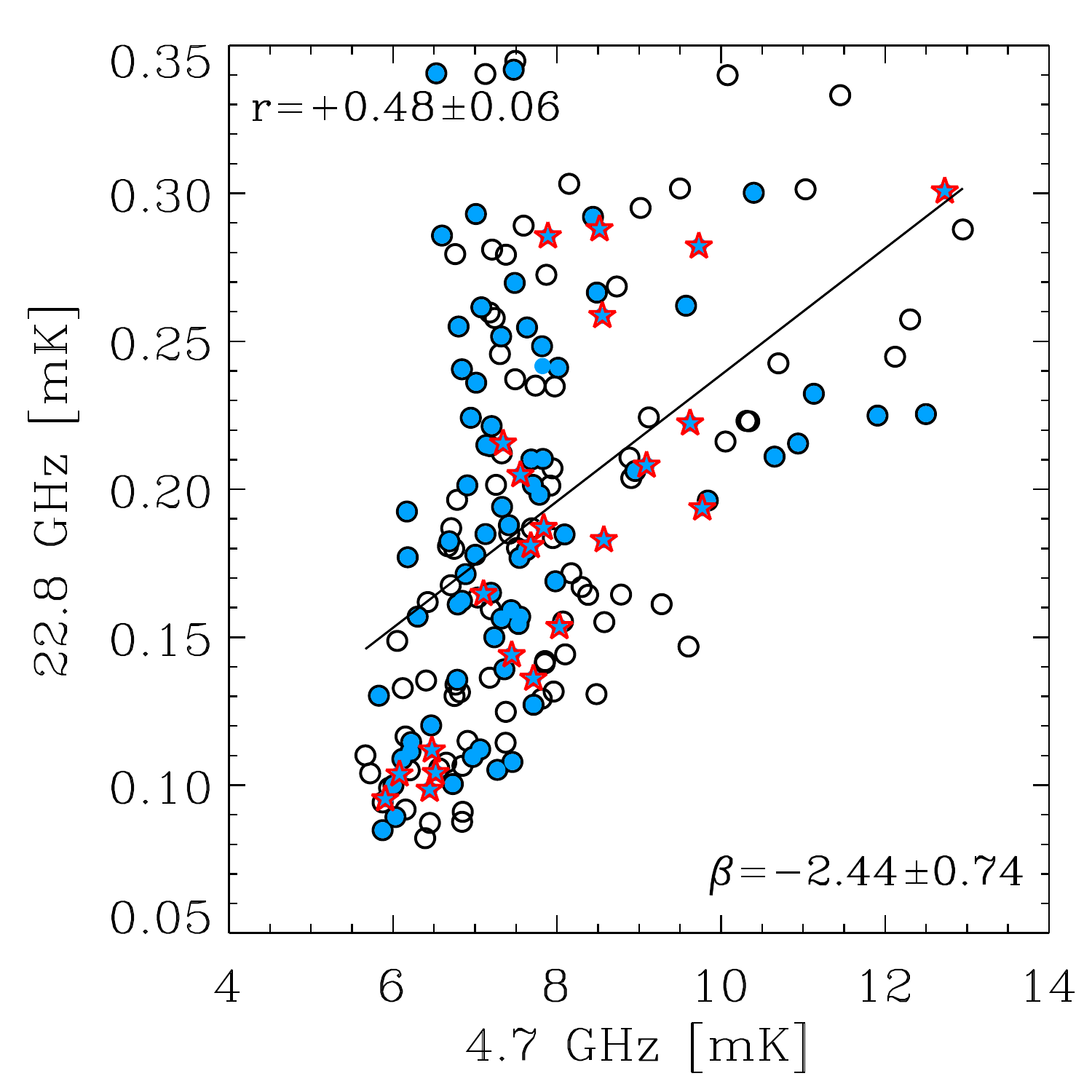}
\includegraphics[width=0.24\textwidth,angle=0]{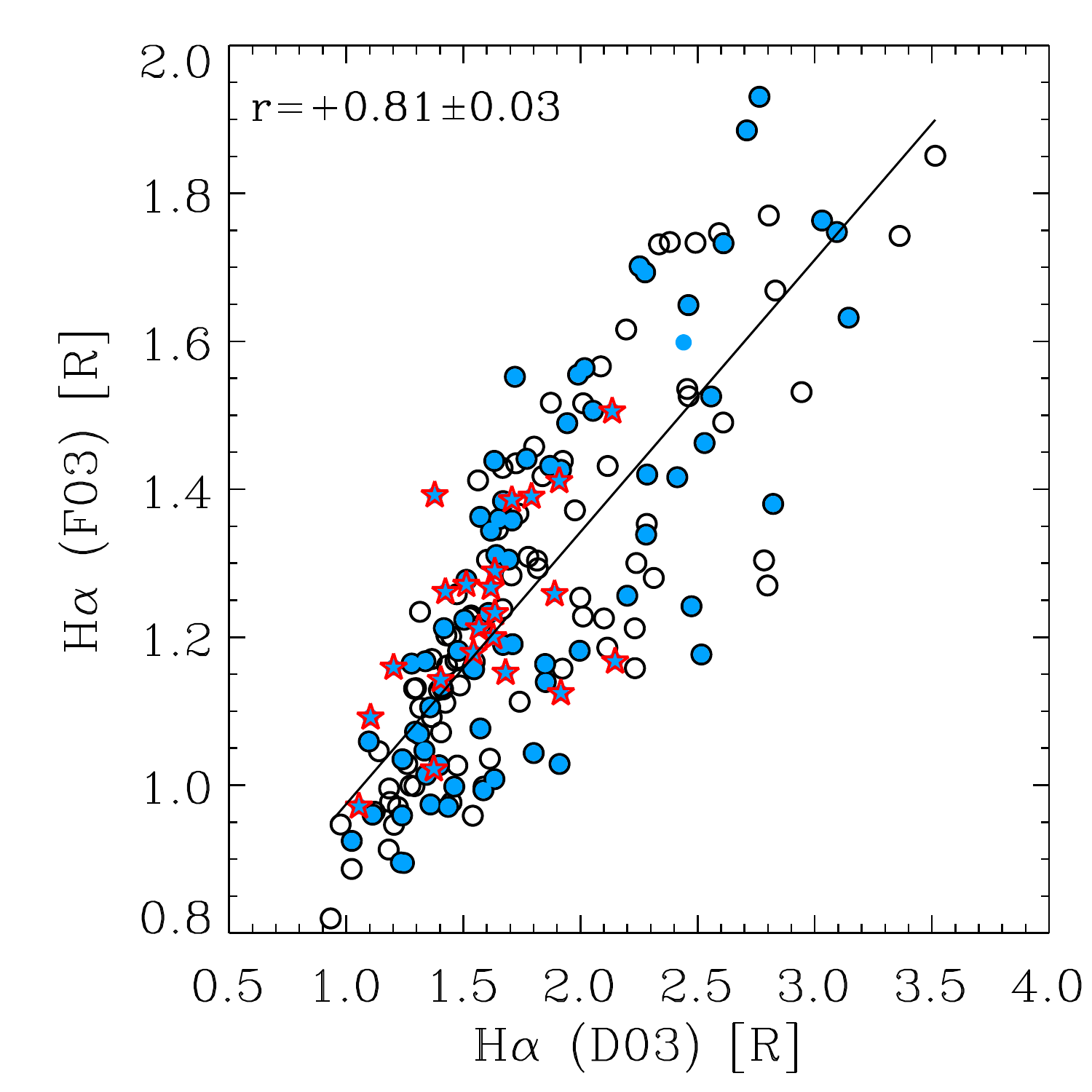}
\includegraphics[width=0.24\textwidth,angle=0]{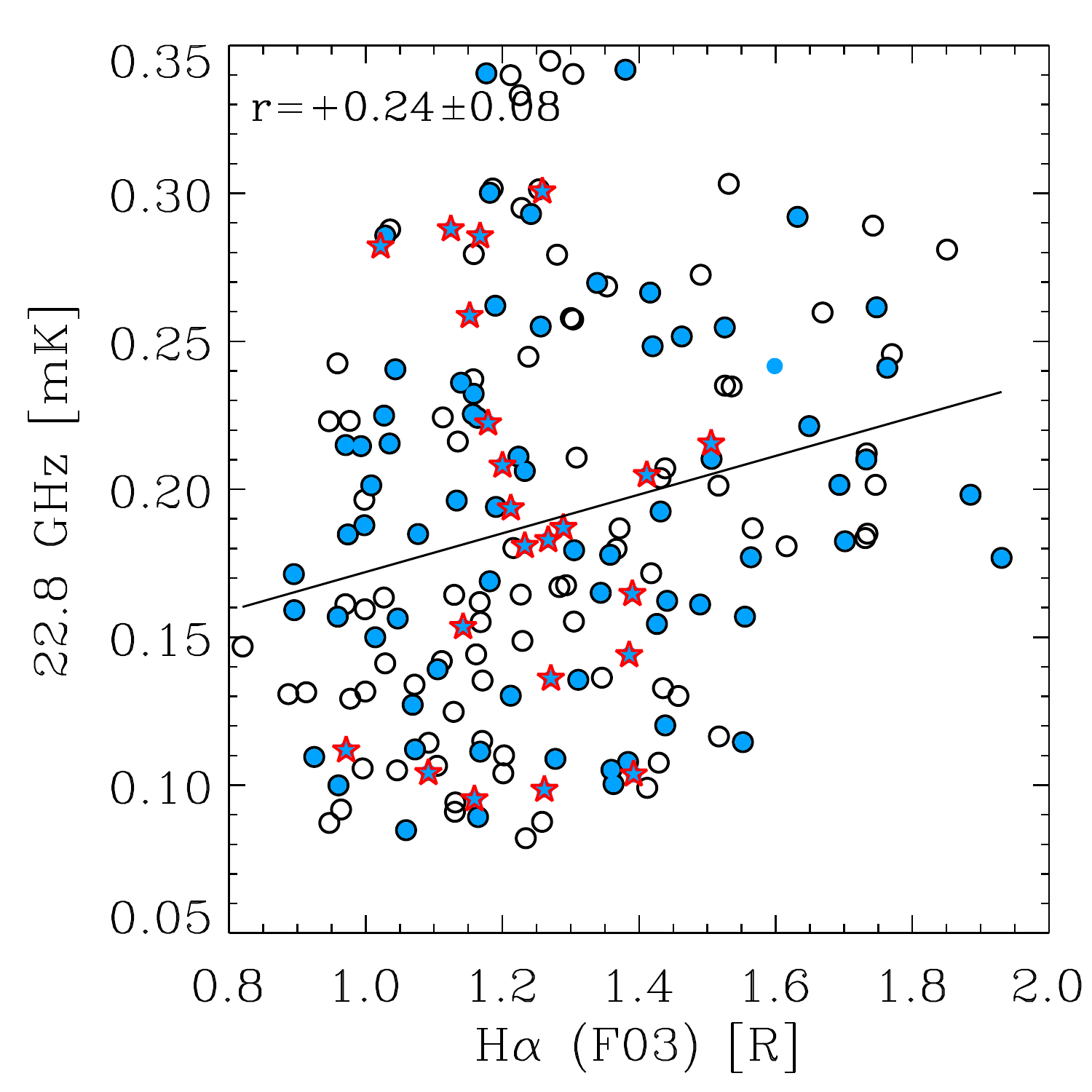}
\includegraphics[width=0.24\textwidth,angle=0]{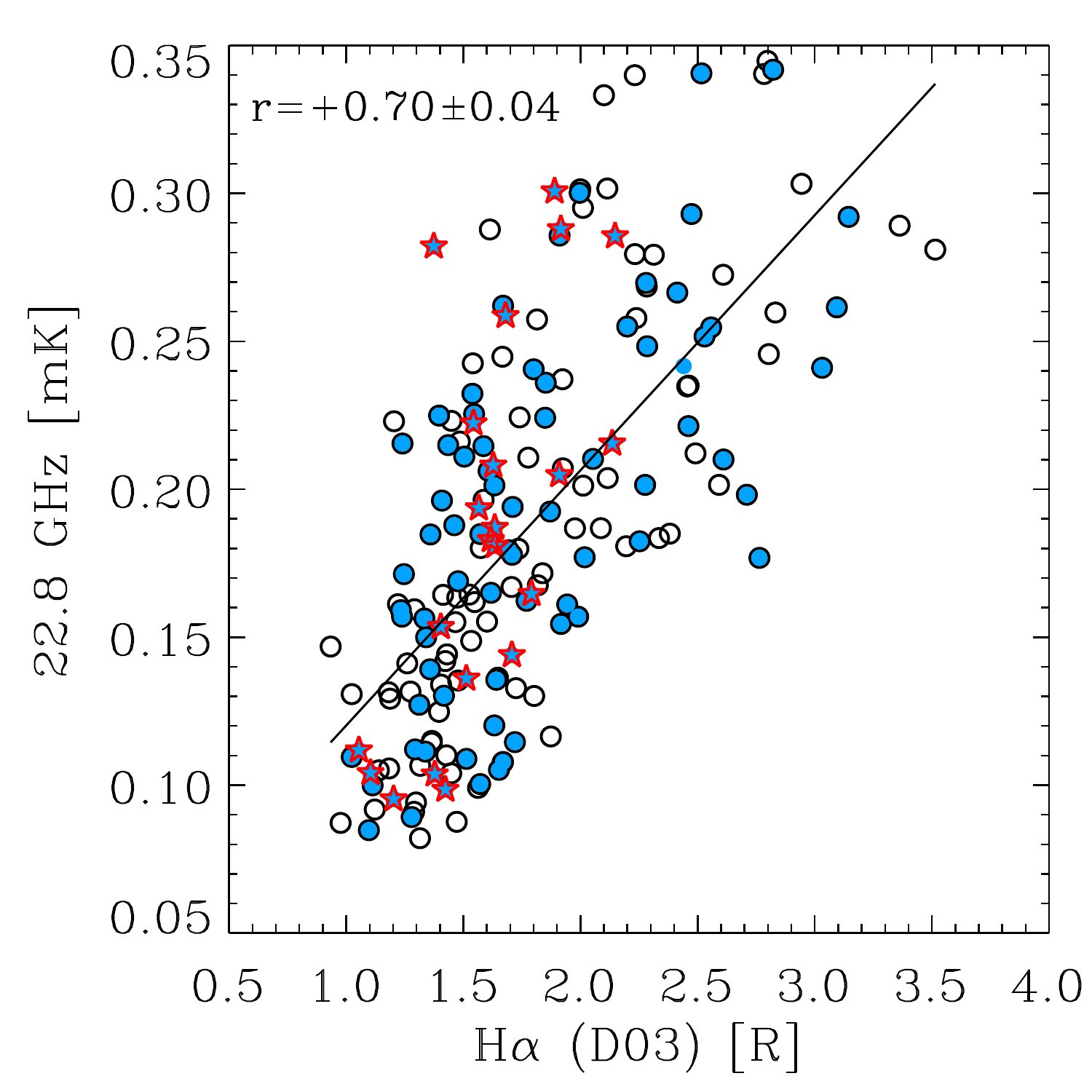}
\includegraphics[width=0.24\textwidth,angle=0]{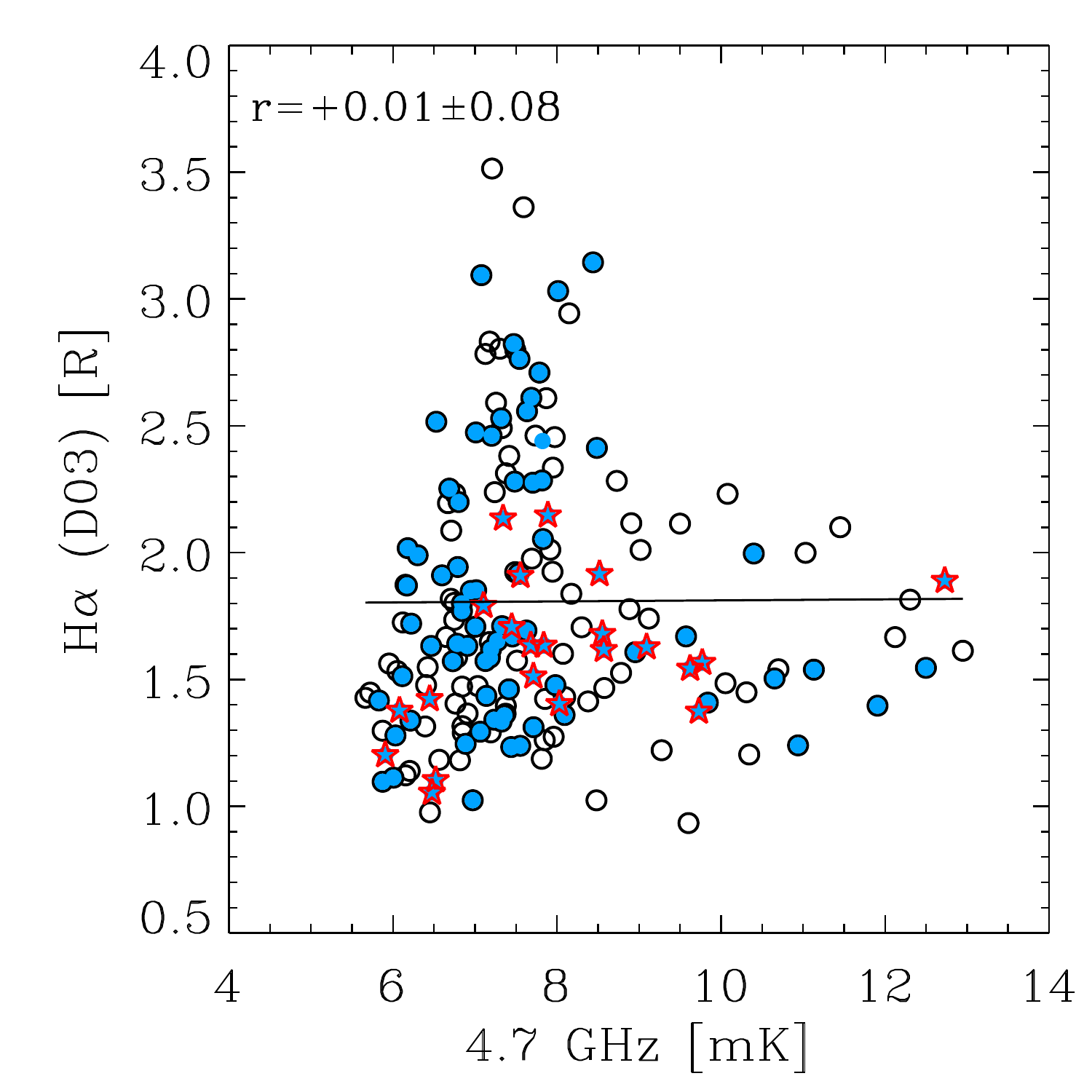}
\includegraphics[width=0.24\textwidth,angle=0]{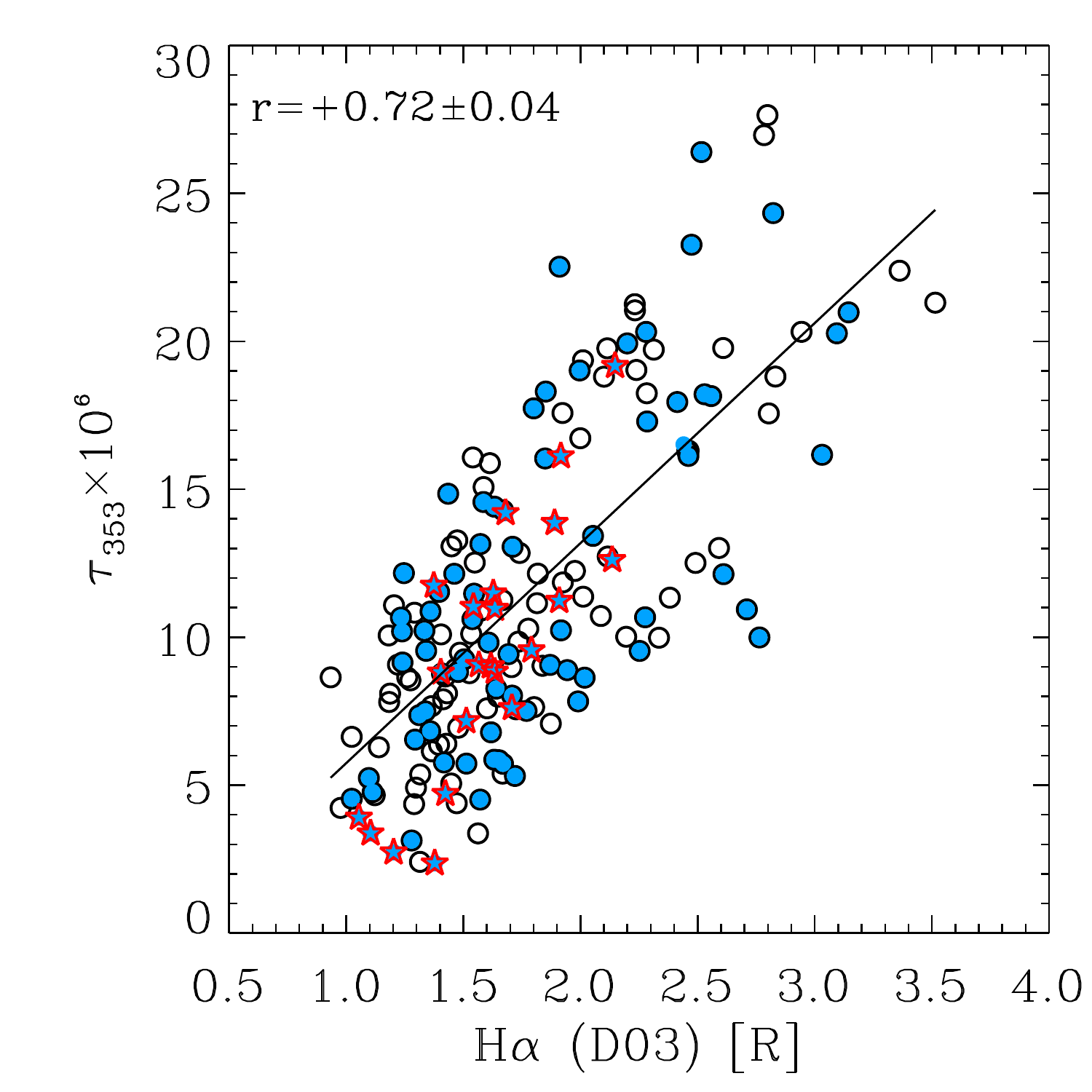}
\includegraphics[width=0.24\textwidth,angle=0]{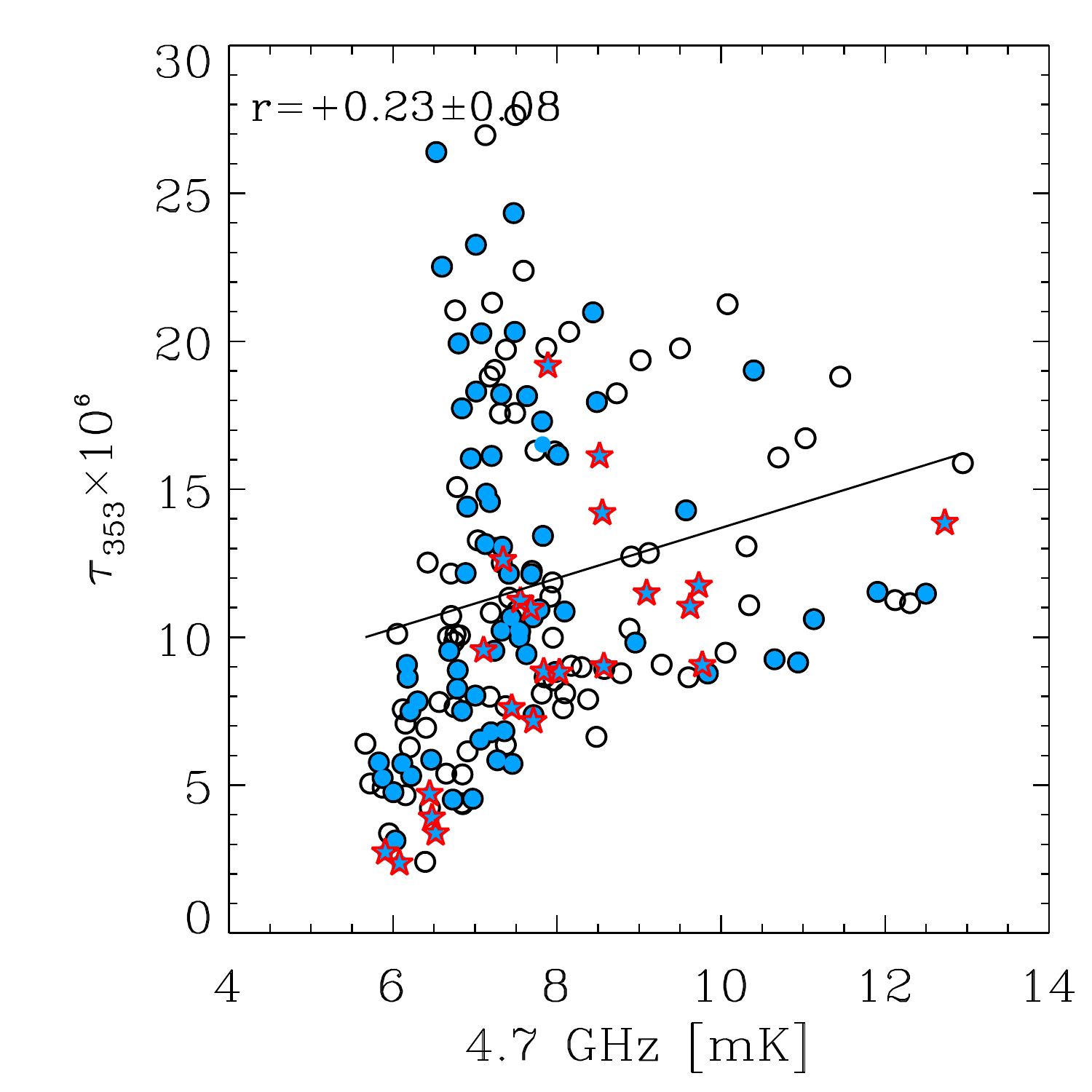}
\includegraphics[width=0.24\textwidth,angle=0]{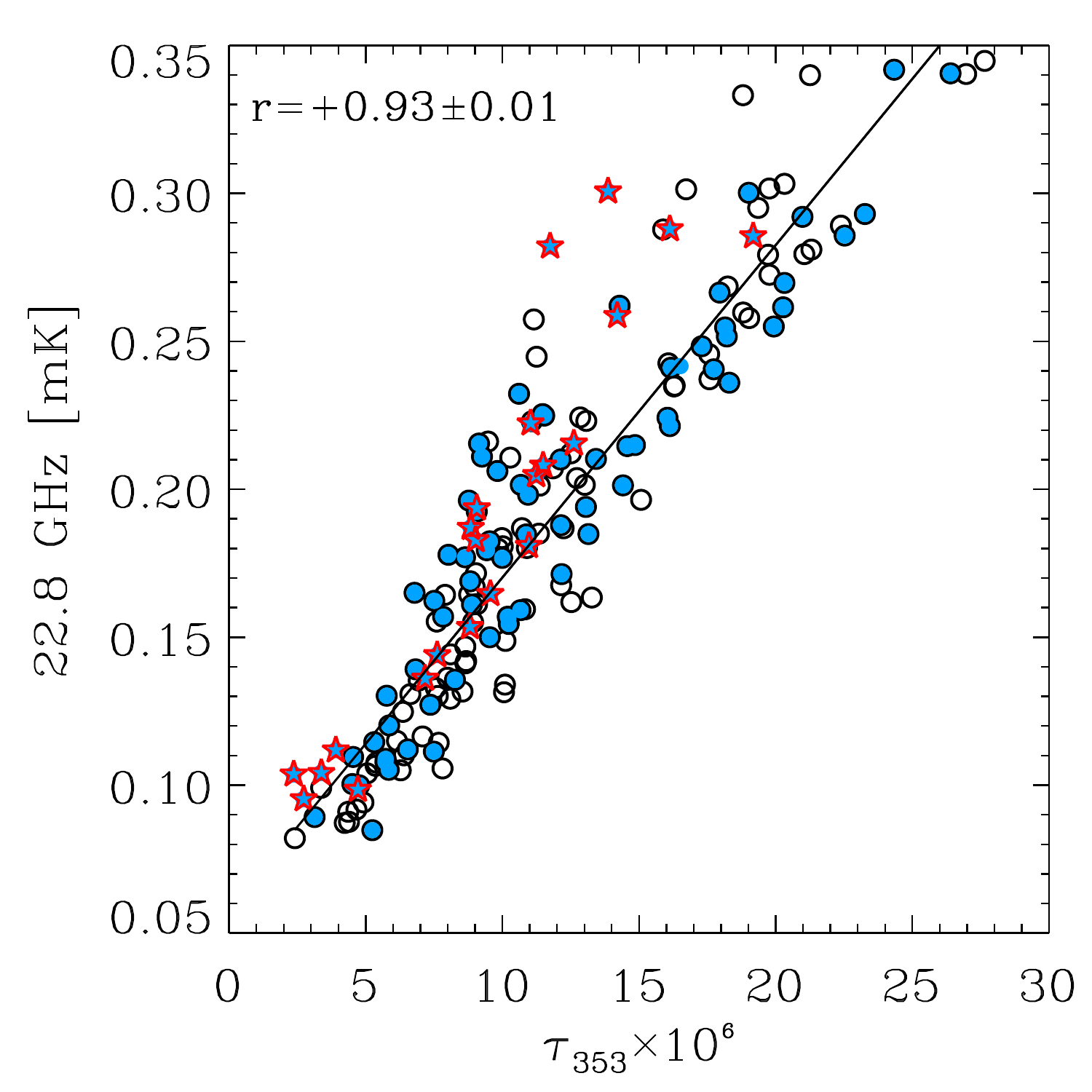}
\includegraphics[width=0.24\textwidth,angle=0]{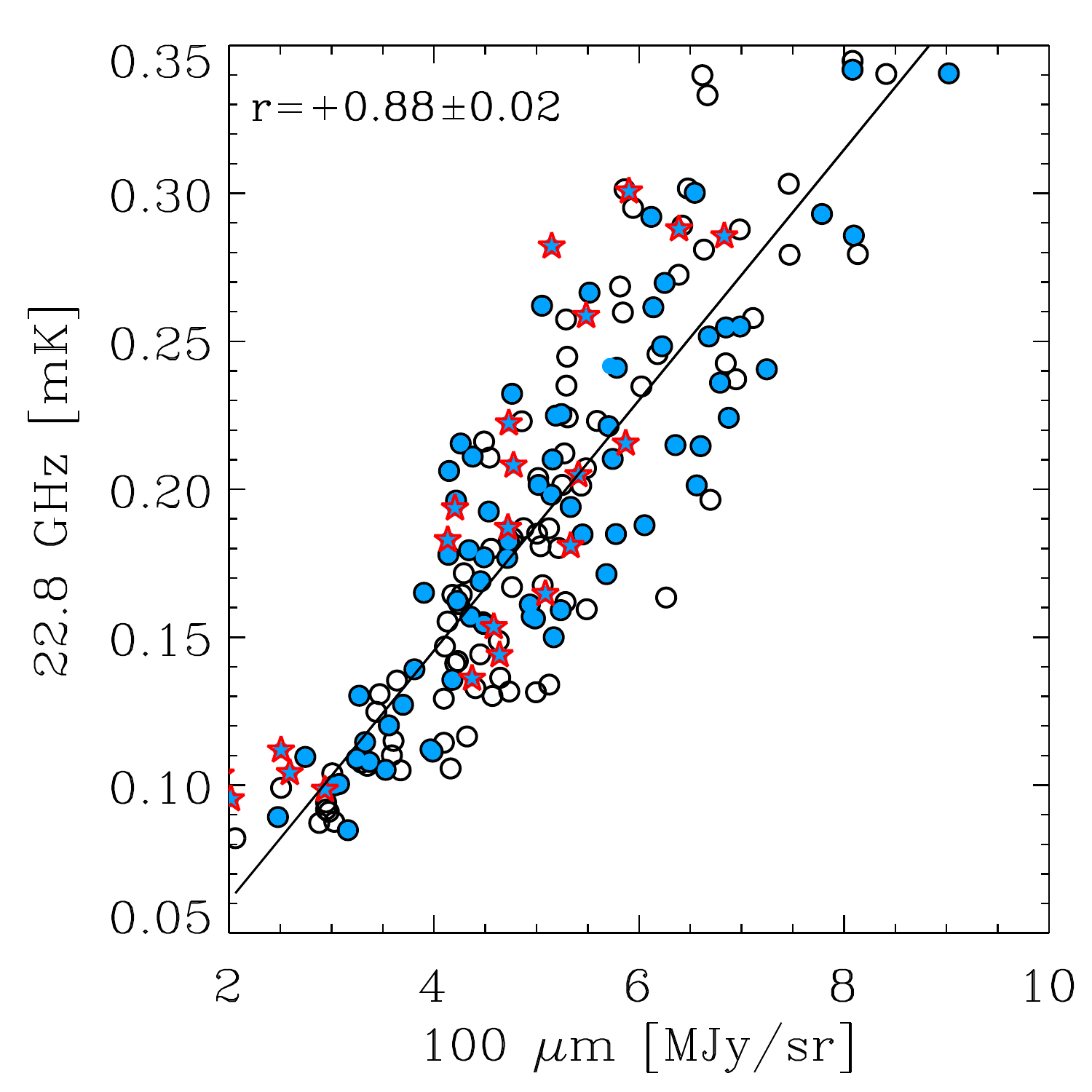}
\caption{T-T plots of the NCP region ($\delta > +83^{\circ}$) for various combinations of maps. The {\it circles} are for pixels used for template fitting. {\it Blue filled circles} are pixels that contain a weak ($< 600$\,mJy) extragalactic source from the \protect\cite{Mingaliev2007} survey. {\it Red stars} are pixels that contain a bright ($>600$\,mJy) source and are excluded from the analysis and fits. The line is the best-fitting straight line to the data after masking ({\it unfilled} and {\it filled blue circles}). The corresponding spectral index $\beta$ (where relevant) and Pearson correlation coefficient $r$ are also shown in each plot.}
\label{fig:TTplots}
\end{center}
\end{figure*}

\subsection{Multi-frequency maps of the NCP region at $1^{\circ}$ resolution}
\label{sec:multimaps}

We now compare the \mbox{C-BASS} data with multi-frequency radio, microwave and FIR data. The datasets are summarized in Table~\ref{tab:data}. The absolute calibration uncertainties are those assumed later in the analysis. Data from {\it WMAP} 9-year \citep{Bennett2013} and {\it Planck} 2018 results (PR3) \mbox{\citep{Planck2018_I}} are the primary data being analysed. We assume a conservative minimum 3\,\% (5\,\%) uncertainty for {\it WMAP}/{\it Planck} LFI (HFI) data primarily due to not applying colour corrections. The uncertainty encompasses any additional errors such as those due to beam asymmetries and other potential low ($<1\,\%$) level errors.

We include low-frequency radio data from the \cite{Haslam1982} 408\,MHz and 1.42\,GHz \cite{Reich1986} surveys. Both maps have been filtered to remove the brightest point sources and stripes due to scanning artefacts. At 408\,MHz, we choose to use the improved destriped and desourced version of \cite{Remazeilles2015} although the results are consistent between the two. For the 1.42\,GHz data we use the full-sky version, which has been destriped and desourced \citep{Reich1982,Reich1986,Reich2001}. The 1.42\,GHz map has been multiplied by a factor of 1.55 to place it on the ``full-beam'' scale, appropriate for diffuse foregrounds (W. Reich, priv. comm.). We assume a nominal 10\,\% calibration uncertainty for both maps. 

As tracers of free-free emission, we use the full-sky H$\alpha$ maps of \cite{Dickinson2003} (D03) and \cite{Finkbeiner2003} (F03), both of which are based on Wisconsin H-Alpha Mapper (WHAM; \citealt{Haffner2003}) data. Both versions contain nominal corrections for dust extinction, which are modest ($\lesssim 0.2$\,mag) at intermediate and high Galactic latitudes, and therefore possible errors in the corrections are not critical.

As tracers of dust (thermal dust and AME), we use several different template maps, listed in Table~\ref{tab:data}. Our primary dust template is the map of the thermal dust optical depth at 353\,GHz, $\tau_{353}$, based on {\it Planck} data \citep{Planck2013_XI,PIP_XLVIII}, which is known to be a good tracer of the dust column density and is a good tracer of AME \citep{Planck2015_XXV}. As will be discussed in Sect.\,\ref{sec:discussion_ame}, this was found to be the best tracer of AME in the NCP region. We also use several other standard dust tracers including the IRIS reprocessed IRAS 100\,$\mu$m map by \cite{Miville-Deschenes2005}, {\it Planck} HFI maps at 353, 545 and 857\,GHz, the map of dust radiance $\Re$ \citep{Planck2013_XI,PIP_XLVIII}, and the FDS model 8 map at 94\,GHz by \cite{Finkbeiner1999}.

To facilitate the comparison, we smooth all the maps to a common $1^{\circ}$\,FWHM (Gaussian point spread function) angular resolution by smoothing the data with the appropriate Gaussian kernel (or the more accurate beam window functions in the case of {\it Planck} data) and re-sampled the maps to a common \mbox{$N_{\rm side}=256$} HEALPix grid. Fig.~\ref{fig:maps} displays $1^{\circ}$-smoothed maps of the NCP region. Strong Galactic emission is seen towards the bottom of the maps in the direction of the Galactic plane. Point sources are evident, especially at the lower frequencies. AME is directly visible at frequencies $\sim 20$--40\,GHz near the centre of the map, with morphology that resembles thermal dust emission at higher frequencies (e.g., 545\,GHz), FIR (e.g., 100\,$\mu$m), and $\tau_{353}$.

The low-frequency radio maps at 0.408 and 1.42\,GHz show clear residual scanning artefacts. These are visible as striations emanating radially outwards from the NCP. This is particularly noticeable in the 0.408\,GHz map. The \mbox{C-BASS} 4.7\,GHz map on the other hand shows no obvious visible artefacts. The 1.42\,GHz map also appears to contain diffuse structure near the NCP that is not visible in either of the 0.408 or 4.7\,GHz maps, nor in H{\sc i} 21\,cm maps \citep{Winkel2016}, and is likely to be an artefact. Nevertheless, none of the low-frequency maps, including \mbox{C-BASS} 4.7\,GHz, show emission that resembles the AME seen at 20--40\,GHz. Given the higher frequency of the \mbox{C-BASS} data, and improved map fidelity, it should be a much more reliable foreground template for synchrotron emission. This will be quantified in detail in Section~\ref{sec:template_fitting}.

The H$\alpha$ maps contain low-level fluctuations at the level of $\approx 1$\,R. However, the structure in the two versions of the map (D03 and F03)  differs in detail, particularly in the central region where the AME is most prominent. Although the original data were the same, there were significant differences in the processing of the data. In particular, removal of stellar residuals in the H$\alpha$ maps is difficult and can result in artefacts at the level of $\approx 1$\,R. We will compare the results of both maps to see how different they are and which one is most consistent with theoretical expectations.

To understand how the structure in these maps correlates, we begin by plotting the pixel intensities of each map against other maps, i.e., T-T plots \citep{Turtle1962}. The data are first smoothed to a common $1^{\circ}$\,FWHM resolution and downgraded to \mbox{$N_{\rm side}=64$} ($\approx 55$\,arcmin pixels) to reduce correlations and make each pixel quasi-independent. For clarity and to reduce the correlations from the nearby Galactic plane, we only plot pixels at $\delta > +83^{\circ}$. This helps isolate the AME clouds away from the plane and reduces the number of plotted symbols in the figures.

Fig.~\ref{fig:TTplots} shows T-T plots for several combinations of radio data, microwave data (specifically the 22.8\,GHz {\it WMAP} map), {\it Planck} 545\,GHz map, {\it Planck} thermal dust optical depth map ($\tau_{353}$), and H$\alpha$ (D03 and F03). Pixels that are masked for the brightest ($>600$\,mJy) sources are marked as red stars while pixels containing known sources above 200\,mJy at 4.8\,GHz are marked as filled blue circles. It is reassuring to see that there is good correlation of the radio maps at 0.408, 1.42 and 4.7\,GHz. However, there is significant scatter (Pearson correlation coefficient, $r \approx 0.7$) which would not be expected given the high signal-to-noise ratio of these maps. Part of this may be due to variations in the synchrotron spectral index across the map and free-free emission, or to differences in source subtraction in the maps. However, inspection of the maps (Fig.~\ref{fig:maps}) clearly reveals artefacts in the maps that are likely responsible for the majority of the scatter. The brightest sources (red stars in Fig.\,\ref{fig:TTplots}) can be seen to have some effect in some of the radio maps, pushing the intensities to larger values, but in general they are not a major issue; the \mbox{C-BASS} data have been source-subtracted and those pixels containing sources above 600\,mJy are masked in the analysis.

The best-fitting straight line, $y=mx+c$, is plotted for each combination of datasets, taking into account uncertainties in both coordinates using the {\sc mpfitexy}\footnote{\url{http://purl.org/mike/mpfitexy}} routine \citep{Markwardt2009,Williams2010}. Only unmasked pixels are included in the fit. The slope, $m$, of each T-T plot between frequencies $\nu_1$ and $\nu_2$, is related to the spectral index by $\beta={\rm ln}(m)/{\rm ln}(\nu_1/\nu_2)$. The spectral index\footnote{Unless otherwise stated, uncertainties in spectral indices include an absolute calibration error term, given in Table~\ref{tab:data}, which is added in quadrature with the intrinsic noise uncertainty.} between 0.408 and 1.42\,GHz is $\beta=-2.95\pm0.65$, which is indicative of steep synchrotron radiation. Note that we have rescaled the uncertainties to take into account the scatter in the data by scaling the uncertainties until $\chi^2_r=\chi^2/\nu=1$,\footnote{We use the {\sc /reduce} option in the {\sc mpfitexy} code.} where $\nu$ is the number of degrees of freedom. For the cases where there is significant scatter, this increases the fitted uncertainties by a factor of several. 

The spectral index between 0.408 and 4.7\,GHz is $\beta=-2.79\pm0.17$ and between 1.42 and 4.7\,GHz it is $\beta=-2.57\pm0.77$. These are consistent with typical values at these frequencies of $\beta\approx -2.8$ with variations of $\Delta \beta \approx 0.2$ \citep{Reich1988,Platania1998,Davies1996}. There is a hint of a slight flattening at higher frequencies, but at less than $1\sigma$ confidence level. If this were the case, we would expect to see this reflected in the cross-correlation analysis (Sect.~\ref{sec:template_fitting}) when using the \mbox{C-BASS} data as a synchrotron template, which will be discussed further in Sect.\,\ref{sec:discussion_synch}. Note that the T-T spectral index between 22.8\,GHz and 44.1\,GHz is $\beta=-3.01\pm0.06$, which is similar to that of synchrotron radiation. However, as we show in  Sects.~\ref{sec:template_fitting}\,and\,\ref{sec:discussion} below, it is actually primarily due to AME. Indeed, the T-T spectral index from 4.7\,GHz to 22.8\,GHz has a flatter value of $-2.44\pm0.74$ which suggests a new component is contributing, while the large scatter (and hence larger uncertainty) shows that the two maps are not tightly correlated, with a Pearson correlation coefficient of $r=0.48\pm0.06$.

The lack of perfect correlation between the two versions of the H$\alpha$ map, D03 and F03 ($r=0.81\pm0.03$), given that they are constructed essentially from the same data, confirms the significant differences between them already noted. The correlation between D03 and the microwave maps at 20--40\,GHz is much stronger  ($r=0.70\pm0.04$ at 22.8\,GHz) than for the F03 map ($r=0.24\pm0.08$). Given that the maps were created independently of the microwave data at a completely different wavelength (in this case, at optical wavelengths), this suggests that the D03 map may be more reliable, at least for this region. We will therefore use the D03 H$\alpha$ map for the main results, but will also consider F03 to test how sensitive the results are to changes in the free-free template (see Sect.\,\ref{sec:discussion_ff}).

The most important result here is that the dust-correlated AME emission clearly visible at 22.8\,GHz (and frequencies in the range $\approx 20$--40\,GHz) is not visible at 4.7\,GHz. The T-T plot (Fig.~\ref{fig:TTplots}) shows that there is some correlation between 4.7\,GHz and 22.8\,GHz ($r=0.48\pm0.06$), but 4.7\,GHz is much less correlated with FIR dust emission than the microwave frequencies are. For example, the correlation coefficient between 22.8\,GHz and $100\,\mu$m is $r=0.88\pm0.02$, and 22.8\,GHz and $\tau_{353}$ gives an even tighter correlation of $r=0.93\pm0.01$, while the correlation between 4.7\,GHz and $\tau_{353}$ is only $r=0.23\pm 0.08$. As noted previously \citep[e.g.,][]{Tibbs2013,Planck2015_XXV} the AME appears to correlate better with $\tau_{353}$ (which is approximately proportional to the line-of-sight column density) than with the FIR such as $100\,\mu$m. 

In summary, the AME visible at 20--40\,GHz does not appear to be related to the emission at 4.7\,GHz and below, i.e., synchrotron or free-free radiation. This will be quantified further in Section~\ref{sec:template_fitting}. Nevertheless, even without further analysis, these morphological comparisons suggest that unaccounted synchrotron and free-free emission cannot be responsible for the majority of the AME in the NCP region.


\section{Template fitting}
\label{sec:template_fitting}

\subsection{Template fitting method}
\label{sec:template_fitting_method}

To separate the contributions of the diffuse foreground components, we will use the different spatial morphologies, as traced by foreground template maps. Table~\ref{tab:data} summarizes the data sets that are used and the ancillary data on which the foreground templates are based. The template fitting method is well-known (see, e.g., \cite{Ghosh2012} and references therein). Briefly, we assume the data vector ${\mathbf d}$ at a given frequency is the sum of each template map vector ${\mathbf t_i}$ multiplied by a template correlation coefficient, $\alpha_i$. The data are then corrupted by noise ${\mathbf n}$, which can consist of various terms including instrumental noise, CMB fluctuations or point sources. For $N$ templates (e.g., 3 foreground components and an offset term), the data vector reads:
\begin{equation}
{\mathbf d} = \sum_{i}^{N} \alpha_i {\mathbf t}_{i} + {\mathbf n} ~.
\end{equation}
The $\chi^2$ for this model, allowing for correlations in noise between pixels is given by:
\begin{equation}
\chi^{2}\ =\ ({\mathbf d}\ -\ \alpha_i {\mathbf t}_{i})^{T}\, {\mathbf\mathrm {C}}^{-1}\, ({\mathbf d} \ -\ \alpha_i {\mathbf t}_{i}) ~.
\label{eqn:chi2}
\end{equation}
Here, the covariance matrix ${\mathbf \mathrm C}$ contains all the sources of noise. In our analysis this can be instrumental noise ($C_{\rm noise}$), CMB fluctuations ($C_{\rm CMB}$), or extragalactic point sources ($C_{\rm PS}$), i.e., 
\begin{equation}
C = C_{\rm noise} + C_{\rm CMB} + C_{\rm PS}~.
\end{equation}

For white, independent Gaussian noise, only the diagonal elements will be non-zero, and are equal to the variance $\sigma^2$. We choose to degrade the maps to \mbox{$N_{\rm side}=64$} (pixels $\approx 55$\,arcmin on-a-side) for the cross-correlation analysis, which means that the instrumental noise is approximately diagonal. To take into account CMB fluctuations, we can add in the CMB covariance matrix, with prior knowledge of the CMB power spectrum $C_{\ell}$. The CMB fluctuations are very close to Gaussian, allowing the CMB covariance matrix to be calculated from the power spectrum alone. It is given by:
\begin{equation}
{\mathbf\mathrm C_{\rm CMB}}\ =\ \frac{1}{4\pi} \sum_{\ell}\  (2\ell + 1) \ C_{\ell}\ P_{\ell}(\cos \theta)\ W_{\ell}~,  
\end{equation}
where $P_{\ell}(\cos\theta)$ is the Legendre polynomial order and the sum runs from $\ell=2$ (monopole and dipole are assumed to have been removed) to \mbox{$\ell_{\rm max}=3N_{\rm side}-1$}. $W_{\ell}$ is the window function, which accounts for the beam $b_{\ell}$ and pixel $p_{\ell}$ window functions. However, treating the CMB in this statistical way increases the uncertainties typically by a factor of $\sim 5$ or more. We therefore do not consider the non-subtracted maps further in this paper. For our final results, we instead directly remove the CMB by subtracting CMB maps produced using component separation algorithms and add in an additional term in the covariance matrix to account for the accuracy to which the CMB map is known (see Sect.~\ref{sec:results}).

By minimizing the $\chi^2$ with respect to $\alpha_i$, the coupling constant $\alpha_i$ can be estimated:
\begin{equation}
\alpha_i\ =\ \frac{{\mathbf t}_{i}\, {\mathbf\mathrm {C}}^{-1}\, {\mathbf d}}{{\mathbf t}_{i}\, {\mathbf\mathrm {C}}^{-1}\, {\mathbf t}_{i}}
\end{equation}
\noindent and its Gaussian uncertainty is given by
\begin{equation}
\sigma^{2}(\alpha_i) =\  ({\mathbf t}_{i}\, {\mathbf\mathrm {C}}^{-1}\, {\mathbf t}_{i})^{-1}~.
\end{equation}
The contribution of component $i$ to the observed sky is the map $\alpha_i T_i$, where $T_i$ is in the natural units of the foreground template map
so that this scaled quantity is in mK. 


\subsection{Contribution of point sources}
\label{sec:sources2}

As discussed in Sect.~\ref{sec:sources}, there is a significant contribution from extragalactic point sources, both in the low-frequency radio maps and, to a lesser extent, the microwave maps. In this paper we are interested in the diffuse Galactic emission and therefore this contribution must be quantified and mitigated. Our main strategy for point-source mitigation is to mask the brightest sources in the analysis. Smoothing the data to $1^{\circ}$ resolution also reduces the impact of point-like sources. 

We use the 4.8\,GHz measurements of \cite{Mingaliev2007}, who measured all sources brighter than 200\,mJy in the 1.4\,GHz NVSS catalogue \citep{Condon1998}. Pixels that have any part of their area within $0.\!^{\circ}7$ of a source with $S_{4.8}>600$\,mJy are masked. We varied this cut-off and found only a slight dependence when including pixels contaminated with sources above 600\,mJy; below this cut-off, the results were consistent within the uncertainties. We also mask $\delta>+89^{\circ}$ because the measurements do not cover the very highest declinations ($\delta >+88^{\circ}$), where \mbox{C-BASS} appears to see at least one relatively bright ($\sim 500$\,mJy) source very close to the NCP. 

For the sources in the higher frequency maps (22.8\,GHz and above), we can take the fainter sources into account statistically by including additional terms in the covariance matrix. Assuming sources are Poisson distributed, in the limit of a large number of sources $N$ the map fluctuations tend to a Gaussian distribution, and the point source covariance matrix, $C_{\rm PS}$ becomes diagonal. The power spectrum of point sources is given from the source counts $dN/dS$ by 
\begin{equation}
C_{\ell}^{\rm PS} = g(\nu)^2 \int_{0}^{S_{\rm max}} S^2 \frac{dN}{dS} dS~,
\end{equation}
where $g(\nu) = 2k \nu^2/c^2$ converts flux density $S$ (in Jy) to brightness temperature units (K). 

There are a number of measurements of source counts at frequencies near 20\,GHz \citep[e.g.,][]{deZotti2010}. We choose to use the simple power-law fit of \cite{Davies2011} to 15.7\,GHz data from the 9C/10C surveys because they measure sources down to millijansky levels, well below our cut-off flux density limit of 0.6\,Jy. They measure $dN/dS \approx 48 S^{-2.13}$\,Jy$^{-1}$\,sr$^{-1}$ in the flux range from 2.2\,mJy to 1\,Jy. Integrating this function from 2.2\,mJy to our nominal flux cut-off of 600\,mJy (there were no sources brighter than this limit not being masked in the {\it Planck} point source catalogues) gives $\Delta T=11.8\,\mu$K rms per \mbox{$N_{\rm side}=64$} pixel at 22.8\,GHz. The contribution of faint sources is therefore small but not completely negligible and so we include this contribution in the noise covariance matrix in our analysis. 

We choose not to scale the source counts from 15.7\,GHz to the higher frequencies. The brighter sources are typically flat spectrum ($\alpha=0$) at frequencies relevant to AME \citep[see, e.g.,][]{deZotti2010}, and hence scaling of these is not necessary. The fainter sources are typically steep spectrum ($\alpha \approx -0.5$) and hence these give a smaller contribution to the source power at higher frequencies. This means that our uncertainties  for this contribution will be slightly over-estimated at higher frequencies, but this has negligible impact on the results.

\begin{table*}
\caption{Cross-correlation template fitting results for $\delta_{\rm lim}>+80^{\circ}$. Each entry is a correlation coefficient in units given in the second column. The two halves of the table are for when the Haslam 408\,MHz ({\it top}) and C-BASS 4.7\,GHz ({\it bottom}) maps are used as the synchrotron template. CMB fluctuations have been removed by direct subtraction of the {\it Planck} SMICA map. Uncertainties do not include absolute calibration errors.}
\begin{tabular}{lcccccc}
\hline
Template  &Unit &{\it WMAP} &{\it Planck}  &{\it WMAP}   &{\it WMAP}   &{\it Planck}  \\
         &     &22.8\,GHz  &28.4\,GHz     &33.0\,GHz    &40.7\,GHz    &44.1\,GHz   \\ \hline
Synchrotron (Haslam) &$\mu$K\,K$^{-1}$ &$9.19\pm0.39$ &$5.22\pm0.27$ &$3.19\pm0.25$ &$1.70\pm0.23$ &$1.54\pm0.22$ \\
H$\alpha$ (D03) &$\mu$K\,R$^{-1}$ &$\!\!\!10.2\pm1.9$ &$\!\!\!5.66\pm1.6$ &$\!\!\!5.09\pm1.6$ &$\!\!\!2.83\pm1.6$ &$\!\!\!3.84\pm1.5$ \\
Dust ($\tau_{353}$) &K &$9.93\pm0.35$ &$4.77\pm0.21$ &$3.11\pm0.18$ &$1.57\pm0.16$ &$1.11\pm0.15$ \\
\hline
Synchrotron (C-BASS) &$\mu$K\,mK$^{-1}$ &$10.1\pm0.41$ &$5.64\pm0.29$ &$3.48\pm0.26$ &$1.80\pm0.24$ &$1.67\pm0.23$ \\
H$\alpha$ (D03) &$\mu$K\,R$^{-1}$ &$\!\!\!13.1\pm1.9$ &$\!\!\!7.30\pm1.6$ &$\!\!\!6.12\pm1.6$ &$\!\!\!3.36\pm1.6$ &$\!\!\!4.33\pm1.5$ \\
Dust ($\tau_{353}$) &K &$9.52\pm0.34$ &$4.56\pm0.21$ &$2.98\pm0.18$ &$1.51\pm0.16$ &$1.05\pm0.15$ \\
\hline
\end{tabular}
\label{tab:results_cmbsub}
\end{table*}


\subsection{Template fitting of {\it WMAP/Planck} data}
\label{sec:results}

We apply the template fitting method to maps at \mbox{$N_{\rm side}=64$}, at which resolution the pixels are close to independent. The data vectors include pixels above a declination limit $\delta_{\rm lim}$, which we vary in the range $+75^{\circ} < \delta_{\rm lim} < +85^{\circ}$ to test the robustness of the results against the precise sky area. We quote the primary results for $\delta_{\rm lim}=+80^{\circ}$, which allows the two bright dust clouds in the region to be included as well as some of the brighter emission at lower Galactic latitudes. The results for different declinations were generally consistent within the uncertainties ($<2\sigma$), except for the synchrotron coefficient which varies by $\approx 4\sigma$ when considering just the inner portion of the map at $\delta>+83^{\circ}$; this is discussed in Sect.~\ref{sec:discussion_synch}. 

The best-fitting template coefficients are listed in Table~\ref{tab:results_cmbsub}. The coefficients represent the amount of emissivity at each {\it WMAP/Planck} frequency, relative to the template foreground component.\footnote{Template coefficient units are typically brightness (e.g., $\mu$K) per unit template (e.g., K or MJy\,sr$^{-1}$). In the case of $\tau_{353}$, which is dimensionless, the coefficient is just brightness (K).} We do not include absolute calibration uncertainties when quoting correlation coefficients or rms brightness temperatures. We focus on the lower channels of {\it WMAP/Planck} (22.8--44.1\,GHz) where AME is strongest. We quote the results after direct removal of the CMB using the SMICA CMB map described in \mbox{\cite{Planck2018_IV}}. Using alternative CMB maps from {\it Planck}  (Commander, SEVEM, NILC) gave consistent results within the uncertainties. In this case we have assumed a conservative CMB residual ``noise'' per pixel of $10\,\mu$K to account for the fact that the component separation is not perfect. This is informed by inspection of the four CMB maps described in \cite{Planck2015_IX,Planck2018_IV} where typical differences (away from bright sources) are of order $5\,\mu$K. This is comparable to the typical instrumental rms noise in the {\it WMAP}/{\it Planck} data at frequencies $\sim 20$--40\,GHz at \mbox{$N_{\rm side}=64$}.

We fit for a synchrotron component using either the \cite{Haslam1982} map (top part of Table~\ref{tab:results_cmbsub}) or the \mbox{C-BASS} 4.7\,GHz map (bottom part). This allows us to test for a flatter-spectrum component of synchrotron radiation, that would be better traced by the higher frequency template from C-BASS. 

The H$\alpha$-correlated (free-free) template is the D03 all-sky H$\alpha$ map \citep{Dickinson2003}, with a correction for dust extinction (which is small at intermediate and high Galactic latitudes). Similar AME/synchrotron results are obtained when using the H$\alpha$ map of \cite{Finkbeiner2003} but with small (but significant) differences for the free-free component. As discussed earlier, the stronger correlation of microwave data with the D03 map suggests this version of the map is a better tracer of free-free emission. This will be discussed further in Sect.~\ref{sec:discussion_ff}. Although the free-free emission is a small fraction of the total emission at 4.7\,GHz, we subtract a model of free-free emission based on the D03 H$\alpha$ template, scaled with a value of 0.32\,mK\,R$^{-1}$ appropriate for this frequency \citep{Dickinson2003}. This has minimal impact on the results.

\begin{figure*}
\begin{center}
\includegraphics[width=0.3\textwidth,angle=0]{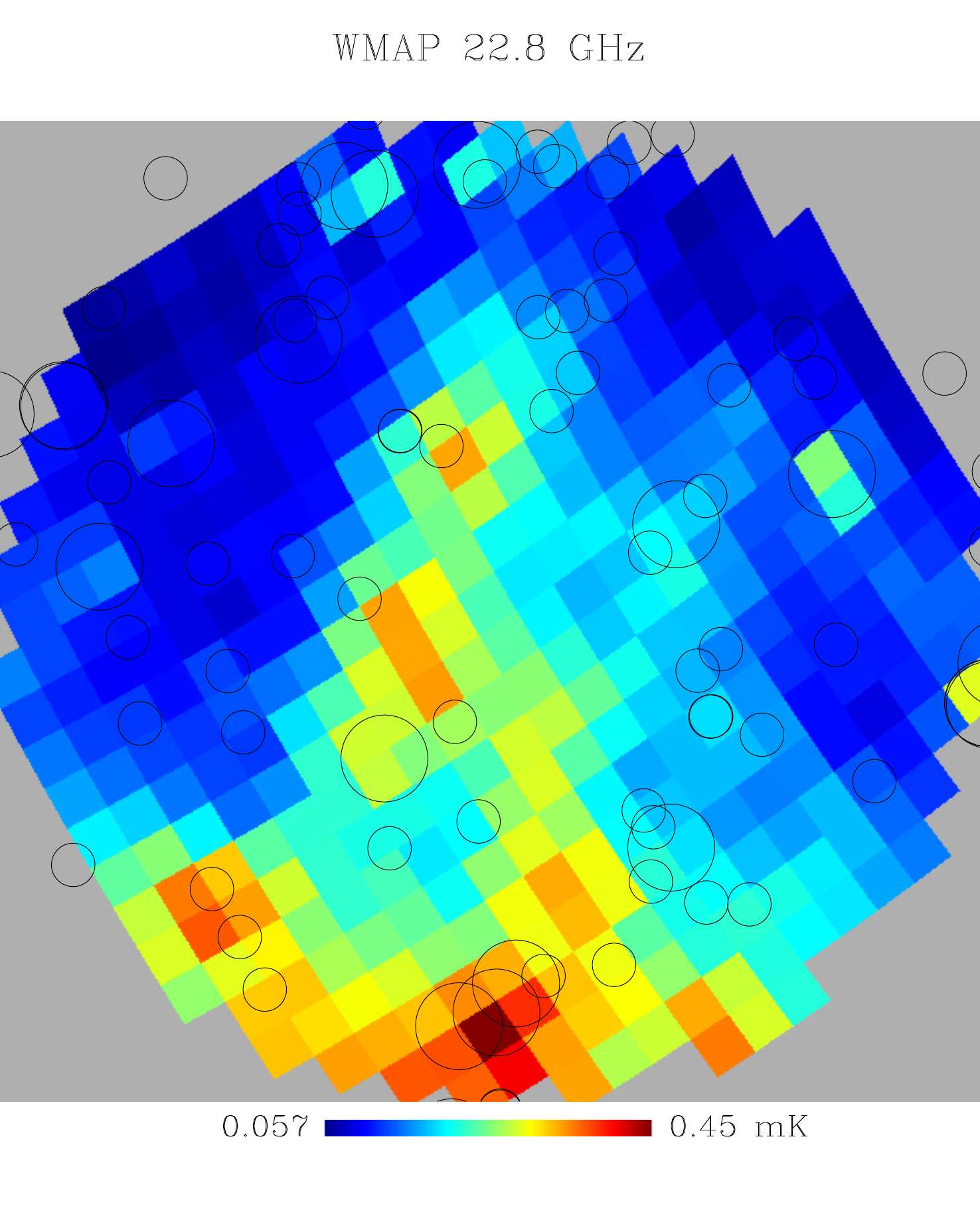}
\includegraphics[width=0.3\textwidth,angle=0]{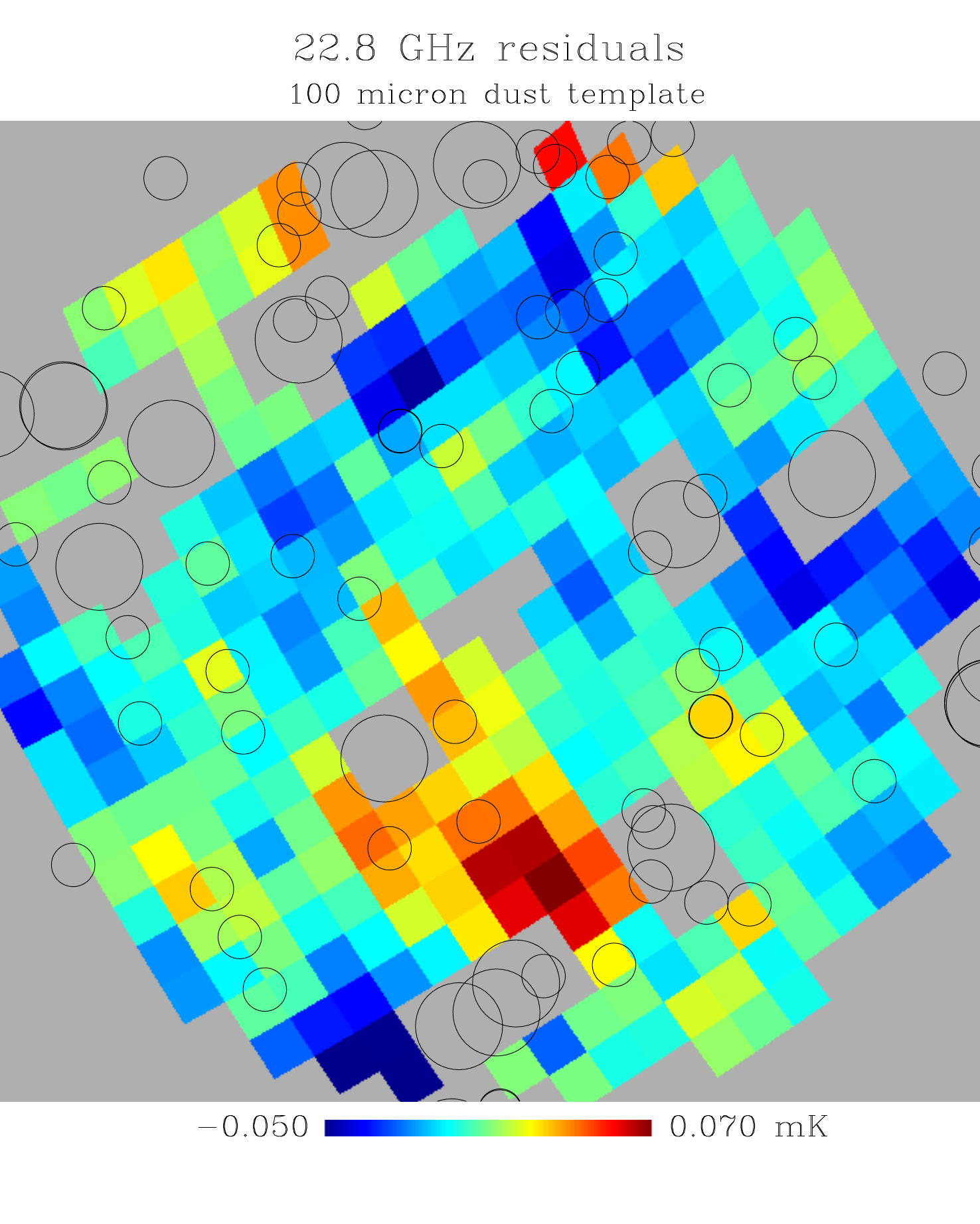}
\includegraphics[width=0.3\textwidth,angle=0]{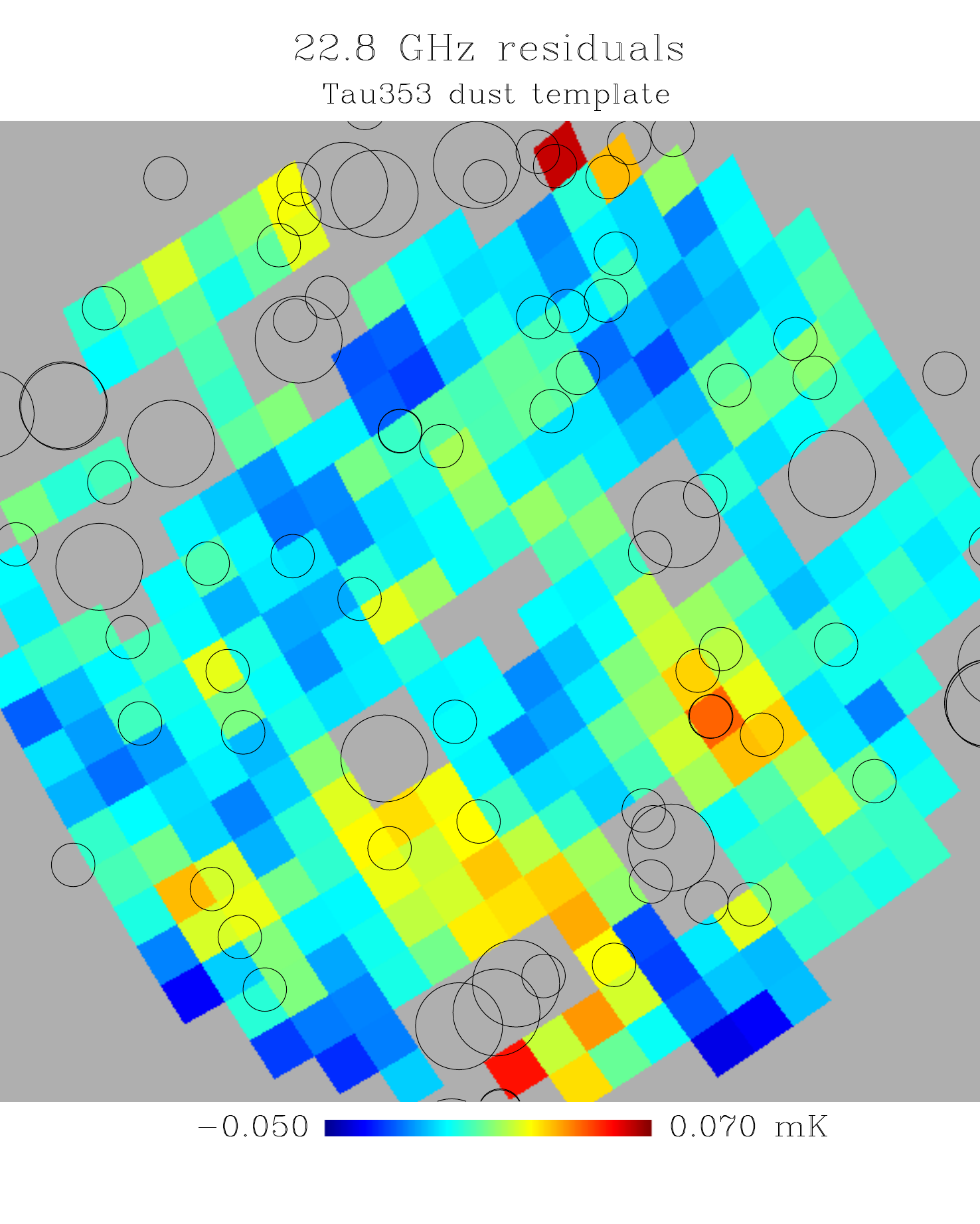}
\caption{WMAP $22.8$\,GHz map ({\it left}) and residual maps at $22.8$\,GHz after template fitting when using the $100\,\mu$m dust template ({\it middle}) and the $\tau_{353}$ dust template ({\it right}). Maps are at $1^{\circ}$ FWHM angular resolution and \mbox{$N_{\rm side}=64$}. The coordinate system is the same as in Fig.\,\ref{fig:maps}. The colour scales are linear but cover reduced temperature scale (by a factor of $\approx 7$) in the residual maps, shown by the minimum/maximum values to the side of the colour bar. There are larger dust-correlated residuals remaining when using the $100\,\mu$m compared to the $\tau_{353}$ as a tracer of dust.}
\label{fig:residmaps}
\end{center}
\end{figure*}

For the dust-correlated component we use a range of dust templates listed in Table~\ref{tab:data}. We quote the results for the best-fitting dust template, the {\it Planck} thermal dust optical depth map at 353\,GHz, $\tau_{353}$, although similar (but not identical) results are found with the rest. Fig.\,\ref{fig:residmaps} shows the 22.8\,GHz map alongside residual maps after template fitting, when using the $100\,\mu$m and $\tau_{353}$ maps as dust templates. One can clearly see significant residuals when using the $100\,\mu$m map. These will be discussed further in Sect.~\ref{sec:discussion_ame}. For the best-fitting templates, the residual 22.8\,GHz map after subtraction of the sky components has an rms of $15\,\mu$K ($21\,\mu$K when using the $100\,\mu$m dust template), which is comparable to the noise level in this region of 5--10\,$\mu$K rms; no obvious large-scale emission is evident. The residuals are mostly due to small contribution ($\sim 5\%$ rms compared to the total rms in the map) from extragalactic sources that have not been masked. This is reflected in the $\chi_r^2$ values, but has been shown to have little impact on the results when trying different flux level cuts (Sect.~\ref{sec:sources2}).

We fit for an offset term since the zero levels, particularly in the low-frequency radio data, are not well determined (the {\it WMAP}/{\it Planck} data have had a correction for the monopole term, but residual monopoles will exist at some level). This term accounts for this difference and means that the \mbox{C-BASS} zero-level is irrelevant. We also tried fitting for a 2-dimensional plane across the field to account for the large-scale Galactic gradient across the field, which could potentially dominate the correlation. We found that the results with or without this term were consistent within the uncertainties. This suggests that either the large-scale emission is not dominant, or it has a spectrum similar to that in the middle of the NCP field.

The results, and in particular the variations of the three components with different templates and assumptions, will be discussed further in Section~\ref{sec:discussion}.


\subsection{Foreground SEDs}
\label{sec:seds}

We can convert the template coefficients derived in Section~\ref{sec:results} for each template-correlated foreground into real units (e.g., $\mu$K rms) for each {\it WMAP}/{\it Planck} map that has been fitted to. This is achieved by multiplying the rms in each template (in the natural units of the template) by the corresponding template coefficients ($\mu$K per template unit), to give the absolute value of rms fluctuations (in $\mu$K, Rayleigh-Jeans) for this particular region of sky and angular resolution. These values can then be used to form a spectral energy distribution (SED) for each component.

\begin{figure}
\begin{center}
\hspace{-5mm}
\includegraphics[width=0.5\textwidth,angle=0]{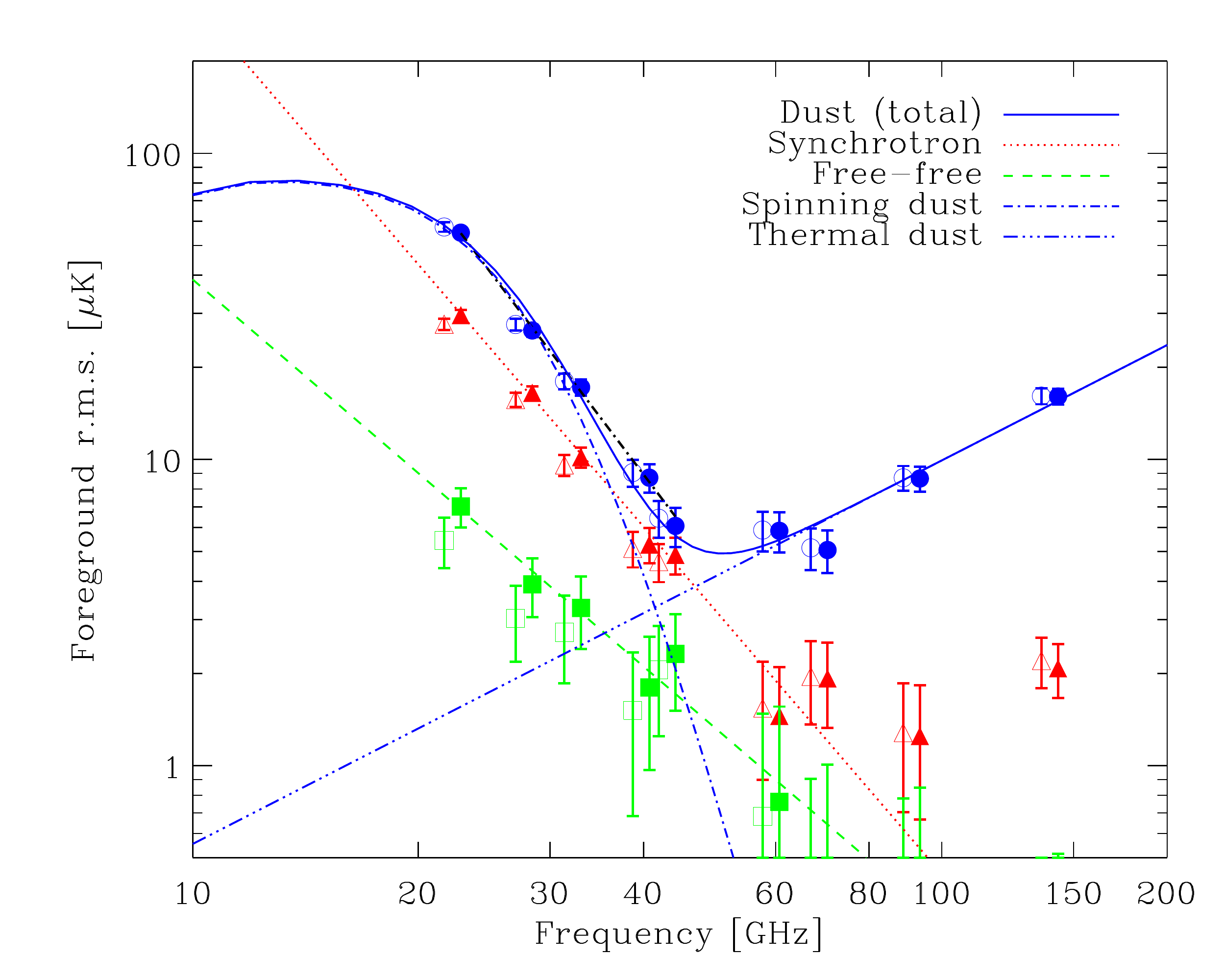}
\caption{SEDs of the three template-correlated foreground components in the NCP ($\delta > +80^{\circ}$) region: dust ({\it blue circles}), synchrotron ({\it red triangles}) and free-free ({\it green squares}). Filled symbols are when the \mbox{C-BASS} 4.7\,GHz map is used as a synchrotron template, while unfilled is when the Haslam 408\,MHz map is  used; the data points have been shifted slightly to the left for clarity. Coloured straight lines are power-law fits to each component of emission. The {\it black dot-dashed} line is a power-law fit to the dust-correlated coefficients between 22.8\,GHz and 44.1\,GHz and the fits were made to the filled circle data points. All data points were evaluated after direct subtraction of the CMB anisotropies using the {\it Planck} SMICA map.}
\label{fig:seds}
\end{center}
\end{figure}

Fig.~\ref{fig:seds} shows the SEDs for the three template-correlated foreground components fitted for all of the {\it WMAP} and {\it Planck} channels for $\delta > +80^{\circ}$, based on the results given in Table~\ref{tab:results_cmbsub}. Filled and open circles are for results using 0.408 and 4.7\,GHz, respectively, as the synchrotron template.

It can be seen that the dust-correlated emission dominates the spectrum at all frequencies. At high frequencies ($\gtrsim 80$\,GHz) the dust-correlated component is due to the low-frequency tail of thermal dust emission while at lower frequencies ($\lesssim 50$\,GHz) it is due to AME. The 100/217\,GHz {\it Planck} data points are artificially high because of the contribution of CO lines in the {\it Planck} data \citep{Planck2013_CO} and have not been plotted. The synchrotron emission is significant and has a steep ($\beta \approx -3.0$) spectrum while free-free emission is the weakest. One of the major results of this work is that the dust-correlated (AME) emission at 20--40\,GHz is not substantially changed when using the higher frequency \mbox{C-BASS} 4.7\,GHz map instead of the Haslam 408\,MHz map. The \mbox{C-BASS} data constrain the synchrotron emission to be about half of the dust-correlated emission at frequencies $\sim 20$--40\,GHz. However, it is interesting to see that the synchrotron emission appears to be slightly brighter at 20--40\,GHz when using the \mbox{C-BASS} map as the synchrotron template. This will be discussed further in Section~\ref{sec:discussion}. 

\section{Discussion}
\label{sec:discussion}

We now discuss in turn the fitted components, beginning with the synchrotron and free-free components, followed by the dust-correlated (AME) component at $\sim 20$--40\,GHz. Unless otherwise stated, quoted results are from the template fitting results presented in Sect.\,\ref{sec:results}, for the Haslam and \mbox{C-BASS} templates. In this section we will also include specific results from additional analyses. 

\subsection{Synchrotron emission}
\label{sec:discussion_synch}

From inspection of the synchrotron templates it appears that synchrotron emission is not dominant in the NCP region at frequencies $\sim 20$--40\,GHz. Nevertheless, using the cross-correlation technique we are able to detect synchrotron fluctuations with high significance, particularly at frequencies below $\sim 40$\,GHz. Synchrotron emission accounts for $29.6\pm1.2\,\mu$K rms at 22.8\,GHz, which is $\approx 33\,\%$ of the total emission at 22.8\,GHz (see Fig.~\ref{fig:seds}). The spectral fit suggests that synchrotron emission becomes dominant below $\approx 15$\,GHz. 

An important result of this analysis is that the amplitude and spectrum of synchrotron emission is consistent when using either the Haslam or \mbox{C-BASS} maps as a template, resulting in little change in the separated AME component. However, the synchrotron amplitudes are consistently $\approx 10$\,\% higher when the \mbox{C-BASS} map is used. This could be due to the fact that the higher frequency is detecting more emission with a flatter spectral index, or, that the improved fidelity of the \mbox{C-BASS} map gives a better correlation with the data. It cannot be due to contamination by free-free emission because this is negligible at 408\,MHz, while at 4.7\,GHz we subtract the small contribution of free-free emission using the H$\alpha$ template. The fact that the the goodness-of-fit is better when using C-BASS ($\chi_r^2=1.54$) compared to Haslam ($\chi_r^2=2.04$) suggests that the C-BASS data traces the sky fluctuations better than the Haslam map; the residual rms at 22.8\,GHz goes from $17.6\,\mu$K to $15.3\,\mu$K. The best-fitting power-law to the synchrotron coefficients below 44.1\,GHz for the \mbox{C-BASS} template (shown as a red dotted line in Fig.~\ref{fig:seds}) yields $\beta=-2.85\pm 0.14$. 

The template coefficients can be converted directly to a spectral index from the template frequency (e.g., 408\,MHz or 4.7\,GHz) to the frequency of the data being fitted to (e.g., {\it WMAP}/{\it Planck}). For a power-law, $T \propto \nu^{\beta}$, the spectral index between two frequencies $\nu_1$ and $\nu_2$ is given by $\beta={\rm log}(\Delta T_1/\Delta T_2)/{\rm log}(\nu_1/\nu_2)$. The template coefficient is effectively the ratio of variations in temperatures, $\Delta T_1$ and $\Delta T_2$. Table\,\ref{tab:synch_indices} summarizes the main synchrotron spectral index measurements from our analyses. It can be seen that consistent values are obtained over the entire frequency range and using different methods. The spectral slope at 20--40\,GHz is consistent with the spectral indices derived solely from the radio data as well as the coefficients directly, suggesting that a power-law with $\beta=-2.9$ is a good approximation for synchrotron emission from $\sim 1$\,GHz up to tens of GHz in this region of the sky. Although not formally significant in the NCP region, there is a hint of steepening with frequency, consistent with earlier results \citep[e.g.,][]{Davies2006,Strong2011,Kogut2012}. This is in contrast to \cite{Peel2012} who found a hint of flattening from 23 to 41\,GHz ($\Delta \beta \approx 0.05$), but again, at the $\sim 1\sigma$ significance level. 

\begin{table}
\centering
\caption{Derived synchrotron spectral indices over various frequency ranges and methods. The methods used are T-T plots (TT), template correlation coefficients (CC), and power-law fitting of the derived spectrum (PL). The uncertainties include absolute calibration uncertainties. }
\begin{tabular}{lcc}
\hline
Frequency range		&Method 			& Synchrotron spectral index $\beta$         	\\
\hline
0.4--4.7\,GHz			&TT				&$-2.79\pm0.17$	\\
0.4--4.7\,GHz			&CC				&$-2.85\pm0.05$ 	\\
0.4--22.8\,GHz			&CC				&$-2.88\pm0.03$	\\
4.7--22.8\,GHz			&CC				&$-2.91\pm0.04$ 	\\
22.8--44\,GHz			&PL				&$-2.85\pm0.14$ 	\\
\hline
\end{tabular}
\label{tab:synch_indices}
\end{table}
\normalsize

One caveat to this simple picture is that when considering only the inner regions of the map at the highest declinations, the synchrotron spectrum appears to flatten slightly. For the case of $\delta>+83^{\circ}$, although the other coefficients are consistent to better than $2\sigma$, the \mbox{C-BASS} synchrotron coefficient at 22.8\,GHz increases to $13.6\pm0.8\,\mu$K\,mK$^{-1}$, corresponding to a synchrotron spectral index of $\beta=-2.72\pm0.05$. The other components do not change as significantly because the dust-correlated emission dominates the fluctuations at 23\,GHz. This suggests that the synchrotron emission in and around the two main AME dust clouds and NCP is slightly flatter than the large-scale emission from the Galactic plane. Nevertheless, the synchrotron emission remains a small component relative to the total emission at 22.8\,GHz. The equivalent spectral index from 0.4\,GHz to 22.8\,GHz remains consistent between the two sky areas, so if the flattening is real it is occurring above a frequency of a few GHz and then steepening again at higher frequencies (otherwise we would see flatter indices above 4.7\,GHz).

An important consideration for the interpretation of synchrotron emission is the presence of free-free emission in the low-frequency radio data. At high Galactic latitudes, the H$\alpha$ intensities are typically a few R at most. The H$\alpha$ map of the NCP region (Fig.~\ref{fig:maps}) shows emission at a level of $\sim 1$\,R with a peak of 3.7\,R on a background of $\approx 0.5$--1\,R; see Section~\ref{sec:discussion_ff}.  A typical high-latitude H$\alpha$ intensity of 1\,R corresponds to 51\,mK at 408\,MHz and 0.3\,mK at 4.7\,GHz \citep{Dickinson2003}. Therefore the contribution from free-free at high latitudes is negligible at 408\,MHz (typical fluctuations of $\sim 10$\,K) and a small contribution at 4.7\,GHz (typical fluctuations $\approx 5$\,mK). Even though it had little impact on the quoted results, we subtracted the free-free contribution at 4.7\,GHz using the H$\alpha$ map and a conversion factor of 0.32\,mK\,R$^{-1}$. As a further test, we performed template-fitting of the \mbox{C-BASS} map itself, using the 408\,MHz and H$\alpha$ maps as tracers of the synchrotron and free-free emission, respectively. We  do not detect H$\alpha$-correlated emission in the NCP region at 4.7\,GHz, with a template coefficient of $0.29\pm0.31$\,mK\,R$^{-1}$, which is consistent with theoretical expectations. Virtually all of the signal at 4.7\,GHz can be accounted for by the 408\,MHz template, with a coefficient of $935\pm72$\,mK\,K$^{-1}$, corresponding to $\beta=-2.85\pm0.05$. This can be contrasted with emission from the Galactic disk, which emits a much larger fraction of free-free emission at 4.7\,GHz \citep{Irfan2015}. 

Table~\ref{tab:rms} lists the rms values for each component from the cross-correlation analysis for $\delta>+80^{\circ}$, calculated by multiplying the cross-correlation template coefficients by the rms fluctuations in the template map for this region. Taking into account correlations between the templates gives a total rms of $89.4\pm4.1\,\mu$K at 22.8\,GHz. We also list the rms values from the corresponding {\it Planck} 2015 {\sc Commander} component separation products \citep{Planck2015_X}, scaled to 22.8\,GHz. We can see that in the cross-correlation analysis using \mbox{C-BASS}, although the AME dominates the fluctuations, the synchrotron emission contributes almost three times more rms fluctuations compared to the {\it Planck} {\sc Commander} solution. This is due to the fact that the {\it Planck}  {\sc Commander} analysis used only one low-frequency data point at 408\,MHz, which meant that the synchrotron spectrum had to be effectively fixed. Their model had an effective spectral index of $\beta \approx -3.05\pm0.05$ above 1\,GHz \citep{Planck2015_XXV}, while the \mbox{C-BASS} data prefers a slightly flatter index of $\beta\approx-2.9$.

\setlength{\tabcolsep}{6pt}
\begin{table}
\centering
\caption{RMS fluctuations at 22.8\,GHz of each component for $\delta>+80^{\circ}$ from the cross-correlation (CC) analysis and from the {\it Planck} 2015 {\sc Commander} component separation products \protect\citep{Planck2015_X}. Units are $\mu$K.}
\begin{tabular}{lcc}
\hline
Component			&CC					&{\it Planck} 2015 	\\
\hline
Synchrotron			&$29.6\pm1.2$			&11.8			\\		
Free-free				&\phantom{2}$7.0\pm1.0$	&46.9	 		\\
AME / dust			&$55.0\pm2.0$			&45.4			\\
Thermal dust			&\ldots				&1.2				\\
Total foreground		&$88.8\pm3.6$			&84.0		  	\\
\hline
\end{tabular}
\label{tab:rms}
\end{table}
\normalsize
\setlength{\tabcolsep}{6pt}

Finally, we comment on the apparent upturn in the synchrotron spectrum above $\sim 90$\,GHz (Fig.~\ref{fig:seds}). At these frequencies the synchrotron component accounts for $\ll10\%$ of the total emission and therefore is difficult to separate from the much brighter dust emission and is also partially (spatially) correlated. This is partially reflected in the larger error bars and therefore we do not believe the upturn to be a real effect.


\subsection{Free-free emission}
\label{sec:discussion_ff}

The H$\alpha$-correlated component is expected to be due to free-free emission \citep{Dickinson2003}. For an electron temperature $T_e=8000$\,K, typical of the diffuse ISM, we would expect to see H$\alpha$ template coefficients of 11.4, 7.8, 5.23, 3.28 and 1.11\,$\mu$K\,R$^{-1}$, for frequencies of 22.8, 28.4, 33.0, 40.7, 44.1\,GHz, respectively. It also assumes that the correction for dust extinction along the line of sight has been done accurately, which renders the standard H$\alpha$ templates useless at very low Galactic latitudes \citep{Dickinson2003}. The theoretical values also assume local thermodynamic equilibrium, 8 per cent contribution from helium, and that there is no scattering of H$\alpha$ from elsewhere off dust grains. The amount of scattered H$\alpha$ light is not clear, with earlier predictions in the range 5--20\,\% \citep{Wood1999}, while more recent work has suggest that it could be more significant along some lines of sight \citep{Seon2012,Brandt2012} and possibly up to 50\% or more \citep{Witt2010}. This would in turn reduce these coefficients by up to a half, giving better agreement with theory \citep{Banday2003,Davies2006,Dobler2009,Ghosh2012}. 

From our analysis, the free-free emission is very weak in the NCP region. The free-free brightness has an rms of $7.0\pm1.0\,\mu$K at 22.8\,GHz (Table~\ref{tab:rms}), corresponding to $\approx 6\%$ of the total emission. Nevertheless, the H$\alpha$-correlated values in Table~\ref{tab:results_cmbsub} are consistent with theoretical expectations; at 22.8\,GHz we expect $\approx 11\,\mu$K\,R$^{-1}$ for typical electron temperatures \citep{Dickinson2003}. Moreover, while the uncertainties are relatively large, the independent coefficients plotted in Fig.\,\ref{fig:seds} precisely follow the spectral dependence expected for free-free emission ($\beta \approx -2.1$) up to 44\,GHz and higher. This is good reassurance that the template fits are yielding physically meaningful results. A previous analysis of the H$\alpha$ fluctuations in the NCP region ($\delta>+81^{\circ}$) by \cite{Gaustad1996} found an upper limit of 0.5\,R on $1^{\circ}$ scales, corresponding to $<6\,\mu$K at 22.8\,GHz. This is consistent with our analysis at the $1\sigma$ level. We note that the {\it Planck} 2015 free-free map gives a much larger rms amplitude of $46.9\,\mu$K (Table~\ref{tab:rms}), at the expense of both the AME and synchrotron components. This is the largest discrepancy between the two analyses. As discussed and demonstrated by \cite{Planck2015_XXV}, the {\sc Commander} free-free solution appears to be over-estimated (typically by a factor of several) due to aliasing of the low frequency (synchrotron or free-free or AME) components by the spectrum alone.

The H$\alpha$ coefficients from our analysis are consistent with expectations from theory, for electron temperatures in the range $T_e \approx 7000$--10000\,K; this indicates that scattered H$\alpha$ is not a major issue in this region of sky. They are also consistent when using either of the synchrotron templates. Note that we chose to subtract the free-free component from the \mbox{C-BASS} 4.7\,GHz map to ensure it was dominated by synchrotron emission. When using the raw 4.7\,GHz without a free-free correction, we naturally found a lower value for the template coefficient of $9.9\pm1.9\,\mu$K\,R$^{-1}$ at 22.8\,GHz, indicating that residual free-free emission in the 4.7\,GHz map is having a small impact, at least on the free-free solution. Fortunately, it has negligible impact on the other template results, with the values changing by less than $0.2\sigma$ for both the synchrotron- and dust-correlated coefficients. This highlights that the free-free emission is relatively weak at all microwave frequencies.

The two H$\alpha$ templates (F03 and D03) also give slightly different results (but consistent within the uncertainties), which can be attributed largely to stellar residuals in the H$\alpha$ maps that have been treated differently. The H$\alpha$-correlated  template coefficient at 22.8\,GHz when using F03 was $18.2\pm2.9\,\mu$K\,R$^{-1}$, slightly higher than for D03, which is expected due to the small fluctuations in the F03 version of the map. This also had a minor impact on the synchrotron and dust coefficients but is consistent to within $1.5\,\sigma$. The dust reddening $E(B-V)$ map of \cite{Schlegel1998} can be used to estimate the maximum level of dust extinction assuming all the dust is in front of the ionized gas \citep{Dickinson2003}. At the original 6.1\,arcmin resolution, the main dust feature in this region has extinction values in the range $\approx 0.2$--0.7\,mag; at $1^{\circ}$ resolution the two dust emission peaks are at a value of $\approx 0.45$\,mag, corresponding to a maximum dust absorbing factor of $\approx 1.5$. It is therefore possible that the H$\alpha$ intensity is under-estimated in this region by $\approx 20$--30\,\% (about half of the maximum value). This would not have a significant effect on the synchrotron and AME results because the free-free component is sub-dominant, being $\approx 7\,\%$ of the total emission.

In summary, the D03 H$\alpha$ template appears to trace free-free emission more reliably than the F03 map, with an amplitude that is as expected for typical electron temperatures of $T_e \approx 7000$--10000\,K. The free-free emission at AME frequencies ($\approx 20$--40\,GHz) in the NCP region appears to be a small portion of the total emission, and therefore assumptions about the template and dust extinction have little effect on the AME amplitude.


\subsection{AME}
\label{sec:discussion_ame}

Our results show a significant detection of dust-correlated emission, particularly in the $\sim$20--40\,GHz range. The dust-correlated emission (AME) is much brighter than can be accounted for by thermal dust (which would have a falling spectrum with decreasing frequency) and synchrotron/free-free emission as traced by the foreground templates. The rms of the dust-correlated component at 22.8\,GHz is $55.0\pm2.0\,\mu$K at 22.8\,GHz, which accounts for $\approx 60\%$ of the total rms (Table\,\ref{tab:rms}).  

The main result of this paper is the lack of change in the AME signal relative to the FIR data, when using the higher frequency \mbox{C-BASS} 4.7\,GHz template compared to the traditional Haslam 408\,MHz map. This is a similar result to those previously found when using a 2.3\,GHz map as a synchrotron template \citep{Peel2012}, who found the AME amplitude changed by only 7\%. For our baseline fit, when using the $\tau_{353}$ map as the dust template, we find coefficients at 22.8\,GHz of $9.93\pm0.35$\,K and $9.52\pm0.34$\,K per unit $\tau_{353}$ (Table\,\ref{tab:results_cmbsub}), when using the Haslam and \mbox{C-BASS} maps as synchrotron templates, respectively, i.e. they are consistent at the $\approx 1\,\sigma$ level with just a 4\% change in amplitude. 

Our main result is consistent with the spectral indices for synchrotron emission that are almost constant across the entire radio/microwave band (see Sect.~\ref{sec:discussion_synch}). An additional flatter (harder) spectrum component of synchrotron is not detected at 4.7\,GHz. For synchrotron emission to explain the bulk of the AME, it would  have to have a much flatter spectrum; a value of $\beta \approx -2.3$ would be needed to extrapolate the $\sim 3$\,mK rms fluctuations at 4.7\,GHz to the $\sim 80\,\mu$K fluctuations observed at 22.8\,GHz (after removing the CMB and free-free contribution). However, in practice it would need to be even flatter than this, since the observed emission at 4.7\,GHz does not correlate well with the AME, and hence a flat synchrotron AME component would need to be subdominant at 4.7\,GHz. 

\setlength{\tabcolsep}{3pt}
\begin{table}
\centering
\caption{AME cross-correlation template coefficients at 22.8\,GHz and $\chi_r^2$ values for different dust templates, ordered in terms of decreasing goodness-of-fit.}
\begin{tabular}{lccc}
\hline
Template				&Correlation Coeff.			&Unit						&$\chi_r^2$      \\
\hline
$\tau_{353}$			&\phantom{2}$9.52\pm0.34$	&K							&1.54		\\
{\it Planck} 545\,GHz		&$67\pm2$				&$\mu$K\,(MJy\,sr$^{-1}$)$^{-1}$	&1.60		\\
FDS94				&\phantom{2}$5.16\pm0.19$	&mK\,mK$^{-1}$				&1.62		\\
{\it Planck} 857\,GHz		&$22.4\pm0.8$				&$\mu$K\,(MJy\,sr$^{-1}$)$^{-1}$	&1.63		\\
{\it Planck} 353\,GHz		&$875\pm32$				&$\mu$K\,mK$^{-1}$				&1.63 		\\
Dust radiance $\Re$		&$563\pm20$				& K\,(W\,m$^{-2}$\,sr$^{-1}$)$^{-1}$	&2.12		\\
IRIS\,$100\,\mu$m		&$31.0\pm1.1$				&$\mu$K\,(MJy\,sr$^{-1}$)$^{-1}$	&2.90		\\	
\hline
\end{tabular}
\label{tab:dust_coeff}
\end{table}
\normalsize
\setlength{\tabcolsep}{6pt}

We find similar levels of AME in the NCP region, relative to the standard dust templates, to those measured previously. Table\,\ref{tab:dust_coeff} lists the AME cross-correlation template coefficients at 22.8\,GHz and $\chi_r^2$ values for each dust template. We adopt $\tau_{353}$  as our default template since it formally provided the best fit to the microwave data, with map residuals of $15\,\mu$K rms. Our best value of $9.52\pm0.34$\,K compares well with the high-latitude ($|b|>10^{\circ}$) value of $9.7\pm1.0$\,K \citep{Planck2015_XXV}. \cite{Hensley2016} found a value of $7.9\pm2.6$\,K at 30\,GHz, which is also consistent. We find slightly larger fluctuations in the AME at 22.8\,GHz ($55\pm2\,\mu$K) compared to the {\it Planck} products ($45.4\,\mu$K), but they are comparable (Table\,\ref{tab:rms}).  This shows that even with very different component separation techniques, the AME is a strong component of the emission at frequencies $\approx 20$--40\,GHz.

We tried several other standard dust templates, listed in Tables~\ref{tab:data} and \ref{tab:dust_coeff}.  The worst template is the $100\,\mu$m template, which is confirmed by the Pearson correlation coefficients at 22.8\,GHz (where AME is dominant), as shown in Sect.~\ref{sec:multimaps}. It is also evident in the residual maps at 22.8\,GHz presented in Fig.~\,\ref{fig:residmaps}; the residual map rms at 22.8\,GHz increases from $15\,\mu$K to $21\,\mu$K rms. This is presumably due to variations in the dust composition or temperature, which significantly modulates the response at wavelengths near the peak of the spectrum at $\sim 100\,\mu$m but has negligible effect at longer wavelengths \citep[e.g.,][]{Finkbeiner2004,Tibbs2012b}. Nevertheless, we measured coupling coefficients of $31.0\pm1.1\,\mu$K\,(MJy\,sr$^{-1}$)$^{-1}$ at 22.8 GHz and $9.9\pm0.7\,\mu$K\,(MJy\,sr$^{-1}$)$^{-1}$ at 33.0\,GHz. This compares well with the $\approx 10\,\mu$K\,(MJy\,sr$^{-1}$)$^{-1}$ at 33.0\,GHz that has been observed before \citep{Banday2003,deOliveira-Costa2004,Davies2006}. This can also be compared with the dust-correlation measured originally by \cite{Leitch1997}, which when extrapolated to 22.8\,GHz, corresponds to $77\,\mu$K\,(MJy\,sr$^{-1}$)$^{-1}$. This value is higher because it is a direct (single template) correlation, therefore neglecting any dust-correlated synchrotron/free-free emission. Also, their best-fitting spectral index of $-2.2$ is flatter than most measurements since then, which increases the relative amplitude at higher frequencies. The interpretation is that about half of the total dust-correlated signal at 20--40\,GHz  is due to AME. 

Interestingly, unlike the results of \cite{Hensley2016} the map of dust radiance ($\Re$) was found not to trace AME as well as $\tau_{353}$ or any of the {\it Planck} HFI maps. Nevertheless, we found similar amplitudes to those estimated at high latitudes. For example, for the case of the dust radiance template, we found a coefficient of  $563\pm20\,$K\,(W\,m$^{-2}$\,sr$^{-1}$)$^{-1}$ compared to $226\pm44\,$K\,(W\,m$^{-2}$\,sr$^{-1}$)$^{-1}$ at 30\,GHz. Surprisingly, the much older FDS94 model \citep{Finkbeiner1999} provided almost as good a fit as $\tau_{353}$. Our coefficient of $5.16\pm0.19$\,mK\,mK$^{-1}$ is slightly lower than the equivalent region 6 of \cite{Davies2006} who found $6.7\pm0.7$\,mK\,mK$^{-1}$. When adopting the Haslam synchrotron template we obtained a consistent value of $5.38\pm0.19$\,mK\,mK$^{-1}$. The slight difference is mostly due to the slightly different sky areas and masking used.

From Fig.~\ref{fig:seds} it is clear that using the much higher frequency synchrotron template from \mbox{C-BASS} makes little difference to the dust-correlated (AME) component, at least in this region of the sky. This indicates that there is no strong component of flat-spectrum synchrotron emission. The spectrum of the dust-correlated component also appears to be well approximated at $\sim 20$--40\,GHz by a power-law with a best-fitting slope of $\beta=-3.23\pm0.13$, which is slightly steeper but remarkably close to the spectrum of the synchrotron component ($\beta \approx -2.9$). We fitted a spinning dust model, using the {\sc spdust2} code \citep{Ali-Hamoud2009,Silsbee2011}, adopting the parameters for the Cold Neutral Medium (CNM) as suggested by \cite{Draine1998a}. This gave a poor fit because this particular model spectrum peaks at 33\,GHz (in flux density units). We therefore included a frequency shift (in log-space) of the CNM spinning dust spectrum, which provided a good fit with a peak frequency of 23\,GHz, as shown in Fig.~\ref{fig:seds}. However, there is an indication that there is excess emission in the range $\approx 50$--100\,GHz. This is likely a failure of the simple single component CNM spinning dust model. 

In reality, there will be a range of dust particles/environments along the line-of-sight that will tend to broaden the spectrum, and which motivated the two-component model by \cite{Planck2015_X,Planck2015_XXV}. Note that spinning dust models generally predict broadly peaked emission spectra with a FWHM in frequency approximately twice the peak frequency; for example, the WNM model of \cite{Draine1998b} has a peak at 22\,GHz in flux density, with half-max points at 13\,GHz and 37\,GHz. The location of the peak is determined by a combination of the nanoparticle size distribution, the electric dipole moment, and the excitation conditions. 

Using a single component model results in a thermal dust emissivity index of $\beta=+1.25\pm0.13$, which is significantly flatter than has been measured for the majority of the sky with $\beta \approx 1.6$ \citep[e.g.,][]{Planck2013_XI}. The low value is within the range measured by {\it Planck} but is sensitive to which CMB map is used and over which frequencies are being fitted. It does not change the results for AME, which dominates at lower frequencies where the thermal dust has a minimal contribution. Given the lack of alternatives, the spinning dust mechanism appears to be the most viable explanation for the diffuse AME observed in the NCP region.


\section{Conclusions}
\label{sec:conclusions}

We have re-analysed the NCP region ($\delta > +80^{\circ}$) in {\it WMAP}/{\it Planck} data to study the diffuse foregrounds. We used template maps of synchrotron (low-frequency radio maps), free-free (H$\alpha$ maps) and dust emission (sub-mm/FIR maps) to fit to {\it WMAP/Planck} data at 22.8\,GHz and above. In addition to the standard Haslam 408\,MHz map, we used new \mbox{C-BASS} data at 4.7\,GHz as a tracer of synchrotron emission. \mbox{C-BASS} data are of higher quality than the \cite{Haslam1982} data, which are known to contain significant striations and other calibration issues. More importantly, at a frequency of $\approx 4.7$\,GHz, \mbox{C-BASS} data are much closer to the frequencies observed by {\it WMAP/Planck} and therefore should be more representative of foregrounds at microwave frequencies. In particular, if there is a significant hard (flat spectrum) component of synchrotron radiation, then the \mbox{C-BASS} maps should be a more reliable synchrotron template.

We have found that the dust-correlated AME component accounts for the bulk ($\approx 60\%$) of the foreground rms in the NCP region and that it does not change significantly when including the \mbox{C-BASS} 4.7\,GHz map. The synchrotron emission has a spectrum close to $\beta=-2.90\pm0.05$ when considering the low frequency data alone via T-T plots, the cross-correlation coefficients directly, or when fitting to the SED of coefficients in the range 20--40\,GHz. This indicates that a power-law is a good model for the synchrotron emission from a frequency of $0.4$\,GHz up to tens of GHz and that there is no strong component of flat-spectrum (harder) synchrotron emission, at least in the NCP region. The synchrotron component accounts for approximately half the rms brightness at 22.8\,GHz. We find that the D03 version of the H$\alpha$ map correlates better with microwave data than the F03 map, due to low level residual artifacts in the maps. Fortunately, free-free emission is relatively weak at high latitudes and accounts for $\approx 6\,\%$ of the total signal at 22.8\,GHz.

We preferred the thermal dust optical depth at 353\,GHz as our tracer of AME since it gave the best overall fit. We found a best-fitting template coefficient of $9.52\pm0.34$\,K per unit $\tau_{353}$ at 22.8\,GHz, which agrees well with previous measurements on different regions of the sky. Other templates, such as dust radiance and FDS94 gave similar results but with slightly larger residuals. The most discrepant of the dust tracers was the IRIS\,$100\,\mu$m map, due to variations in dust temperature within the region, which modulates the spatial morphology at wavelengths near $100\,\mu$m

A power-law provides a good fit to the AME spectrum above 20\,GHz. Alternatively, it can be well-fitted by a shifted spinning dust model, with a peak frequency (flux density units) around 23\,GHz. However, in this case the thermal dust emissivity index flattens to $\approx +1.3$. This is likely a failure of a simple single component spinning dust model, which is inevitably narrower than the true distribution of dust particles and environments.

The observations for the northern \mbox{C-BASS} survey are now complete and the southern survey is just beginning. With full-sky \mbox{C-BASS} maps we will be able to investigate the AME across the whole sky and to investigate the possible contribution of a harder spectrum of synchrotron radiation. Future papers will include applying T-T plot (Jew et al., in prep.) and template fitting (Harper et al., in prep.) techniques to the high latitude sky. We will also use more advanced component separation techniques such as parametric fitting \citep[e.g.,][]{Eriksen2008b}. With the wealth of high precision data covering a wide range of frequencies, including new data from S-PASS \citep{Krachmalnicoff2018} and QUIJOTE \citep{Genova-Santos2015a}, we should be able to fit for all these components as a function of position on the sky.


\section*{Acknowledgments}

The \mbox{C-BASS} project is a collaboration between Oxford and Manchester Universities in the UK, the California Institute of Technology in the U.S.A., Rhodes University, UKZN and the South African Radio Observatory in South Africa, and the King Abdulaziz City for Science and Technology (KACST) in Saudi Arabia. It has been supported by the NSF awards AST-0607857, AST-1010024, AST-1212217, and AST-1616227, and NASA award NNX15AF06G, the University of Oxford, the Royal Society, STFC, and the other participating institutions. This research was also supported by the South African Radio Astronomy Observatory, which is a facility of the National Research Foundation, an agency of the Department of Science and Technology. We would like to thank Russ Keeney for technical help at OVRO. CD acknowledges support from an ERC Starting (Consolidator) Grant (no.~307209). CD also thanks the California Institute of Technology for their hospitality and hosting during several extended visits. MWP acknowledges funding from a FAPESP Young Investigator fellowship, grant 2015/19936-1. We make use of the HEALPix package \citep{Gorski2005},  IDL astronomy library and Python astropy, matplotlib, numpy, healpy, and scipy packages. This research has made use of the NASA/IPAC Extragalactic Database (NED) which is operated by the Jet Propulsion Laboratory, California Institute of Technology, under contract with the National Aeronautics and Space Administration. The \mbox{C-BASS} collaboration would like to remember the late Prof. Rodney D. Davies and Prof. Richard J. Davis, who were both strong supporters of the \mbox{C-BASS} project (\url{http://cbass.web.ox.ac.uk}).


\bibliographystyle{mn2e}
\bibliography{cbassbiblio,refs}

\begin{thebibliography}{116}
\expandafter\ifx\csname natexlab\endcsname\relax\def\natexlab#1{#1}\fi

\bibitem[{{Ali-Ha{\"\i}moud}, {Hirata} \& {Dickinson}(2009){Ali-Ha{\"\i}moud},
  {Hirata}, \& {Dickinson}}]{Ali-Hamoud2009}
{Ali-Ha{\"\i}moud} Y., {Hirata} C.~M., {Dickinson} C., 2009, \mnras, 395, 1055

\bibitem[{{Alves} {et~al}\mbox{.}(2012){Alves}, {Davies}, {Dickinson},
  {Calabretta}, {Davis}, \& {Staveley-Smith}}]{Alves2012}
{Alves} M.~I.~R., {Davies} R.~D., {Dickinson} C., {Calabretta} M., {Davis} R.,
  {Staveley-Smith} L., 2012, \mnras, 422, 2429

\bibitem[{{AMI Consortium} {et~al}\mbox{.}(2011){AMI Consortium}, {Davies},
  {Franzen}, {Waldram}, {Grainge}, {Hobson}, {Hurley-Walker}, {Lasenby},
  {Olamaie}, {Pooley}, {Riley}, {Rodr{\'{\i}}guez-Gonz{\'a}lvez}, {Saunders},
  {Scaife}, {Schammel}, {Scott}, {Shimwell}, {Titterington}, \&
  {Zwart}}]{Davies2011}
{AMI Consortium} {et~al.}, 2011, \mnras, 415, 2708

\bibitem[{{Armitage-Caplan} {et~al}\mbox{.}(2012){Armitage-Caplan}, {Dunkley},
  {Eriksen}, \& {Dickinson}}]{Armitage-Caplan2012}
{Armitage-Caplan} C., {Dunkley} J., {Eriksen} H.~K., {Dickinson} C., 2012,
  \mnras, 424, 1914

\bibitem[{{Banday} {et~al}\mbox{.}(2003){Banday}, {Dickinson}, {Davies},
  {Davis}, \& {G{\'o}rski}}]{Banday2003}
{Banday} A.~J., {Dickinson} C., {Davies} R.~D., {Davis} R.~J., {G{\'o}rski}
  K.~M., 2003, \mnras, 345, 897

\bibitem[{{Bennett} {et~al}\mbox{.}(2003){Bennett}, {Hill}, {Hinshaw}, {Nolta},
  {Odegard}, {Page}, {Spergel}, {Weiland}, {Wright}, {Halpern}, {Jarosik},
  {Kogut}, {Limon}, {Meyer}, {Tucker}, \& {Wollack}}]{Bennett2003b}
{Bennett} C.~L. {et~al.}, 2003, \apjs, 148, 97

\bibitem[{{Bennett} {et~al}\mbox{.}(2013){Bennett}, {Larson}, {Weiland},
  {Jarosik}, {Hinshaw}, {Odegard}, {Smith}, {Hill}, {Gold}, {Halpern},
  {Komatsu}, {Nolta}, {Page}, {Spergel}, {Wollack}, {Dunkley}, {Kogut},
  {Limon}, {Meyer}, {Tucker}, \& {Wright}}]{Bennett2013}
{Bennett} C.~L. {et~al.}, 2013, \apjs, 208, 20

\bibitem[{{Brandt} \& {Draine}(2012)}]{Brandt2012}
{Brandt} T.~D., {Draine} B.~T., 2012, \apj, 744, 129

\bibitem[{{Casassus} {et~al}\mbox{.}(2006){Casassus}, {Cabrera}, {F{\"o}rster},
  {Pearson}, {Readhead}, \& {Dickinson}}]{Casassus2006}
{Casassus} S., {Cabrera} G.~F., {F{\"o}rster} F., {Pearson} T.~J., {Readhead}
  A.~C.~S., {Dickinson} C., 2006, \apj, 639, 951

\bibitem[{{Casassus} {et~al}\mbox{.}(2008){Casassus}, {Dickinson}, {Cleary},
  {Paladini}, {Etxaluze}, {Lim}, {White}, {Burton}, {Indermuehle}, {Stahl}, \&
  {Roche}}]{Casassus2008}
{Casassus} S. {et~al.}, 2008, \mnras, 391, 1075

\bibitem[{{Condon} {et~al}\mbox{.}(1998){Condon}, {Cotton}, {Greisen}, {Yin},
  {Perley}, {Taylor}, \& {Broderick}}]{Condon1998}
{Condon} J.~J., {Cotton} W.~D., {Greisen} E.~W., {Yin} Q.~F., {Perley} R.~A.,
  {Taylor} G.~B., {Broderick} J.~J., 1998, \aj, 115, 1693

\bibitem[{{Davies} {et~al}\mbox{.}(2006){Davies}, {Dickinson}, {Banday},
  {Jaffe}, {G{\'o}rski}, \& {Davis}}]{Davies2006}
{Davies} R.~D., {Dickinson} C., {Banday} A.~J., {Jaffe} T.~R., {G{\'o}rski}
  K.~M., {Davis} R.~J., 2006, \mnras, 370, 1125

\bibitem[{{Davies}, {Watson} \& {Gutierrez}(1996){Davies}, {Watson}, \&
  {Gutierrez}}]{Davies1996}
{Davies} R.~D., {Watson} R.~A., {Gutierrez} C.~M., 1996, \mnras, 278, 925

\bibitem[{{de Oliveira-Costa} {et~al}\mbox{.}(2004){de Oliveira-Costa},
  {Tegmark}, {Davies}, {Guti{\'e}rrez}, {Lasenby}, {Rebolo}, \&
  {Watson}}]{deOliveira-Costa2004}
{de Oliveira-Costa} A., {Tegmark} M., {Davies} R.~D., {Guti{\'e}rrez} C.~M.,
  {Lasenby} A.~N., {Rebolo} R., {Watson} R.~A., 2004, \apjl, 606, L89

\bibitem[{{de Zotti} {et~al}\mbox{.}(2010){de Zotti}, {Massardi}, {Negrello},
  \& {Wall}}]{deZotti2010}
{de Zotti} G., {Massardi} M., {Negrello} M., {Wall} J., 2010, \aapr, 18, 1

\bibitem[{{Delabrouille} \& {Cardoso}(2009)}]{Delabrouille2009}
{Delabrouille} J., {Cardoso} J., 2009, in Lecture Notes in Physics, Berlin
  Springer Verlag, Vol. 665, Data Analysis in Cosmology,
  {V.~J.~Mart{\'{\i}}nez, E.~Saar, E.~Mart{\'{\i}}nez-Gonz{\'a}lez, \&
  M.-J.~Pons-Border{\'{\i}}a}, ed., pp. 159--205

\bibitem[{{Dickinson} {et~al}\mbox{.}(2018){Dickinson}, {Ali-Ha{\"i}moud},
  {Barr}, {Battistelli}, {Bell}, {Bernstein}, {Casassus}, {Cleary}, {Draine},
  {G{\'e}nova-Santos}, {Harper}, {Hensley}, {Hill-Valler}, {Hoang}, {Israel},
  {Jew}, {Lazarian}, {Leahy}, {Leech}, {L{\'o}pez-Caraballo}, {McDonald},
  {Murphy}, {Onaka}, {Paladini}, {Peel}, {Perrott}, {Poidevin}, {Readhead},
  {Rubi{\~n}o-Mart{\'{\i}}n}, {Taylor}, {Tibbs}, {Todorovi{\'c}}, \&
  {Vidal}}]{Dickinson2018}
{Dickinson} C. {et~al.}, 2018, \nar, 80, 1

\bibitem[{{Dickinson} {et~al}\mbox{.}(2010){Dickinson}, {Casassus}, {Davies},
  {Allison}, {Bustos}, {Cleary}, {Davis}, {Jones}, {Pearson}, {Readhead},
  {Reeves}, {Taylor}, {Tibbs}, \& {Watson}}]{Dickinson2010}
{Dickinson} C. {et~al.}, 2010, \mnras, 407, 2223

\bibitem[{{Dickinson} {et~al}\mbox{.}(2009){Dickinson}, {Davies}, {Allison},
  {Bond}, {Casassus}, {Cleary}, {Davis}, {Jones}, {Mason}, {Myers}, {Pearson},
  {Readhead}, {Sievers}, {Taylor}, {Todorovi{\'c}}, {White}, \&
  {Wilkinson}}]{Dickinson2009a}
{Dickinson} C. {et~al.}, 2009, \apj, 690, 1585

\bibitem[{{Dickinson}, {Davies} \& {Davis}(2003){Dickinson}, {Davies}, \&
  {Davis}}]{Dickinson2003}
{Dickinson} C., {Davies} R.~D., {Davis} R.~J., 2003, \mnras, 341, 369

\bibitem[{{Dickinson}, {Peel} \& {Vidal}(2011){Dickinson}, {Peel}, \&
  {Vidal}}]{Dickinson2011}
{Dickinson} C., {Peel} M., {Vidal} M., 2011, \mnras, 418, L35

\bibitem[{{Dobler}, {Draine} \& {Finkbeiner}(2009){Dobler}, {Draine}, \&
  {Finkbeiner}}]{Dobler2009}
{Dobler} G., {Draine} B., {Finkbeiner} D.~P., 2009, \apj, 699, 1374

\bibitem[{{Draine} \& {Hensley}(2013)}]{Draine2013}
{Draine} B.~T., {Hensley} B., 2013, \apj, 765, 159

\bibitem[{{Draine} \& {Lazarian}(1998{\natexlab{a}})}]{Draine1998a}
{Draine} B.~T., {Lazarian} A., 1998{\natexlab{a}}, \apjl, 494, L19

\bibitem[{{Draine} \& {Lazarian}(1998{\natexlab{b}})}]{Draine1998b}
{Draine} B.~T., {Lazarian} A., 1998{\natexlab{b}}, \apj, 508, 157

\bibitem[{{Draine} \& {Lazarian}(1999)}]{Draine1999}
{Draine} B.~T., {Lazarian} A., 1999, \apj, 512, 740

\bibitem[{{Dunkley} {et~al}\mbox{.}(2009{\natexlab{a}}){Dunkley}, {Amblard},
  {Baccigalupi}, {Betoule}, {Chuss}, {Cooray}, {Delabrouille}, {Dickinson},
  {Dobler}, {Dotson}, {Eriksen}, {Finkbeiner}, {Fixsen}, {Fosalba}, {Fraisse},
  {Hirata}, {Kogut}, {Kristiansen}, {Lawrence}, {Magalha\~{}Es},
  {Miville-Deschenes}, {Meyer}, {Miller}, {Naess}, {Page}, {Peiris},
  {Phillips}, {Pierpaoli}, {Rocha}, {Vaillancourt}, \& {Verde}}]{Dunkley2009a}
{Dunkley} J. {et~al.}, 2009{\natexlab{a}}, in American Institute of Physics
  Conference Series, Vol. 1141, American Institute of Physics Conference
  Series, {Dodelson} S., {Baumann} D., {Cooray} A., {Dunkley} J., {Fraisse} A.,
  {Jackson} M.~G., {Kogut} A., {Krauss} L., {Zaldarriaga} M., {Smith} K., eds.,
  pp. 222--264

\bibitem[{{Dunkley} {et~al}\mbox{.}(2009{\natexlab{b}}){Dunkley}, {Spergel},
  {Komatsu}, {Hinshaw}, {Larson}, {Nolta}, {Odegard}, {Page}, {Bennett},
  {Gold}, {Hill}, {Jarosik}, {Weiland}, {Halpern}, {Kogut}, {Limon}, {Meyer},
  {Tucker}, {Wollack}, \& {Wright}}]{Dunkley2009b}
{Dunkley} J. {et~al.}, 2009{\natexlab{b}}, \apj, 701, 1804

\bibitem[{{Erickson}(1957)}]{Erickson1957}
{Erickson} W.~C., 1957, \apj, 126, 480

\bibitem[{{Eriksen} {et~al}\mbox{.}(2008){Eriksen}, {Jewell}, {Dickinson},
  {Banday}, {G{\'o}rski}, \& {Lawrence}}]{Eriksen2008b}
{Eriksen} H.~K., {Jewell} J.~B., {Dickinson} C., {Banday} A.~J., {G{\'o}rski}
  K.~M., {Lawrence} C.~R., 2008, \apj, 676, 10

\bibitem[{{Errard} \& {Stompor}(2012)}]{Errard2012}
{Errard} J., {Stompor} R., 2012, \prd, 85, 083006

\bibitem[{{Finkbeiner}(2003)}]{Finkbeiner2003}
{Finkbeiner} D.~P., 2003, \apjs, 146, 407

\bibitem[{{Finkbeiner}(2004)}]{Finkbeiner2004}
{Finkbeiner} D.~P., 2004, \apj, 614, 186

\bibitem[{{Finkbeiner}, {Davis} \& {Schlegel}(1999){Finkbeiner}, {Davis}, \&
  {Schlegel}}]{Finkbeiner1999}
{Finkbeiner} D.~P., {Davis} M., {Schlegel} D.~J., 1999, \apj, 524, 867

\bibitem[{{Finkbeiner} {et~al}\mbox{.}(2002){Finkbeiner}, {Schlegel}, {Frank},
  \& {Heiles}}]{Finkbeiner2002}
{Finkbeiner} D.~P., {Schlegel} D.~J., {Frank} C., {Heiles} C., 2002, \apj, 566,
  898

\bibitem[{{Gaustad}, {McCullough} \& {van Buren}(1996){Gaustad}, {McCullough},
  \& {van Buren}}]{Gaustad1996}
{Gaustad} J.~E., {McCullough} P.~R., {van Buren} D., 1996, \pasp, 108, 351

\bibitem[{{G{\'e}nova-Santos} {et~al}\mbox{.}(2017){G{\'e}nova-Santos},
  {Rubi{\~n}o-Mart{\'{\i}}n}, {Pel{\'a}ez-Santos}, {Poidevin}, {Rebolo},
  {Vignaga}, {Artal}, {Harper}, {Hoyland}, {Lasenby},
  {Mart{\'{\i}}nez-Gonz{\'a}lez}, {Piccirillo}, {Tramonte}, \&
  {Watson}}]{Genova-Santos2017}
{G{\'e}nova-Santos} R. {et~al.}, 2017, \mnras, 464, 4107

\bibitem[{{G{\'e}nova-Santos} {et~al}\mbox{.}(2015){G{\'e}nova-Santos},
  {Rubi{\~n}o-Mart{\'{\i}}n}, {Rebolo}, {Aguiar}, {G{\'o}mez-Re{\~n}asco},
  {Guti{\'e}rrez}, {Hoyland}, {L{\'o}pez-Caraballo}, {Pel{\'a}ez-Santos},
  {P{\'e}rez-de-Taoro}, {Poidevin}, {S{\'a}nchez de la Rosa}, {Tramonte},
  {Vega-Moreno}, {Viera-Curbelo}, {Vignasa}, {Mart{\'{\i}}nez-Gonz{\'a}lez},
  {Barreiro}, {Casaponsa}, {Casas}, {Diego}, {Fern{\'a}ndez-Cobos}, {Herranz},
  {L{\'o}pez-Caniego}, {Ortiz}, {Vielva}, {Artal}, {Aja}, {Cagigas}, {Cano},
  {de la Fuente}, {Mediavilla}, {Ter{\'a}n}, {Villa}, {Piccirillo}, {Davies},
  {Davis}, {Dickinson}, {Grainge}, {Harper}, {Maffei}, {McCulloch}, {Melhuish},
  {Pisano}, {Watson}, {Lasenby}, {Ashdown}, {Hobson}, {Perrott},
  {Razavi-Ghods}, {Saunders}, {Titterington}, \& {Scott}}]{Genova-Santos2015a}
{G{\'e}nova-Santos} R. {et~al.}, 2015, in Highlights of Spanish Astrophysics
  VIII, {Cenarro} A.~J., {Figueras} F., {Hern{\'a}ndez-Monteagudo} C.,
  {Trujillo Bueno} J., {Valdivielso} L., eds., pp. 207--212 [arXiv:1504.03514]

\bibitem[{{Ghosh} {et~al}\mbox{.}(2012){Ghosh}, {Banday}, {Jaffe}, {Dickinson},
  {Davies}, {Davis}, \& {Gorski}}]{Ghosh2012}
{Ghosh} T., {Banday} A.~J., {Jaffe} T., {Dickinson} C., {Davies} R., {Davis}
  R., {Gorski} K., 2012, \mnras, 422, 3617

\bibitem[{{Gold} {et~al}\mbox{.}(2011){Gold}, {Odegard}, {Weiland}, {Hill},
  {Kogut}, {Bennett}, {Hinshaw}, {Chen}, {Dunkley}, {Halpern}, {Jarosik},
  {Komatsu}, {Larson}, {Limon}, {Meyer}, {Nolta}, {Page}, {Smith}, {Spergel},
  {Tucker}, {Wollack}, \& {Wright}}]{Gold2011}
{Gold} B. {et~al.}, 2011, \apjs, 192, 15

\bibitem[{{G{\'o}rski} {et~al}\mbox{.}(2005){G{\'o}rski}, {Hivon}, {Banday},
  {Wandelt}, {Hansen}, {Reinecke}, \& {Bartelmann}}]{Gorski2005}
{G{\'o}rski} K.~M., {Hivon} E., {Banday} A.~J., {Wandelt} B.~D., {Hansen}
  F.~K., {Reinecke} M., {Bartelmann} M., 2005, \apj, 622, 759

\bibitem[{{Gregory} {et~al}\mbox{.}(1996){Gregory}, {Scott}, {Douglas}, \&
  {Condon}}]{Gregory1996}
{Gregory} P.~C., {Scott} W.~K., {Douglas} K., {Condon} J.~J., 1996, \apjs, 103,
  427

\bibitem[{{Haffner} {et~al}\mbox{.}(2003){Haffner}, {Reynolds}, {Tufte},
  {Madsen}, {Jaehnig}, \& {Percival}}]{Haffner2003}
{Haffner} L.~M., {Reynolds} R.~J., {Tufte} S.~L., {Madsen} G.~J., {Jaehnig}
  K.~P., {Percival} J.~W., 2003, \apjs, 149, 405

\bibitem[{{Hargrave} \& {McEllin}(1975)}]{Hargrave1975}
{Hargrave} P.~J., {McEllin} M., 1975, \mnras, 173, 37

\bibitem[{{Haslam} {et~al}\mbox{.}(1982){Haslam}, {Salter}, {Stoffel}, \&
  {Wilson}}]{Haslam1982}
{Haslam} C.~G.~T., {Salter} C.~J., {Stoffel} H., {Wilson} W.~E., 1982, \aaps,
  47, 1

\bibitem[{{Healey} {et~al}\mbox{.}(2009){Healey}, {Fuhrmann}, {Taylor},
  {Romani}, \& {Readhead}}]{Healey2009}
{Healey} S.~E., {Fuhrmann} L., {Taylor} G.~B., {Romani} R.~W., {Readhead}
  A.~C.~S., 2009, \aj, 138, 1032

\bibitem[{{Hensley}, {Draine} \& {Meisner}(2016){Hensley}, {Draine}, \&
  {Meisner}}]{Hensley2016}
{Hensley} B.~S., {Draine} B.~T., {Meisner} A.~M., 2016, \apj, 827, 45

\bibitem[{{Hoang} \& {Lazarian}(2016)}]{Hoang2016}
{Hoang} T., {Lazarian} A., 2016, \apj, 821, 91

\bibitem[{{Holler} {et~al}\mbox{.}(2013){Holler}, {Taylor}, {Jones}, {King},
  {Muchovej}, {Stevenson}, {Wylde}, {Copley}, {Davis}, {Pearson}, \&
  {Readhead}}]{Holler2013}
{Holler} C.~M. {et~al.}, 2013, IEEE Transactions on Antennas and Propagation,
  61, 117

\bibitem[{Irfan(2014)}]{Irfan2014}
Irfan M.~O., 2014, PhD thesis, The University of Manchester, Oxford Road,
  Manchester, M13 9PL, U.K.

\bibitem[{{Irfan} {et~al}\mbox{.}(2015){Irfan}, {Dickinson}, {Davies},
  {Copley}, {Davis}, {Ferreira}, {Holler}, {Jonas}, {Jones}, {King}, {Leahy},
  {Leech}, {Leitch}, {Muchovej}, {Pearson}, {Peel}, {Readhead}, {Stevenson},
  {Sutton}, {Taylor}, \& {Zuntz}}]{Irfan2015}
{Irfan} M.~O. {et~al.}, 2015, \mnras, 448, 3572

\bibitem[{{Jaffe} {et~al}\mbox{.}(2011){Jaffe}, {Banday}, {Leahy}, {Leach}, \&
  {Strong}}]{Jaffe2011}
{Jaffe} T.~R., {Banday} A.~J., {Leahy} J.~P., {Leach} S., {Strong} A.~W., 2011,
  \mnras, 416, 1152

\bibitem[{{Jonas}, {Baart} \& {Nicolson}(1998){Jonas}, {Baart}, \&
  {Nicolson}}]{Jonas1998}
{Jonas} J.~L., {Baart} E.~E., {Nicolson} G.~D., 1998, \mnras, 297, 977

\bibitem[{{Jones} {et~al}\mbox{.}(2018){Jones}, {Taylor}, {Aich}, {Copley},
  {Chiang}, {Davis}, {Dickinson}, {Grumitt}, {Hafez}, {Heilgendorff}, {Holler},
  {Irfan}, {Jew}, {John}, {Jonas}, {King}, {Leahy}, {Leech}, {Leitch},
  {Muchovej}, {Pearson}, {Peel}, {Readhead}, {Sievers}, {Stevenson}, \&
  {Zuntz}}]{Jones2018}
{Jones} M.~E. {et~al.}, 2018, \mnras, 480, 3224

\bibitem[{{King} {et~al}\mbox{.}(2010){King}, {Copley}, {Davies}, {Davis},
  {Dickinson}, {Hafez}, {Holler}, {John}, {Jonas}, {Jones}, {Leahy},
  {Muchovej}, {Pearson}, {Readhead}, {Stevenson}, \& {Taylor}}]{King2010}
{King} O.~G. {et~al.}, 2010, in Society of Photo-Optical Instrumentation
  Engineers (SPIE) Conference Series, Vol. 7741, Society of Photo-Optical
  Instrumentation Engineers (SPIE) Conference Series

\bibitem[{{King} {et~al}\mbox{.}(2014){King}, {Jones}, {Blackhurst}, {Copley},
  {Davis}, {Dickinson}, {Holler}, {Irfan}, {John}, {Leahy}, {Leech},
  {Muchovej}, {Pearson}, {Stevenson}, \& {Taylor}}]{King2014}
{King} O.~G. {et~al.}, 2014, \mnras, 438, 2426

\bibitem[{{Kogut}(2012)}]{Kogut2012}
{Kogut} A., 2012, \apj, 753, 110

\bibitem[{{Kogut} {et~al}\mbox{.}(1996){Kogut}, {Banday}, {Bennett}, {Gorski},
  {Hinshaw}, {Smoot}, \& {Wright}}]{Kogut1996}
{Kogut} A., {Banday} A.~J., {Bennett} C.~L., {Gorski} K.~M., {Hinshaw} G.,
  {Smoot} G.~F., {Wright} E.~I., 1996, \apjl, 464, L5

\bibitem[{{Kogut} {et~al}\mbox{.}(2007){Kogut}, {Dunkley}, {Bennett},
  {Dor{\'e}}, {Gold}, {Halpern}, {Hinshaw}, {Jarosik}, {Komatsu}, {Nolta},
  {Odegard}, {Page}, {Spergel}, {Tucker}, {Weiland}, {Wollack}, \&
  {Wright}}]{Kogut2007}
{Kogut} A. {et~al.}, 2007, \apj, 665, 355

\bibitem[{{Kogut} {et~al}\mbox{.}(2011){Kogut}, {Fixsen}, {Levin}, {Limon},
  {Lubin}, {Mirel}, {Seiffert}, {Singal}, {Villela}, {Wollack}, \&
  {Wuensche}}]{Kogut2011}
{Kogut} A. {et~al.}, 2011, \apj, 734, 4

\bibitem[{{Krachmalnicoff} {et~al}\mbox{.}(2018){Krachmalnicoff}, {Carretti},
  {Baccigalupi}, {Bernardi}, {Brown}, {Gaensler}, {Haverkorn}, {Kesteven},
  {Perrotta}, {Poppi}, \& {Staveley-Smith}}]{Krachmalnicoff2018}
{Krachmalnicoff} N. {et~al.}, 2018, \aap, 618, A166

\bibitem[{{K\"{u}hr} {et~al}\mbox{.}(1981{\natexlab{a}}){K\"{u}hr},
  {Pauliny-Toth}, {Witzel}, \& {Schmidt}}]{Kuhr1981b}
{K\"{u}hr} H., {Pauliny-Toth} I.~I.~K., {Witzel} A., {Schmidt} J.,
  1981{\natexlab{a}}, \aj, 86, 854

\bibitem[{{K\"{u}hr} {et~al}\mbox{.}(1981{\natexlab{b}}){K\"{u}hr}, {Witzel},
  {Pauliny-Toth}, \& {Nauber}}]{Kuhr1981a}
{K\"{u}hr} H., {Witzel} A., {Pauliny-Toth} I.~I.~K., {Nauber} U.,
  1981{\natexlab{b}}, \aaps, 45, 367

\bibitem[{{Lagache}(2003)}]{Lagache2003}
{Lagache} G., 2003, \aap, 405, 813

\bibitem[{{Lawson} {et~al}\mbox{.}(1987){Lawson}, {Mayer}, {Osborne}, \&
  {Parkinson}}]{Lawson1987}
{Lawson} K.~D., {Mayer} C.~J., {Osborne} J.~L., {Parkinson} M.~L., 1987,
  \mnras, 225, 307

\bibitem[{{Leach} {et~al}\mbox{.}(2008){Leach}, {Cardoso}, {Baccigalupi},
  {Barreiro}, {Betoule}, {Bobin}, {Bonaldi}, {Delabrouille}, {de Zotti},
  {Dickinson}, {Eriksen}, {Gonz{\'a}lez-Nuevo}, {Hansen}, {Herranz}, {Le
  Jeune}, {L{\'o}pez-Caniego}, {Mart{\'{\i}}nez-Gonz{\'a}lez}, {Massardi},
  {Melin}, {Miville-Desch{\^e}nes}, {Patanchon}, {Prunet}, {Ricciardi},
  {Salerno}, {Sanz}, {Starck}, {Stivoli}, {Stolyarov}, {Stompor}, \&
  {Vielva}}]{Leach2008}
{Leach} S.~M. {et~al.}, 2008, \aap, 491, 597

\bibitem[{{Leitch} {et~al}\mbox{.}(1997){Leitch}, {Readhead}, {Pearson}, \&
  {Myers}}]{Leitch1997}
{Leitch} E.~M., {Readhead} A.~C.~S., {Pearson} T.~J., {Myers} S.~T., 1997,
  \apjl, 486, L23

\bibitem[{{Liu} {et~al}\mbox{.}(2014){Liu}, {Cui}, {Liu}, {Ding}, \&
  {Song}}]{Liu2014}
{Liu} J., {Cui} L., {Liu} X., {Ding} Z., {Song} H.-G., 2014, Journal of
  Astrophysics and Astronomy

\bibitem[{{L{\'o}pez-Caraballo} {et~al}\mbox{.}(2011){L{\'o}pez-Caraballo},
  {Rubi{\~n}o-Mart{\'{\i}}n}, {Rebolo}, \&
  {G{\'e}nova-Santos}}]{Lopez-Caraballo2011}
{L{\'o}pez-Caraballo} C.~H., {Rubi{\~n}o-Mart{\'{\i}}n} J.~A., {Rebolo} R.,
  {G{\'e}nova-Santos} R., 2011, \apj, 729, 25

\bibitem[{{Macellari} {et~al}\mbox{.}(2011){Macellari}, {Pierpaoli},
  {Dickinson}, \& {Vaillancourt}}]{Macellari2011}
{Macellari} N., {Pierpaoli} E., {Dickinson} C., {Vaillancourt} J.~E., 2011,
  \mnras, 418, 888

\bibitem[{{Markwardt}(2009)}]{Markwardt2009}
{Markwardt} C.~B., 2009, in Astronomical Society of the Pacific Conference
  Series, Vol. 411, Astronomical Data Analysis Software and Systems XVIII,
  {Bohlender} D.~A., {Durand} D., {Dowler} P., eds., p. 251

\bibitem[{{Mingaliev} {et~al}\mbox{.}(2007){Mingaliev}, {Sotnikova}, {Bursov},
  {Kardashev}, \& {Larionov}}]{Mingaliev2007}
{Mingaliev} M.~G., {Sotnikova} Y.~V., {Bursov} N.~N., {Kardashev} N.~S.,
  {Larionov} M.~G., 2007, Astronomy Reports, 51, 343

\bibitem[{{Miville-Desch{\^e}nes} \& {Lagache}(2005)}]{Miville-Deschenes2005}
{Miville-Desch{\^e}nes} M., {Lagache} G., 2005, \apjs, 157, 302

\bibitem[{{Oni{\'c}}(2013)}]{Onic2013}
{Oni{\'c}} D., 2013, \apss, 346, 3

\bibitem[{{Orlando} \& {Strong}(2013)}]{Orlando2013}
{Orlando} E., {Strong} A., 2013, \mnras, 436, 2127

\bibitem[{{Peel} {et~al}\mbox{.}(2012){Peel}, {Dickinson}, {Davies}, {Banday},
  {Jaffe}, \& {Jonas}}]{Peel2012}
{Peel} M.~W., {Dickinson} C., {Davies} R.~D., {Banday} A.~J., {Jaffe} T.~R.,
  {Jonas} J.~L., 2012, \mnras, 424, 2676

\bibitem[{{Planck Collaboration} {et~al}\mbox{.}(2014{\natexlab{a}}){Planck
  Collaboration}, {Abergel}, {Ade}, {Aghanim}, {Alves}, {Aniano},
  {Armitage-Caplan}, {Arnaud}, {Ashdown}, {Atrio-Barandela}, \&
  et~al.}]{Planck2013_XI}
{Planck Collaboration} {et~al.}, 2014{\natexlab{a}}, \aap, 571, A11

\bibitem[{{Planck Collaboration} {et~al}\mbox{.}(2011{\natexlab{a}}){Planck
  Collaboration}, {Abergel}, {Ade}, {Aghanim}, {Arnaud}, {Ashdown}, {Aumont},
  {Baccigalupi}, {Balbi}, {Banday}, {Barreiro}, {Bartlett}, {Battaner},
  {Benabed}, {Beno{\^i}t}, {Bernard}, {Bersanelli}, {Bhatia}, {Bock},
  {Bonaldi}, {Bond}, {Borrill}, {Bouchet}, {Boulanger}, {Bucher}, {Burigana},
  {Cabella}, {Cardoso}, {Catalano}, {Cay{\'o}n}, {Challinor}, {Chamballu},
  {Chiang}, {Chiang}, {Christensen}, {Colombi}, {Couchot}, {Coulais}, {Crill},
  {Cuttaia}, {Dame}, {Danese}, {Davies}, {Davis}, {de Bernardis}, {de
  Gasperis}, {de Rosa}, {de Zotti}, {Delabrouille}, {Delouis}, {D{\'e}sert},
  {Dickinson}, {Donzelli}, {Dor{\'e}}, {D{\"o}rl}, {Douspis}, {Dupac},
  {Efstathiou}, {En{\ss}lin}, {Finelli}, {Forni}, {Frailis}, {Franceschi},
  {Galeotta}, {Ganga}, {Giard}, {Giardino}, {Giraud-H{\'e}raud},
  {Gonz{\'a}lez-Nuevo}, {G{\'o}rski}, {Gratton}, {Gregorio}, {Grenier},
  {Gruppuso}, {Hansen}, {Harrison}, {Henrot-Versill{\'e}}, {Herranz},
  {Hildebrandt}, {Hivon}, {Hobson}, {Holmes}, {Hovest}, {Hoyland},
  {Huffenberger}, {Jaffe}, {Jaffe}, {Jones}, {Juvela}, {Keih{\"a}nen},
  {Keskitalo}, {Kisner}, {Kneissl}, {Knox}, {Kurki-Suonio}, {Lagache},
  {L{\"a}hteenm{\"a}ki}, {Lamarre}, {Lasenby}, {Laureijs}, {Lawrence}, {Leach},
  {Leonardi}, {Leroy}, {Lilje}, {Linden-V{\o}rnle}, {L{\'o}pez-Caniego},
  {Lubin}, {Mac{\'{\i}}as-P{\'e}rez}, {MacTavish}, {Maffei}, {Mandolesi},
  {Mann}, {Maris}, {Marshall}, {Mart{\'{\i}}nez-Gonz{\'a}lez}, {Masi},
  {Matarrese}, {Matthai}, {Mazzotta}, {McGehee}, {Meinhold}, {Melchiorri},
  {Mendes}, {Mennella}, {Miville-Desch{\^e}nes}, {Moneti}, {Montier},
  {Morgante}, {Mortlock}, {Munshi}, {Murphy}, {Naselsky}, {Natoli},
  {Netterfield}, {N{\o}rgaard-Nielsen}, {Noviello}, {Novikov}, {Novikov},
  {Osborne}, {Pajot}, {Paladini}, {Pasian}, {Patanchon}, {Perdereau},
  {Perotto}, {Perrotta}, {Piacentini}, {Piat}, {Plaszczynski}, {Pointecouteau},
  {Polenta}, {Ponthieu}, {Poutanen}, {Pr{\'e}zeau}, {Prunet}, {Puget},
  {Rachen}, {Reach}, {Rebolo}, {Reich}, {Renault}, {Ricciardi}, {Riller},
  {Ristorcelli}, {Rocha}, {Rosset}, {Rubi{\~n}o-Mart{\'{\i}}n}, {Rusholme},
  {Sandri}, {Santos}, {Savini}, {Scott}, {Seiffert}, {Shellard}, {Smoot},
  {Starck}, {Stivoli}, {Stolyarov}, {Stompor}, {Sudiwala}, {Sygnet}, {Tauber},
  {Terenzi}, {Toffolatti}, {Tomasi}, {Torre}, {Tristram}, {Tuovinen}, {Umana},
  {Valenziano}, {Varis}, {Vielva}, {Villa}, {Vittorio}, {Wade}, {Wandelt},
  {Wilkinson}, {Ysard}, {Yvon}, {Zacchei}, \& {Zonca}}]{PEP_XXI}
{Planck Collaboration} {et~al.}, 2011{\natexlab{a}}, \aap, 536, A21

\bibitem[{{Planck Collaboration} {et~al}\mbox{.}(2016{\natexlab{a}}){Planck
  Collaboration}, {Adam}, {Ade}, {Aghanim}, {Alves}, {Arnaud}, {Ashdown},
  {Aumont}, {Baccigalupi}, {Banday}, \& et~al.}]{Planck2015_X}
{Planck Collaboration} {et~al.}, 2016{\natexlab{a}}, \aap, 594, A10

\bibitem[{{Planck Collaboration} {et~al}\mbox{.}(2016{\natexlab{b}}){Planck
  Collaboration}, {Adam}, {Ade}, {Aghanim}, {Arnaud}, {Ashdown}, {Aumont},
  {Baccigalupi}, {Banday}, {Barreiro}, \& et~al.}]{Planck2015_IX}
{Planck Collaboration} {et~al.}, 2016{\natexlab{b}}, \aap, 594, A9

\bibitem[{{Planck Collaboration} {et~al}\mbox{.}(2014{\natexlab{b}}){Planck
  Collaboration}, {Ade}, {Aghanim}, {Alves}, {Armitage-Caplan}, {Arnaud},
  {Ashdown}, {Atrio-Barandela}, {Aumont}, {Baccigalupi}, \&
  et~al.}]{Planck2013_CO}
{Planck Collaboration} {et~al.}, 2014{\natexlab{b}}, \aap, 571, A13

\bibitem[{{Planck Collaboration} {et~al}\mbox{.}(2013{\natexlab{a}}){Planck
  Collaboration}, {Ade}, {Aghanim}, {Alves}, {Arnaud}, {Ashdown},
  {Atrio-Barandela}, {Aumont}, {Baccigalupi}, {Balbi}, {Banday}, {Barreiro},
  {Bartlett}, {Battaner}, {Bedini}, {Benabed}, {Beno{\^i}t}, {Bernard},
  {Bersanelli}, {Bonaldi}, {Bond}, {Borrill}, {Bouchet}, {Boulanger},
  {Burigana}, {Butler}, {Cabella}, {Cardoso}, {Chen}, {Chiang}, {Christensen},
  {Clements}, {Colombi}, {Colombo}, {Coulais}, {Cuttaia}, {Davies}, {Davis},
  {de Bernardis}, {de Gasperis}, {de Zotti}, {Delabrouille}, {Dickinson},
  {Diego}, {Dobler}, {Dole}, {Donzelli}, {Dor{\'e}}, {Douspis}, {Dupac},
  {En{\ss}lin}, {Finelli}, {Forni}, {Frailis}, {Franceschi}, {Galeotta},
  {Ganga}, {G{\'e}nova-Santos}, {Ghosh}, {Giard}, {Giardino},
  {Giraud-H{\'e}raud}, {Gonz{\'a}lez-Nuevo}, {G{\'o}rski}, {Gregorio},
  {Gruppuso}, {Hansen}, {Harrison}, {Hern{\'a}ndez-Monteagudo}, {Hildebrandt},
  {Hivon}, {Hobson}, {Holmes}, {Hornstrup}, {Hovest}, {Huffenberger}, {Jaffe},
  {Jaffe}, {Juvela}, {Keih{\"a}nen}, {Keskitalo}, {Kisner}, {Knoche}, {Kunz},
  {Kurki-Suonio}, {Lagache}, {L{\"a}hteenm{\"a}ki}, {Lamarre}, {Lasenby},
  {Lawrence}, {Leach}, {Leonardi}, {Lilje}, {Linden-V{\o}rnle}, {Lubin},
  {Mac{\'{\i}}as-P{\'e}rez}, {Maffei}, {Maino}, {Mandolesi}, {Maris},
  {Marshall}, {Martin}, {Mart{\'{\i}}nez-Gonz{\'a}lez}, {Masi}, {Massardi},
  {Matarrese}, {Mazzotta}, {Melchiorri}, {Mennella}, {Mitra},
  {Miville-Desch{\^e}nes}, {Moneti}, {Montier}, {Morgante}, {Mortlock},
  {Munshi}, {Murphy}, {Naselsky}, {Nati}, {Natoli}, {N{\o}rgaard-Nielsen},
  {Noviello}, {Novikov}, {Novikov}, {Osborne}, {Oxborrow}, {Pajot}, {Paladini},
  {Paoletti}, {Peel}, {Perotto}, {Perrotta}, {Piacentini}, {Piat}, {Pierpaoli},
  {Pietrobon}, {Plaszczynski}, {Pointecouteau}, {Polenta}, {Popa}, {Poutanen},
  {Pratt}, {Prunet}, {Puget}, {Rachen}, {Reach}, {Rebolo}, {Reinecke},
  {Renault}, {Ricciardi}, {Ristorcelli}, {Rocha}, {Rosset},
  {Rubi{\~n}o-Mart{\'{\i}}n}, {Rusholme}, {Salerno}, {Sandri}, {Savini},
  {Scott}, {Spencer}, {Stolyarov}, {Sudiwala}, {Suur-Uski}, {Sygnet}, {Tauber},
  {Terenzi}, {Tibbs}, {Toffolatti}, {Tomasi}, {Tristram}, {Valenziano}, {Van
  Tent}, {Varis}, {Vielva}, {Villa}, {Vittorio}, {Wade}, {Wandelt}, {Ysard},
  {Yvon}, {Zacchei}, \& {Zonca}}]{PIP_XII}
{Planck Collaboration} {et~al.}, 2013{\natexlab{a}}, \aap, 557, A53

\bibitem[{{Planck Collaboration} {et~al}\mbox{.}(2016{\natexlab{c}}){Planck
  Collaboration}, {Ade}, {Aghanim}, {Alves}, {Arnaud}, {Ashdown}, {Aumont},
  {Baccigalupi}, {Banday}, {Barreiro}, \& et~al.}]{Planck2015_XXV}
{Planck Collaboration} {et~al.}, 2016{\natexlab{c}}, \aap, 594, A25

\bibitem[{{Planck Collaboration} {et~al}\mbox{.}(2014{\natexlab{c}}){Planck
  Collaboration}, {Ade}, {Aghanim}, {Alves}, {Arnaud}, {Atrio-Barandela},
  {Aumont}, {Baccigalupi}, {Banday}, {Barreiro}, {Battaner}, {Benabed},
  {Benoit-L{\'e}vy}, {Bernard}, {Bersanelli}, {Bielewicz}, {Bobin}, {Bonaldi},
  {Bond}, {Borrill}, {Bouchet}, {Boulanger}, {Burigana}, {Cardoso}, {Casassus},
  {Catalano}, {Chamballu}, {Chen}, {Chiang}, {Chiang}, {Christensen},
  {Clements}, {Colombi}, {Colombo}, {Couchot}, {Crill}, {Cuttaia}, {Danese},
  {Davies}, {Davis}, {de Bernardis}, {de Rosa}, {de Zotti}, {Delabrouille},
  {D{\'e}sert}, {Dickinson}, {}, {Diego}, {Donzelli}, {Dor{\'e}}, {Dupac},
  {En{\ss}lin}, {Eriksen}, {Finelli}, {Forni}, {Franceschi}, {Galeotta},
  {Ganga}, {G{\'e}nova-Santos}, {Ghosh}, {Giard}, {Gonz{\'a}lez-Nuevo},
  {G{\'o}rski}, {Gregorio}, {Gruppuso}, {Hansen}, {Harrison}, {Helou},
  {Hern{\'a}ndez-Monteagudo}, {Hildebrandt}, {Hivon}, {Hobson}, {Hornstrup},
  {Jaffe}, {Jaffe}, {Jones}, {Keih{\"a}nen}, {Keskitalo}, {Kneissl}, {Knoche},
  {Kunz}, {Kurki-Suonio}, {L{\"a}hteenm{\"a}ki}, {Lamarre}, {Lasenby},
  {Lawrence}, {Leonardi}, {Liguori}, {Lilje}, {Linden-V{\o}rnle},
  {L{\'o}pez-Caniego}, {Mac{\'{\i}}as-P{\'e}rez}, {Maffei}, {Maino},
  {Mandolesi}, {Marshall}, {Martin}, {Mart{\'{\i}}nez-Gonz{\'a}lez}, {Masi},
  {Massardi}, {Matarrese}, {Mazzotta}, {Meinhold}, {Melchiorri}, {Mendes},
  {Mennella}, {Migliaccio}, {Miville-Desch{\^e}nes}, {Moneti}, {Montier},
  {Morgante}, {Mortlock}, {Munshi}, {Naselsky}, {Nati}, {Natoli},
  {N{\o}rgaard-Nielsen}, {Noviello}, {Novikov}, {Novikov}, {Oxborrow},
  {Pagano}, {Pajot}, {Paladini}, {Paoletti}, {Patanchon}, {Pearson}, {Peel},
  {Perdereau}, {Perrotta}, {Piacentini}, {Piat}, {Pierpaoli}, {Pietrobon},
  {Plaszczynski}, {Pointecouteau}, {Polenta}, {Ponthieu}, {Popa}, {Pratt},
  {Prunet}, {Puget}, {Rachen}, {Rebolo}, {Reich}, {Reinecke}, {Remazeilles},
  {Renault}, {Ricciardi}, {Riller}, {Ristorcelli}, {Rocha}, {Rosset},
  {Roudier}, {Rubi{\~n}o-Mart{\'{\i}}n}, {Rusholme}, {Sandri}, {Savini},
  {Scott}, {Spencer}, {Stolyarov}, {Sutton}, {Suur-Uski}, {Sygnet}, {Tauber},
  {Tavagnacco}, {Terenzi}, {Tibbs}, {Toffolatti}, {Tomasi}, {Tristram},
  {Tucci}, {Valenziano}, {Valiviita}, {Van Tent}, {Varis}, {Verstraete},
  {Vielva}, {Villa}, {Wandelt}, {Watson}, {Wilkinson}, {Ysard}, {Yvon},
  {Zacchei}, \& {Zonca}}]{PIP_XV}
{Planck Collaboration} {et~al.}, 2014{\natexlab{c}}, \aap, 565, A103

\bibitem[{{Planck Collaboration} {et~al}\mbox{.}(2016{\natexlab{d}}){Planck
  Collaboration}, {Ade}, {Aghanim}, {Arg{\"u}eso}, {Arnaud}, {Ashdown},
  {Aumont}, {Baccigalupi}, {Banday}, {Barreiro}, \& et~al.}]{Planck2015_XXVI}
{Planck Collaboration} {et~al.}, 2016{\natexlab{d}}, \aap, 594, A26

\bibitem[{{Planck Collaboration} {et~al}\mbox{.}(2013{\natexlab{b}}){Planck
  Collaboration}, {Ade}, {Aghanim}, {Arnaud}, {Ashdown}, {Atrio-Barandela},
  {Aumont}, {Baccigalupi}, {Balbi}, {Banday}, {Barreiro}, {Bartlett},
  {Battaner}, {Benabed}, {Beno{\^i}t}, {Bernard}, {Bersanelli}, {Bonaldi},
  {Bond}, {Borrill}, {Bouchet}, {Burigana}, {Cabella}, {Cardoso}, {Catalano},
  {Cay{\'o}n}, {Chary}, {Chiang}, {Christensen}, {Clements}, {Colombo},
  {Coulais}, {Crill}, {Cuttaia}, {Danese}, {D'Arcangelo}, {Davis}, {de
  Bernardis}, {de Gasperis}, {de Rosa}, {de Zotti}, {Delabrouille},
  {Dickinson}, {Diego}, {Dobler}, {Dole}, {Donzelli}, {Dor{\'e}}, {D{\"o}rl},
  {Douspis}, {Dupac}, {Efstathiou}, {En{\ss}lin}, {Eriksen}, {Finelli},
  {Forni}, {Frailis}, {Franceschi}, {Galeotta}, {Ganga}, {Giard}, {Giardino},
  {Gonz{\'a}lez-Nuevo}, {G{\'o}rski}, {Gratton}, {Gregorio}, {Gruppuso},
  {Hansen}, {Harrison}, {Helou}, {Henrot-Versill{\'e}},
  {Hern{\'a}ndez-Monteagudo}, {Hildebrandt}, {Hivon}, {Hobson}, {Holmes},
  {Hornstrup}, {Hovest}, {Huffenberger}, {Jaffe}, {Jagemann}, {Jewell},
  {Jones}, {Juvela}, {Keih{\"a}nen}, {Knoche}, {Knox}, {Kunz}, {Kurki-Suonio},
  {Lagache}, {L{\"a}hteenm{\"a}ki}, {Lamarre}, {Lasenby}, {Lawrence}, {Leach},
  {Leonardi}, {Lilje}, {Linden-V{\o}rnle}, {L{\'o}pez-Caniego}, {Lubin},
  {Mac{\'{\i}}as-P{\'e}rez}, {Maffei}, {Maino}, {Mandolesi}, {Maris},
  {Marshall}, {Martin}, {Mart{\'{\i}}nez-Gonz{\'a}lez}, {Masi}, {Massardi},
  {Matarrese}, {Matthai}, {Mazzotta}, {Meinhold}, {Melchiorri}, {Mendes},
  {Mennella}, {Mitra}, {Moneti}, {Montier}, {Morgante}, {Munshi}, {Murphy},
  {Naselsky}, {Natoli}, {N{\o}rgaard-Nielsen}, {Noviello}, {Novikov},
  {Novikov}, {Osborne}, {Pajot}, {Paladini}, {Paoletti}, {Partridge},
  {Pearson}, {Perdereau}, {Perrotta}, {Piacentini}, {Piat}, {Pierpaoli},
  {Pietrobon}, {Plaszczynski}, {Pointecouteau}, {Polenta}, {Ponthieu}, {Popa},
  {Poutanen}, {Pratt}, {Prunet}, {Puget}, {Rachen}, {Rebolo}, {Reinecke},
  {Renault}, {Ricciardi}, {Riller}, {Ristorcelli}, {Rocha}, {Rosset},
  {Rubi{\~n}o-Mart{\'{\i}}n}, {Rusholme}, {Sandri}, {Savini}, {Schaefer},
  {Scott}, {Smoot}, {Spencer}, {Stivoli}, {Sudiwala}, {Suur-Uski}, {Sygnet},
  {Tauber}, {Terenzi}, {Toffolatti}, {Tomasi}, {Tristram}, {T{\"u}rler},
  {Umana}, {Valenziano}, {Van Tent}, {Vielva}, {Villa}, {Vittorio}, {Wade},
  {Wandelt}, {White}, {Yvon}, {Zacchei}, \& {Zonca}}]{PIP_IX}
{Planck Collaboration} {et~al.}, 2013{\natexlab{b}}, \aap, 554, A139

\bibitem[{{Planck Collaboration} {et~al}\mbox{.}(2011{\natexlab{b}}){Planck
  Collaboration}, {Ade}, {Aghanim}, {Arnaud}, {Ashdown}, {Aumont},
  {Baccigalupi}, {Balbi}, {Banday}, {Barreiro}, \& et~al.}]{PEP_XX}
{Planck Collaboration} {et~al.}, 2011{\natexlab{b}}, \aap, 536, A20

\bibitem[{{Planck Collaboration} {et~al}\mbox{.}(2016{\natexlab{e}}){Planck
  Collaboration}, {Aghanim}, {Ashdown}, {Aumont}, {Baccigalupi}, {Ballardini},
  {Banday}, {Barreiro}, {Bartolo}, {Basak}, {Benabed}, {Bernard}, {Bersanelli},
  {Bielewicz}, {Bonavera}, {Bond}, {Borrill}, {Bouchet}, {Boulanger},
  {Burigana}, {Calabrese}, {Cardoso}, {Carron}, {Chiang}, {Colombo}, {Comis},
  {Couchot}, {Coulais}, {Crill}, {Curto}, {Cuttaia}, {de Bernardis}, {de
  Zotti}, {Delabrouille}, {Di Valentino}, {Dickinson}, {Diego}, {Dor{\'e}},
  {Douspis}, {Ducout}, {Dupac}, {Dusini}, {Elsner}, {En{\ss}lin}, {Eriksen},
  {Falgarone}, {Fantaye}, {Finelli}, {Forastieri}, {Frailis}, {Fraisse},
  {Franceschi}, {Frolov}, {Galeotta}, {Galli}, {Ganga}, {G{\'e}nova-Santos},
  {Gerbino}, {Ghosh}, {Giraud-H{\'e}raud}, {Gonz{\'a}lez-Nuevo}, {G{\'o}rski},
  {Gruppuso}, {Gudmundsson}, {Hansen}, {Helou}, {Henrot-Versill{\'e}},
  {Herranz}, {Hivon}, {Huang}, {Jaffe}, {Jones}, {Keih{\"a}nen}, {Keskitalo},
  {Kiiveri}, {Kisner}, {Krachmalnicoff}, {Kunz}, {Kurki-Suonio}, {Lamarre},
  {Langer}, {Lasenby}, {Lattanzi}, {Lawrence}, {Le Jeune}, {Levrier}, {Lilje},
  {Lilley}, {Lindholm}, {L{\'o}pez-Caniego}, {Ma}, {Mac{\'{\i}}as-P{\'e}rez},
  {Maggio}, {Maino}, {Mandolesi}, {Mangilli}, {Maris}, {Martin},
  {Mart{\'{\i}}nez-Gonz{\'a}lez}, {Matarrese}, {Mauri}, {McEwen}, {Melchiorri},
  {Mennella}, {Migliaccio}, {Miville-Desch{\^e}nes}, {Molinari}, {Moneti},
  {Montier}, {Morgante}, {Moss}, {Natoli}, {Oxborrow}, {Pagano}, {Paoletti},
  {Patanchon}, {Perdereau}, {Perotto}, {Pettorino}, {Piacentini},
  {Plaszczynski}, {Polastri}, {Polenta}, {Puget}, {Rachen}, {Racine},
  {Reinecke}, {Remazeilles}, {Renzi}, {Rocha}, {Rosset}, {Rossetti}, {Roudier},
  {Rubi{\~n}o-Mart{\'{\i}}n}, {Ruiz-Granados}, {Salvati}, {Sandri},
  {Savelainen}, {Scott}, {Sirignano}, {Sirri}, {Soler}, {Spencer}, {Suur-Uski},
  {Tauber}, {Tavagnacco}, {Tenti}, {Toffolatti}, {Tomasi}, {Tristram},
  {Trombetti}, {Valiviita}, {Van Tent}, {Vielva}, {Villa}, {Vittorio},
  {Wandelt}, {Wehus}, {Zacchei}, \& {Zonca}}]{PIP_XLVIII}
{Planck Collaboration} {et~al.}, 2016{\natexlab{e}}, \aap, 596, A109

\bibitem[{{Planck Collaboration} {et~al}\mbox{.}(2018{\natexlab{a}}){Planck
  Collaboration}, {Akrami}, {Arroja}, {Ashdown}, {Aumont}, {Baccigalupi},
  {Ballardini}, {Banday}, {Barreiro}, {Bartolo}, {Basak}, {Battye}, {Benabed},
  {Bernard}, {Bersanelli}, {Bielewicz}, {Bock}, {Bond}, {Borrill}, {Bouchet},
  {Boulanger}, {Bucher}, {Burigana}, {Butler}, {Calabrese}, {Cardoso},
  {Carron}, {Casaponsa}, {Challinor}, {Chiang}, {Colombo}, {Combet},
  {Contreras}, {Crill}, {Cuttaia}, {de Bernardis}, {de Zotti}, {Delabrouille},
  {Delouis}, {D{\'e}sert}, {Di Valentino}, {Dickinson}, {Diego}, {Donzelli},
  {Dor{\'e}}, {Douspis}, {Ducout}, {Dupac}, {Efstathiou}, {Elsner},
  {En{\ss}lin}, {Eriksen}, {Falgarone}, {Fantaye}, {Fergusson},
  {Fernandez-Cobos}, {Finelli}, {Forastieri}, {Frailis}, {Franceschi},
  {Frolov}, {Galeotta}, {Galli}, {Ganga}, {G{\'e}nova-Santos}, {Gerbino},
  {Ghosh}, {Gonz{\'a}lez-Nuevo}, {G{\'o}rski}, {Gratton}, {Gruppuso},
  {Gudmundsson}, {Hamann}, {Handley}, {Hansen}, {Helou}, {Herranz}, {Hivon},
  {Huang}, {Jaffe}, {Jones}, {Karakci}, {Keih{\"a}nen}, {Keskitalo}, {Kiiveri},
  {Kim}, {Kisner}, {Knox}, {Krachmalnicoff}, {Kunz}, {Kurki-Suonio}, {Lagache},
  {Lamarre}, {Langer}, {Lasenby}, {Lattanzi}, {Lawrence}, {Le Jeune}, {Leahy},
  {Lesgourgues}, {Levrier}, {Lewis}, {Liguori}, {Lilje}, {Lilley}, {Lindholm},
  {L{\'o}pez-Caniego}, {Lubin}, {Ma}, {Mac{\'{\i}}as-P{\'e}rez}, {Maggio},
  {Maino}, {Mandolesi}, {Mangilli}, {Marcos-Caballero}, {Maris}, {Martin},
  {Mart{\'{\i}}nez-Gonz{\'a}lez}, {Matarrese}, {Mauri}, {McEwen}, {Meerburg},
  {Meinhold}, {Melchiorri}, {Mennella}, {Migliaccio}, {Millea}, {Mitra},
  {Miville-Desch{\^e}nes}, {Molinari}, {Moneti}, {Montier}, {Morgante}, {Moss},
  {Mottet}, {M{\"u}nchmeyer}, {Natoli}, {N{\o}rgaard-Nielsen}, {Oxborrow},
  {Pagano}, {Paoletti}, {Partridge}, {Patanchon}, {Pearson}, {Peel}, {Peiris},
  {Perrotta}, {Pettorino}, {Piacentini}, {Polastri}, {Polenta}, {Puget},
  {Rachen}, {Reinecke}, {Remazeilles}, {Renzi}, {Rocha}, {Rosset}, {Roudier},
  {Rubi{\~n}o-Mart{\'{\i}}n}, {Ruiz-Granados}, {Salvati}, {Sandri},
  {Savelainen}, {Scott}, {Shellard}, {Shiraishi}, {Sirignano}, {Sirri},
  {Spencer}, {Sunyaev}, {Suur-Uski}, {Tauber}, {Tavagnacco}, {Tenti},
  {Terenzi}, {Toffolatti}, {Tomasi}, {Trombetti}, {Valiviita}, {Van Tent},
  {Vibert}, {Vielva}, {Villa}, {Vittorio}, {Wandelt}, {Wehus}, {White},
  {White}, {Zacchei}, \& {Zonca}}]{Planck2018_I}
{Planck Collaboration} {et~al.}, 2018{\natexlab{a}}, submitted to A\&A
  [arXiv:1807.06205]

\bibitem[{{Planck Collaboration} {et~al}\mbox{.}(2018{\natexlab{b}}){Planck
  Collaboration}, {Akrami}, {Ashdown}, {Aumont}, {Baccigalupi}, {Ballardini},
  {Banday}, {Barreiro}, {Bartolo}, {Basak}, {Benabed}, {Bersanelli},
  {Bielewicz}, {Bond}, {Borrill}, {Bouchet}, {Boulanger}, {Bucher}, {Burigana},
  {Calabrese}, {Cardoso}, {Carron}, {Casaponsa}, {Challinor}, {Colombo},
  {Combet}, {Crill}, {Cuttaia}, {de Bernardis}, {de Rosa}, {de Zotti},
  {Delabrouille}, {Delouis}, {Di Valentino}, {Dickinson}, {Diego}, {Donzelli},
  {Dor{\'e}}, {Ducout}, {Dupac}, {Efstathiou}, {Elsner}, {En{\ss}lin},
  {Eriksen}, {Falgarone}, {Fernandez-Cobos}, {Finelli}, {Forastieri},
  {Frailis}, {Fraisse}, {Franceschi}, {Frolov}, {Galeotta}, {Galli}, {Ganga},
  {G{\'e}nova-Santos}, {Gerbino}, {Ghosh}, {Gonz{\'a}lez-Nuevo}, {G{\'o}rski},
  {Gratton}, {Gruppuso}, {Gudmundsson}, {Handley}, {Hansen}, {Helou},
  {Herranz}, {Huang}, {Jaffe}, {Karakci}, {Keih{\"a}nen}, {Keskitalo},
  {Kiiveri}, {Kim}, {Kisner}, {Krachmalnicoff}, {Kunz}, {Kurki-Suonio},
  {Lagache}, {Lamarre}, {Lasenby}, {Lattanzi}, {Lawrence}, {Le Jeune},
  {Levrier}, {Liguori}, {Lilje}, {Lindholm}, {L{\'o}pez-Caniego}, {Lubin},
  {Ma}, {Mac{\'{\i}}as-P{\'e}rez}, {Maggio}, {Maino}, {Mandolesi}, {Mangilli},
  {Marcos-Caballero}, {Martin}, {Mart{\'{\i}}nez-Gonz{\'a}lez}, {Matarrese},
  {Mauri}, {McEwen}, {Meinhold}, {Melchiorri}, {Mennella}, {Migliaccio},
  {Miville-Desch{\^e}nes}, {Molinari}, {Moneti}, {Montier}, {Morgante},
  {Natoli}, {Oppizzi}, {Pagano}, {Paoletti}, {Partridge}, {Peel}, {Pettorino},
  {Piacentini}, {Polenta}, {Puget}, {Rachen}, {Reinecke}, {Remazeilles},
  {Renzi}, {Rocha}, {Roudier}, {Rubi{\~n}o-Mart{\'{\i}}n}, {Ruiz-Granados},
  {Salvati}, {Sandri}, {Savelainen}, {Scott}, {Seljebotn}, {Sirignano},
  {Spencer}, {Suur-Uski}, {Tauber}, {Tavagnacco}, {Tenti}, {Thommesen},
  {Toffolatti}, {Tomasi}, {Trombetti}, {Valiviita}, {Van Tent}, {Vielva},
  {Villa}, {Vittorio}, {Wandelt}, {Wehus}, {Zacchei}, \&
  {Zonca}}]{Planck2018_IV}
{Planck Collaboration} {et~al.}, 2018{\natexlab{b}}, submitted to A\&A
  [arXiv:1807.06208]

\bibitem[{{Platania} {et~al}\mbox{.}(1998){Platania}, {Bensadoun},
  {Bersanelli}, {de Amici}, {Kogut}, {Levin}, {Maino}, \&
  {Smoot}}]{Platania1998}
{Platania} P., {Bensadoun} M., {Bersanelli} M., {de Amici} G., {Kogut} A.,
  {Levin} S., {Maino} D., {Smoot} G.~F., 1998, \apj, 505, 473

\bibitem[{{Reich} \& {Reich}(1986)}]{Reich1986}
{Reich} P., {Reich} W., 1986, \aaps, 63, 205

\bibitem[{{Reich} \& {Reich}(1988)}]{Reich1988}
{Reich} P., {Reich} W., 1988, \aaps, 74, 7

\bibitem[{{Reich}, {Testori} \& {Reich}(2001){Reich}, {Testori}, \&
  {Reich}}]{Reich2001}
{Reich} P., {Testori} J.~C., {Reich} W., 2001, \aap, 376, 861

\bibitem[{{Reich}(1982)}]{Reich1982}
{Reich} W., 1982, \aaps, 48, 219

\bibitem[{{Remazeilles} {et~al}\mbox{.}(2015){Remazeilles}, {Dickinson},
  {Banday}, {Bigot-Sazy}, \& {Ghosh}}]{Remazeilles2015}
{Remazeilles} M., {Dickinson} C., {Banday} A.~J., {Bigot-Sazy} M.-A., {Ghosh}
  T., 2015, \mnras, 451, 4311

\bibitem[{{Remazeilles} {et~al}\mbox{.}(2016){Remazeilles}, {Dickinson},
  {Eriksen}, \& {Wehus}}]{Remazeilles2016}
{Remazeilles} M., {Dickinson} C., {Eriksen} H.~K.~K., {Wehus} I.~K., 2016,
  \mnras, 458, 2032

\bibitem[{{Ricci} {et~al}\mbox{.}(2013){Ricci}, {Righini}, {Verma}, {Prandoni},
  {Carretti}, {Mack}, {Massardi}, {Procopio}, {Zanichelli}, {Gregorini},
  {Mantovani}, {Gawro{\'n}ski}, \& {Peel}}]{Ricci2013}
{Ricci} R. {et~al.}, 2013, \mnras, 435, 2793

\bibitem[{{Righini} {et~al}\mbox{.}(2012){Righini}, {Carretti}, {Ricci},
  {Zanichelli}, {Mack}, {Massardi}, {Prandoni}, {Procopio}, {Verma},
  {L{\'o}pez-Caniego}, {Gregorini}, \& {Mantovani}}]{Righini2012}
{Righini} S. {et~al.}, 2012, \mnras, 426, 2107

\bibitem[{{Rubi{\~n}o-Mart{\'{\i}}n}
  {et~al}\mbox{.}(2012){Rubi{\~n}o-Mart{\'{\i}}n}, {L{\'o}pez-Caraballo},
  {G{\'e}nova-Santos}, \& {Rebolo}}]{Rubino-Martin2012a}
{Rubi{\~n}o-Mart{\'{\i}}n} J.~A., {L{\'o}pez-Caraballo} C.~H.,
  {G{\'e}nova-Santos} R., {Rebolo} R., 2012, Advances in Astronomy, 2012

\bibitem[{{Scaife} {et~al}\mbox{.}(2009){Scaife}, {Hurley-Walker}, {Green},
  {Davies}, {Franzen}, {Grainge}, {Hobson}, {Lasenby}, {Pooley},
  {Rodr{\'{\i}}guez-Gonz{\'a}lvez}, {Saunders}, {Scott}, {Shimwell},
  {Titterington}, {Waldram}, \& {Zwart}}]{Scaife2009}
{Scaife} A.~M.~M. {et~al.}, 2009, \mnras, 400, 1394

\bibitem[{{Schlegel}, {Finkbeiner} \& {Davis}(1998){Schlegel}, {Finkbeiner}, \&
  {Davis}}]{Schlegel1998}
{Schlegel} D.~J., {Finkbeiner} D.~P., {Davis} M., 1998, \apj, 500, 525

\bibitem[{{Seon} \& {Witt}(2012)}]{Seon2012}
{Seon} K.-I., {Witt} A.~N., 2012, \apj, 758, 109

\bibitem[{{Silsbee}, {Ali-Ha{\"i}moud} \& {Hirata}(2011){Silsbee},
  {Ali-Ha{\"i}moud}, \& {Hirata}}]{Silsbee2011}
{Silsbee} K., {Ali-Ha{\"i}moud} Y., {Hirata} C.~M., 2011, \mnras, 411, 2750

\bibitem[{{Strong}, {Orlando} \& {Jaffe}(2011){Strong}, {Orlando}, \&
  {Jaffe}}]{Strong2011}
{Strong} A.~W., {Orlando} E., {Jaffe} T.~R., 2011, \aap, 534, A54

\bibitem[{{Sutton} {et~al}\mbox{.}(2010){Sutton}, {Zuntz}, {Ferreira}, {Brown},
  {Eriksen}, {Johnson}, {Kusaka}, {N{\ae}ss}, \& {Wehus}}]{Sutton2010}
{Sutton} D. {et~al.}, 2010, \mnras, 407, 1387

\bibitem[{{Tibbs} {et~al}\mbox{.}(2011){Tibbs}, {Flagey}, {Paladini},
  {Compi{\`e}gne}, {Shenoy}, {Carey}, {Noriega-Crespo}, {Dickinson},
  {Ali-Ha{\"\i}moud}, {Casassus}, {Cleary}, {Davies}, {Davis}, {Hirata}, \&
  {Watson}}]{Tibbs2011}
{Tibbs} C.~T. {et~al.}, 2011, \mnras, 418, 1889

\bibitem[{{Tibbs}, {Paladini} \& {Dickinson}(2012){Tibbs}, {Paladini}, \&
  {Dickinson}}]{Tibbs2012b}
{Tibbs} C.~T., {Paladini} R., {Dickinson} C., 2012, Advances in Astronomy,
  2012, 124931

\bibitem[{{Tibbs} {et~al}\mbox{.}(2013){Tibbs}, {Scaife}, {Dickinson},
  {Paladini}, {Davies}, {Davis}, {Grainge}, \& {Watson}}]{Tibbs2013}
{Tibbs} C.~T., {Scaife} A.~M.~M., {Dickinson} C., {Paladini} R., {Davies}
  R.~D., {Davis} R.~J., {Grainge} K.~J.~B., {Watson} R.~A., 2013, \apj, 768, 98

\bibitem[{{Turtle} {et~al}\mbox{.}(1962){Turtle}, {Pugh}, {Kenderdine}, \&
  {Pauliny-Toth}}]{Turtle1962}
{Turtle} A.~J., {Pugh} J.~F., {Kenderdine} S., {Pauliny-Toth} I.~I.~K., 1962,
  \mnras, 124, 297

\bibitem[{{Watson} {et~al}\mbox{.}(2005){Watson}, {Rebolo},
  {Rubi{\~n}o-Mart{\'{\i}}n}, {Hildebrandt}, {Guti{\'e}rrez},
  {Fern{\'a}ndez-Cerezo}, {Hoyland}, \& {Battistelli}}]{Watson2005}
{Watson} R.~A., {Rebolo} R., {Rubi{\~n}o-Mart{\'{\i}}n} J.~A., {Hildebrandt}
  S., {Guti{\'e}rrez} C.~M., {Fern{\'a}ndez-Cerezo} S., {Hoyland} R.~J.,
  {Battistelli} E.~S., 2005, \apjl, 624, L89

\bibitem[{{Weiland} {et~al}\mbox{.}(2011){Weiland}, {Odegard}, {Hill},
  {Wollack}, {Hinshaw}, {Greason}, {Jarosik}, {Page}, {Bennett}, {Dunkley},
  {Gold}, {Halpern}, {Kogut}, {Komatsu}, {Larson}, {Limon}, {Meyer}, {Nolta},
  {Smith}, {Spergel}, {Tucker}, \& {Wright}}]{Weiland2011}
{Weiland} J.~L. {et~al.}, 2011, \apjs, 192, 19

\bibitem[{{Williams}, {Bureau} \& {Cappellari}(2010){Williams}, {Bureau}, \&
  {Cappellari}}]{Williams2010}
{Williams} M.~J., {Bureau} M., {Cappellari} M., 2010, \mnras, 409, 1330

\bibitem[{{Winkel} {et~al}\mbox{.}(2016){Winkel}, {Kerp}, {Fl{\"o}er},
  {Kalberla}, {Ben Bekhti}, {Keller}, \& {Lenz}}]{Winkel2016}
{Winkel} B., {Kerp} J., {Fl{\"o}er} L., {Kalberla} P.~M.~W., {Ben Bekhti} N.,
  {Keller} R., {Lenz} D., 2016, \aap, 585, A41

\bibitem[{{Witt} {et~al}\mbox{.}(2010){Witt}, {Gold}, {Barnes}, {DeRoo},
  {Vijh}, \& {Madsen}}]{Witt2010}
{Witt} A.~N., {Gold} B., {Barnes}, III F.~S., {DeRoo} C.~T., {Vijh} U.~P.,
  {Madsen} G.~J., 2010, \apj, 724, 1551

\bibitem[{{Wood} \& {Reynolds}(1999)}]{Wood1999}
{Wood} K., {Reynolds} R.~J., 1999, \apj, 525, 799

\end{thebibliography}

\label{lastpage}

\end{document}